\documentclass[smallextended]{svjour3}       
\smartqed  
\usepackage{graphicx}
\usepackage{natbib}  
\usepackage[colorlinks=true,citecolor=blue,urlcolor=blue,breaklinks]{hyperref}
\usepackage{longtable}
 \newcommand{\lae}{\mathrel{<\kern-1.0em\lower0.9ex\hbox{$\sim$}}}
 \newcommand{\gae}{\mathrel{>\kern-1.0em\lower0.9ex\hbox{$\sim$}}}

 \newcommand{\mone}{$^{-1}$} 
 \newcommand{\mtwo}{$^{-2}$}

\journalname{The Astronomy and Astrophysics Review}
\begin{document}

\title{Compact steep-spectrum and peaked-spectrum radio sources}



\author{Christopher P.\ O'Dea \and
        D.~J. Saikia
}


\institute{C.~P. O'Dea \at
              Dept of Physics \& Astronomy, University of Manitoba, Winnipeg MB R3T 2N2 Canada \\
              Tel.: +1 204-474-9863 \\
              Fax: +1 204-474-7622 \\
              \email{odeac@umanitoba.ca}           
           \and
           D.~J. Saikia \at
Inter-University Centre for Astronomy and Astrophysics, Savitribai Phule Pune University Campus,  Ganeshkind P.O., Pune  411007, India\\
\email{dhrubasaikia.tifr.ccsu@gmail.com, dhrubasaikia@iucaa.in}
}


\maketitle

\begin{abstract}
Compact steep-spectrum (CSS) and peaked spectrum (PS) radio sources are compact, powerful radio sources. The multi-frequency observational properties and current theories are reviewed with emphasis on developments since the earlier review of O'Dea (1998). There are three main hypotheses for the nature of PS and CSS sources. (1) The PS sources might be very young radio galaxies which will evolve into CSS sources on their way to becoming large radio galaxies. (2) The PS and CSS sources might be compact because they are confined (and enhanced in radio power) by interaction with dense gas in their environments. (3) Alternately, the PS  sources might be transient or intermittent sources. Each of these hypotheses may apply to individual objects. The relative number in each population will have  significant implications for the radio galaxy paradigm. Proper motion studies over long time baselines have helped determine  hotspot speeds for over three dozen sources and establish that these are young objects. Multifrequency polarization observations have demonstrated that many CSS/PS sources are embedded in a dense interstellar medium and vigorously interacting with it. The detection of emission line gas aligned with the radio source, and blue-shifted H{\sc i} absorption and [OIII] emission lines indicates that AGN feedback is present in these objects -- possibly driven by the radio source. Also, CSS/PS sources with evidence of episodic AGN over a large range of time-scales have been discussed. The review closes with a discussion of open questions and prospects for the future. 

\end{abstract}

\setcounter{tocdepth}{3}
\tableofcontents

\section{Introduction}\label{s:intro}
Among the different types of active galactic nuclei (AGN), the radio luminous extragalactic radio sources associated with massive elliptical galaxies and referred to as radio AGN, continue to pose a number of interesting astrophysical questions \citep[e.g.,][]{Padovani2017}. These include the triggering of radio AGN activity, the formation of jets squirting out from the AGN with relativistic speeds, its effect on the host galaxy and its subsequent evolution, and relationship of the different radio and
optical classification schemes to modes of accretion onto the
supermassive black holes which are believed to fuel the AGN activity \citep[e.g.,][]{Tadhunter2016b,Hardcastle2020}. Radio AGN extend in their overall projected linear size from parsec-scales to over a few Mpc, with the largest known radio galaxy so far having a size of about 4.7 Mpc \citep{Machalski2008}. The extended sources inclined at small angles to the line of sight may appear small due to projection effects. Such sources tend to have dominant flat-spectrum radio cores over a large frequency range due to relativistic beaming of the nuclear jets moving at velocities close to that of light. 
A flat-spectrum is defined to have a spectral index $\alpha <0.5$ where flux density, S, varies with frequency $\nu$ as S$\propto \nu ^{-\alpha}$.
The flat and complex spectra of the dominant cores are largely due to synchrotron self absorption in the radio core and knots of emission in the nuclear jets.  Intrinsically small sources are however not generally dominated by their nuclear or core radio emission, and are believed to represent the early stages of evolution of radio AGN while some may be confined to small dimensions by a dense interstellar medium \citep{ODea1998}. 

{\bf 
\begin{table*}
	\caption{Nomenclature for Compact 
	Radio Sources}
	\begin{tabular}{l l  c  l l }
  \hline
 Acronym & Text  &  Definition  &  ref & Notes\\
 \hline
  GPS & GHz-Peaked Spectrum & $0.5 \leq \nu_p \leq 5$ GHz &
  1,2 & a \\
  HFP & High Frequency Peakers & $\nu_p > 5 $ GHz & 3 & a \\
  MPS & MHz-Peaked Spectrum & $\nu_p < 1 $ GHz & 4 & \\
  PS & Peaked Spectrum & All GPS, HFP, MPS &  5 & \\
   & & & & \\
  CSO & Compact Symmetric Object & $LS < 1 $ kpc & 6,7 & b \\
  CSS & Compact Steep Spectrum & $1 \leq LS \leq 20$ kpc & 8,9,10 & c \\
  MSO & Medium Symmetric Object & $1 \leq LS \leq 20$ kpc &  11 & d \\
    \hline
			\end{tabular}
      \label{t:definitions}
      
 Columns. (1) Acronym. (2) Text. (3) Definition. $\nu_p$ is the observed frequency of the peak in the spectrum. LS is projected linear size.
 (4) Reference for initial recognition of this class of sources and/or use of the acronym. (5) Notes. a. The GPS and HFP are also very small ($< 500$ pc in size).  b. The CSOs tend to have a core and two lobes and are preferentially associated with GPS and HFP galaxies, while core-jet morphology is more typical in GPS and HFP quasars. The description  ``compact symmetric object" replaced the earlier description  ``compact double" which was suggested by \citet{Phillips1982}. c. The CSS sources also have a requirement for a steep spectrum at high frequencies ($\alpha \geq 0.5$). 
 d. The MSOs and CSS sources are really the same objects, but the name MSO is meant to emphasize that they are larger versions of the CSOs. 
References. 1. \cite{Spoelstra1985}; 2. \cite{GPS1983}; 3. \cite{Dallacasa2000}; 4. \cite{Coppejans2015}; 5. This work; 6. \cite{Wilkinson1994}; 7. \cite{Readhead1994}; 
8. \cite{Fanti1985}; 9. \cite{Peacock1982}; 10. \cite{Kapahi1981}; 11. \cite{Fanti1995}. 
 
 \end{table*}
 }
 
The radio spectra of these small sources may appear peaked either due to synchrotron self absorption or free-free absorption. However their spectra are steep and optically thin above the peak frequency.
The GHz-Peaked-Spectrum (GPS) sources are selected to have their radio spectra dominated by a peak in the flux density around 1 GHz, 
in practice in the range $\sim 500$ MHz to $\sim 5$ GHz \citep{GPS1983}; though \citet{Stanghellini1998b} used a slightly larger range of 400 MHz to 6 GHz. 
Sources that peak above 5 GHz are called High Frequency Peakers (HFPs)  \citep{Dallacasa2000}. Here we will refer to GPS and HFP sources together as Peaked-Spectrum (PS) sources. See Table \ref{t:definitions} for a summary of common nomenclature.

The frequency of the peak in a synchrotron self-absorbed source scales inversely with the angular size of the source as $\nu_m \propto \theta^{-4/5}$ (see Sect.~\ref{s:radio}). Thus, the 
requirement for a peak at high frequencies selects compact radio sources. 
The PS sources tend to have projected linear sizes less than 500 pc while Compact Steep Spectrum (CSS) sources tend to have sizes between 500 pc and 20 kpc 
\citep[e.g.,][]{ODea1998,Fanti1990}.
Compact radio sources (called Compact Symmetric Objects or CSOs) have been defined to be those with radio lobe emission on both sides of an active nucleus and an overall size less than about a kpc. CSOs are found directly via VLBI imaging surveys  \citep[e.g.,][]{Wilkinson1994,Peck2000b}.  

Sources which peak at frequencies below 400 MHz are the Compact Steep Spectrum (CSS) sources though they are not selected
specifically on the basis of the location of the spectral peak
\citep{Fanti1990}.  
The selection criteria of the Fanti sample of CSS sources include projected linear size less than 20 kpc and flux density at 178 MHz greater than 10 Jy \citep{Fanti1990}. The requirement that the sources are bright  at 178 MHz tends to select sources which peak below $\sim 400$ MHz. Peaked-spectrum or PS sources with a peak below about 1 GHz in the observer's frame have also been referred to as Megahertz Peaked-Spectrum or MPS sources \citep{Coppejans2015, Coppejans2016a, Coppejans2016b}. The MPS sources are likely to be a combination of relatively nearby CSS and/or GPS sources, or more distant, compact high-frequency peakers whose peak frequency has been shifted to low frequencies due to the cosmological redshift.

The turnover in the radio spectrum in CSS and PS sources likely requires an absorption mechanism \citep[e.g.,][]{deKool1989, ODea1998,Tingay2003a}. The physics of the absorption mechanism provides important constraints on physical conditions in and around the radio source which are typically not available for much larger radio sources. 

The compact sizes mean that the radio sources interact with their host galaxy interstellar medium (ISM) on a range of size scales ranging from parsecs to about 10 kpc. Energy input from radio  sources (radio mode feedback) may be important in regulating star formation in massive galaxies and determining the shape of the high end of the galaxy luminosity function and also determining the balance of heating and cooling in the intracluster medium (ICM)
 \citep[e.g.,][]{Croton2006,Fabian2012}.  As the PS and CSS sources are propagating through their host galaxies, they can provide examples of  radio mode feedback \citep[e.g.,][]{Holt2008,Morganti2013,Tadhunter2016a, Hardcastle2020}. Evidence of interaction with the cold component of the interstellar medium in these galaxies via H{\sc i} absorption line studies has been reviewed recently by \citet{Morganti2018}. 

There are three main  hypotheses for the nature of PS and CSS sources. (1) The PS sources might be very young radio galaxies which will evolve into CSS sources on their way to becoming large radio galaxies  \citep[e.g.,][]{Fanti1995,An2012a}.
(2) The PS and CSS sources might be compact because they are confined by interaction with dense gas in their environments \citep[e.g.,][]{vanBreugel1984b,Dicken2012}. (3) Alternately,
the PS sources might be transient or intermittent sources \citep[e.g.,][]{Readhead1993,Reynolds1997}.
Each of these hypotheses may apply to individual objects. Determining which hypotheses apply to which fractions of the population  will have important implications for radio source origin and evolution.

In this review of PS and CSS sources we provide a comprehensive overview of the observational properties and theoretical understanding of these interesting objects. 
We will focus mainly on progress since the review of \citet{ODea1998}.
We discuss the current samples in Sect.~\ref{s:samples}, radio properties in Sect.~\ref{s:radio}, infrared properties in Sect.~\ref{s:IR}, host galaxies in Sect.~\ref{s:hosts}, high energy properties in Sect.~\ref{s:highenergy}. A discussion of current topics of interest and the nature of the PS and CSS sources is presented in Sect.~\ref{s:disc} and potential future work is presented in Sect.~\ref{s:future}.   New topics include the relation of PS and CSS sources with FR0 sources (Sect.~\ref{s:fr0_sources}) and narrow-line Seyfert 1s (Sect.~\ref{s:NLS1}), black hole masses and accretion rates  (Sect.~\ref{s:BH}), $\gamma$-ray emission (\S\ref{s:gamma}), a possible association between compact radio sources and star formation (Sect.~\ref{S:SFR}), and the role of compact sources in AGN feedback (Sect.~\ref{s:feedback}).

We adopt $H_0 = 70$~km~s$^{-1}$~Mpc$^{-1}$, $\Omega_{m_0} = 0.3$, $\Omega_{\Lambda_0} = 0.7$,  and 20 kpc as the limiting size for CSS/PS sources.

\section{Samples}\label{s:samples}

Among the most significant developments in  recent years have been surveys at low radio frequencies using telescopes which are not limited by narrow bandwidths, leading to increased sensitivity and more reliable determination of their low-frequency spectra. These include the GaLactic and Extragalactic All-sky Murchison Widefield Array survey \citep[GLEAM;][]{Wayth2015}, the LOFAR Multifrequency Snapshot Sky Survey \citep[MSSS;][]{Heald2015}, and more recently the LOFAR Two-Metre Sky Survey \citep[LoTSS;][]{Shimwell2017, Shimwell2019}, and the TIFR GMRT Sky Survey \citep[TGSS ADR1;][]{Intema2017}. At higher frequencies complete samples could be constructed in the near future at significantly lower flux density levels than has so far been possible from surveys with upgraded or newly constructed telescopes such as the Very Large Array Sky Survey \citep[VLASS;][]{Myers2019,Lacy2020}; the Evolutionary Map of the Universe \citep[EMU;][]{Norris2011} with the Australian Square Kilometre Array Pathfinder (ASKAP) and MeerKAT International GigaHertz Tiered Extragalactic Exploration \citep[MIGHTEE;][]{Jarvis2016}.

These surveys when they come on-line will lead to much larger samples of CSS and PS sources. For this review we briefly summarize those that have been compiled and/or studied since the review of \citet{ODea1998}. We first describe briefly the different samples, starting with the low-frequency ones, and then discuss what we have learnt from studies of sources in these surveys. 

\paragraph{GLEAM sample of peaked sources}~\\
\citet{Callingham2017} presented a sample of 1483 peaked-spectrum sources identified from the GLEAM extragalactic catalogue consisting of 307,456 sources  with 20 flux density measurements between 72 and 231 MHz. They also used flux densities from the NRAO VLA Sky Survey at 1400 MHz \citep[NVSS;][]{Condon1998} and Sydney University Molonglo Sky Survey at 843 MHz \citep[SUMSS;][]{Bock1999, Mauch2003} to determine the overall spectra reliably and be sensitive to PS sources with peaks between 72 MHz and 843/1400 MHz. 
 \citet{Coppejans2015} found 24 of the Megahertz PS sources with redshift measurements in their sample to have an average redshift of 1.3, with four having a redshift $>$2, and a further four fainter objects without a redshift measurement being possibly even farther. They suggest that highly compact MPS sources could be a useful way of finding high-redshift radio-loud AGN.

\paragraph{B3/VLA sample}~\\
\citet{Fanti2001} presented a new sample of 87 CSS objects selected from the B3-VLA survey with a flux density S$_{\rm 408 MHz} >$ 0.8 Jy. The B3-VLA sample consisting of 1049 sources \citep{Vigotti1989} is expected to contain about 275 CSS/GPS objects, although \citet{Fanti2001} have focussed on the bright subsample as these had more complete spectroscopic information.

\paragraph{Sample from the B2 survey}~\\
A small sample of 19 candidate CSSs from a complete sample of 52 sources of intermediate strength selected from the B2.1 catalogue with flux densities in the range $0.9 \leq$ S$_{\rm 408} <$ 2.5 Jy \citep{Padrielli1981}, was studied by \citet{Saikia2002}.

\paragraph{CORALZ sample}~\\
A sample of Compact Radio sources at Low Redshift (CORALZ) consisting of 28 sources, with S$_{\rm 1400 MHz} >$ 100 mJy and angular size $<$2 arcsec, was compiled by \citet{Snellen2004}. These were selected from the FIRST survey and cross-correlated with the APM Palomar Sky Survey catalogue, and span the redshift range $0.005<z<0.16$. Although these have not been selected on the basis of their spectral indices, over 90 per cent of the objects are either CSS or GPS objects.

\paragraph{Additional FIRST-based samples}~\\
To probe the nature of low-luminosity CSS objects, weaker samples were compiled from the FIRST survey. \citet{Kunert2002} started with the Green Bank (GB) survey \citep{White1992} with $\alpha_{1.4}^{5} > 0.5$ and S$_{\rm 5 GHz} >$ 150 mJy in a region which overlaps with the FIRST survey. They listed 60 CSS sources which are compact with the FIRST beam after excluding those with a peaked spectrum. 

Using a somewhat similar approach, \citet{KB2009} considered the GB6 survey at 4.85 GHz and the FIRST survey and compiled a sample of 44 CSS objects with $\alpha_{1.4}^{4.85} > 0.7$, 70 mJy $\leq$ S$_{\rm 1400 MHz} \leq$ 1 Jy and appearing compact with the FIRST beam. This sample too focused on CSS sources.

\paragraph{Sample of GPS sources based on the Parkes survey}~\\
\citet{Snellen2003} compiled a sample of 49 GPS sources from the PKSCAT90 survey \citep{Otrupcek1991} with S$_{\rm 2700 MHz} >$ 0.5 Jy and located in the declination range $-40^\circ < \delta < +15^\circ$ and Galactic latitude $\mid \rm{b} \mid > 20^\circ$. This is the southern counterpart to GPS samples in the northern hemisphere. All the GPS sources in this sample are identified with galaxies with redshifts in the range from 0.17 to 1.54 \citep{deVries2007}. 

\paragraph{The GPS 1-Jy sample}~\\
A complete sample of 33 GPS sources with S$_{\rm 5000 MHz} >$ 1 Jy, a turn-over frequency in the range 0.4 to 6 GHz with $\alpha_{\rm thin} > 0.5$, and with declination $\delta > -25^\circ$ and $\mid \rm{b} \mid > 10^\circ$ was compiled by \citep{Stanghellini1998b}. 

\paragraph{CSSs from the S4 survey}~\\
\citet{Saikia2001} observed a sample of candidate CSSs from the S4 survey \citep{Pauliny-Toth1978} with S$_{\rm 5000 MHz} \geq $ 0.5 Jy, which were earlier found to be either unresolved or partially resolved with the Westerbork Synthesis Radio Telescope \citep{Kapahi1981}.

\paragraph{The COINS sample}~\\
From large-scale VLBI continuum surveys, \citet{Peck2000a} identified a sample of 52 compact symmetric objects or CSOs observed in the northern sky (COINS) with S$_{\rm 5000 MHz}$ larger than about 100 mJy. This sample is not based on the spectra of the sources but on the observed VLBI-scale structure of the objects.   

\paragraph{The VIPS sample}~\\
From the Very Long Baseline Array (VLBA) Imaging and Polarization Survey \citep[VIPS;][]{Taylor2005,Helmboldt2007} of 1127 sources, \citet{Tremblay2009} compiled a sample of 103 CSOs, making it one of the largest samples.

\paragraph{High-frequency peakers: GB and NVSS/FIRST}~\\
To identify sources where the peak of the spectrum may be higher than a few GHz, sources with a rising spectrum between 1.40 and 4.85 GHz were identified by cross-correlating the 87GB and NVSS catalogues for a `bright' sample \citep{Dallacasa2000}. These were selected to have S$_{\rm 5000 MHz} >$ 300 mJy. Similarly a `faint' sample was compiled by comparing with the FIRST survey at 1400~MHz and limiting it to those with S$_{\rm 5000 MHz} >$ 50 mJy \citep{Dallacasa2002c,Orienti2012,Orienti2020}.

\paragraph{High-frequency peakers: AT20G sample}~\\
From the Australia Telescope Compact Array survey at 20~GHz (AT20G survey) consisting of 5450 radio sources, \citet{Hancock2009a} compiled a sample of 656 radio sources with peaks above 5~GHz. Of these 466 were identified with optical objects, 104 with galaxies and 362 with quasars. HFPs are discussed further in Sect.~\ref{s:mm}.

\medskip
 Several of the surveys summarized here extend the radio power  of the compact sources to significantly lower values since the review by \citet{ODea1998}. For example the log of median radio power  of \citet{ODea1998} samples at 5~GHz in units of W Hz$^{-1}$ consisting of bright 3CR sources \citep{Fanti1990} and PS sources \citep{Stanghellini1998b} is 27.6. For a sample of faint PS sources compiled by \citet{Snellen1998b}, the log of median radio power  at 5~GHz of 26.2 is marginally lower than that for the GLEAM sample which has a value of 26.5. The radio power of the B3-VLA sample selected at 408 MHz has a similar value of 26.6 when extrapolated to 5~GHz assuming a spectral index of 1. As one goes to fainter samples, lower-luminosity sources are selected. For example in the CORALZ sample \citep{Snellen2004}, the log of radio power at 5~GHz in W Hz$^{-1}$ ranges from 22.96 to 25.03 with a median value of about 24.4. The log of median radio power  of the galaxies in the AT20G-6dFGS sample at 1.4~GHz is 24.5 \citep{Sadler2016}, corresponding to 23.95 at 5~GHz assuming a spectral index of 1. The log of radio power  of the FR0 sources many of which are likely to be CSS and PS sources range from 22 to 24 at 1.4~GHz in units of W Hz$^{-1}$ \citep{Capetti2020,Baldi2018}. 

The local 1.4 GHz radio luminosity function of AGN extends down to about $\log_{10} \sim 20.5\mathrm{\ W\ Hz}^{-1}$ \citep{Mauch2007}. The current samples of PS and CSS have radio powers which extend down to about $\log_{10} \sim 23\mathrm{\ W\ Hz}^{-1}$ which is within about 1.5  orders of magnitude of the faintest radio AGN. The distribution of 1.4 GHz radio powers of Seyfert galaxies range from $\log_{10} \sim 20.5$--$24.0\mathrm{\ W\ Hz}^{-1}$ \citep[e.g.,][]{Ulvestad1989}. There is overlap of the PS and CSS radio powers with  the high power end of the Seyfert galaxies. Thus, the physics of the PS and CSS radio sources may extrapolate to the faintest radio AGN, including Seyferts.

\section{Radio properties}\label{s:radio}

\begin{figure*}
	\centering
	\hbox{ 
		\includegraphics[width=6.5cm]{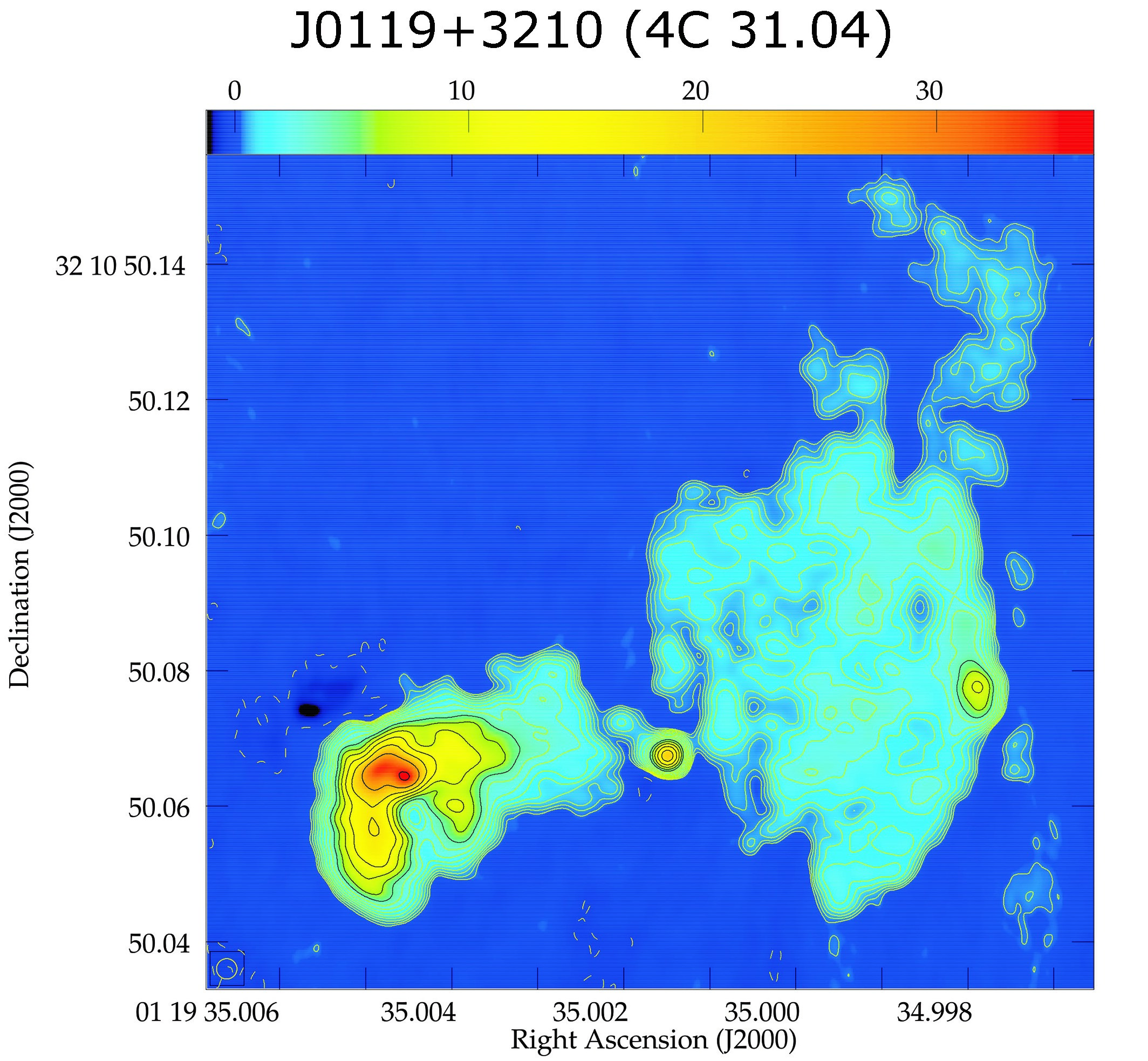}
		\includegraphics[width=6.0cm]{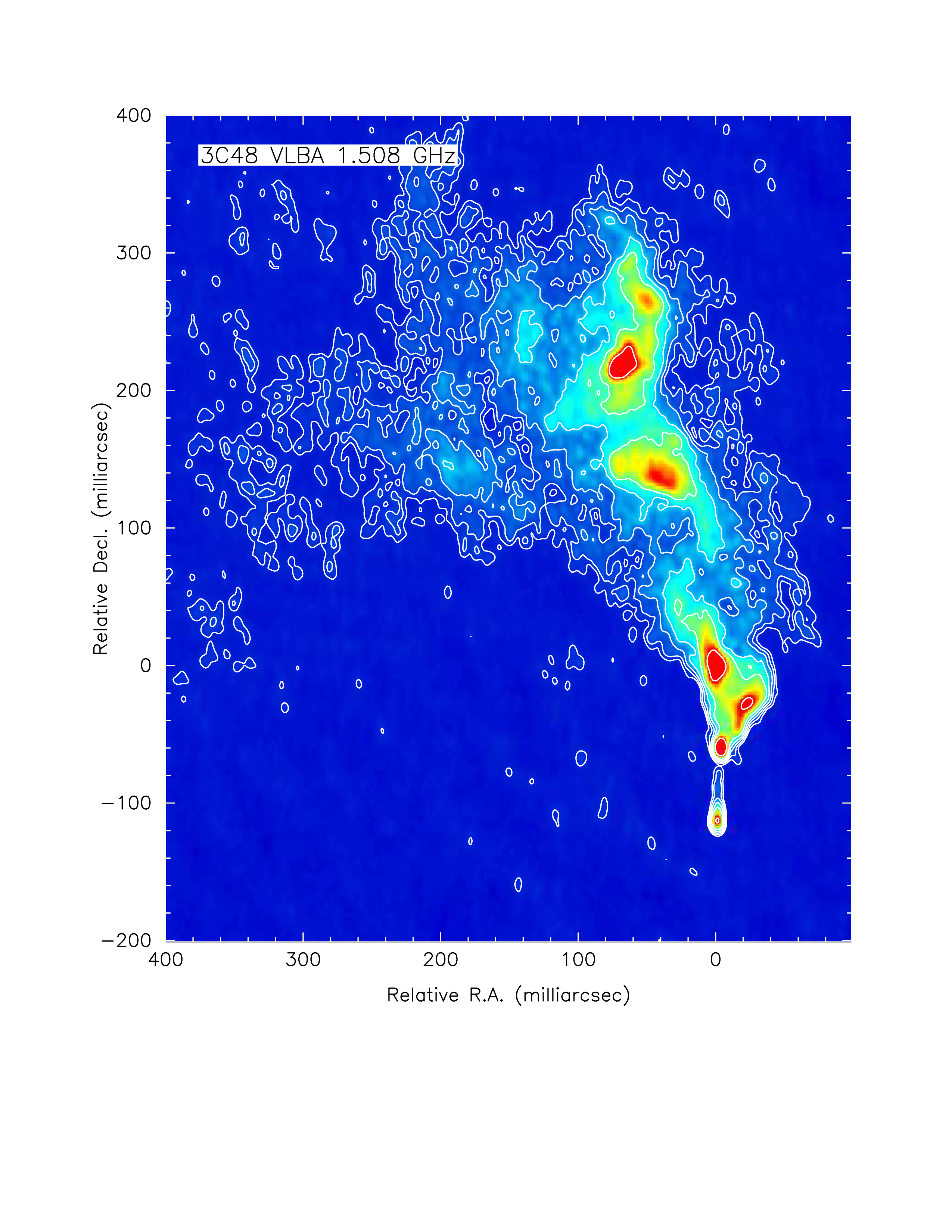}
	}
	\hbox{
		\includegraphics[width=6.0cm]{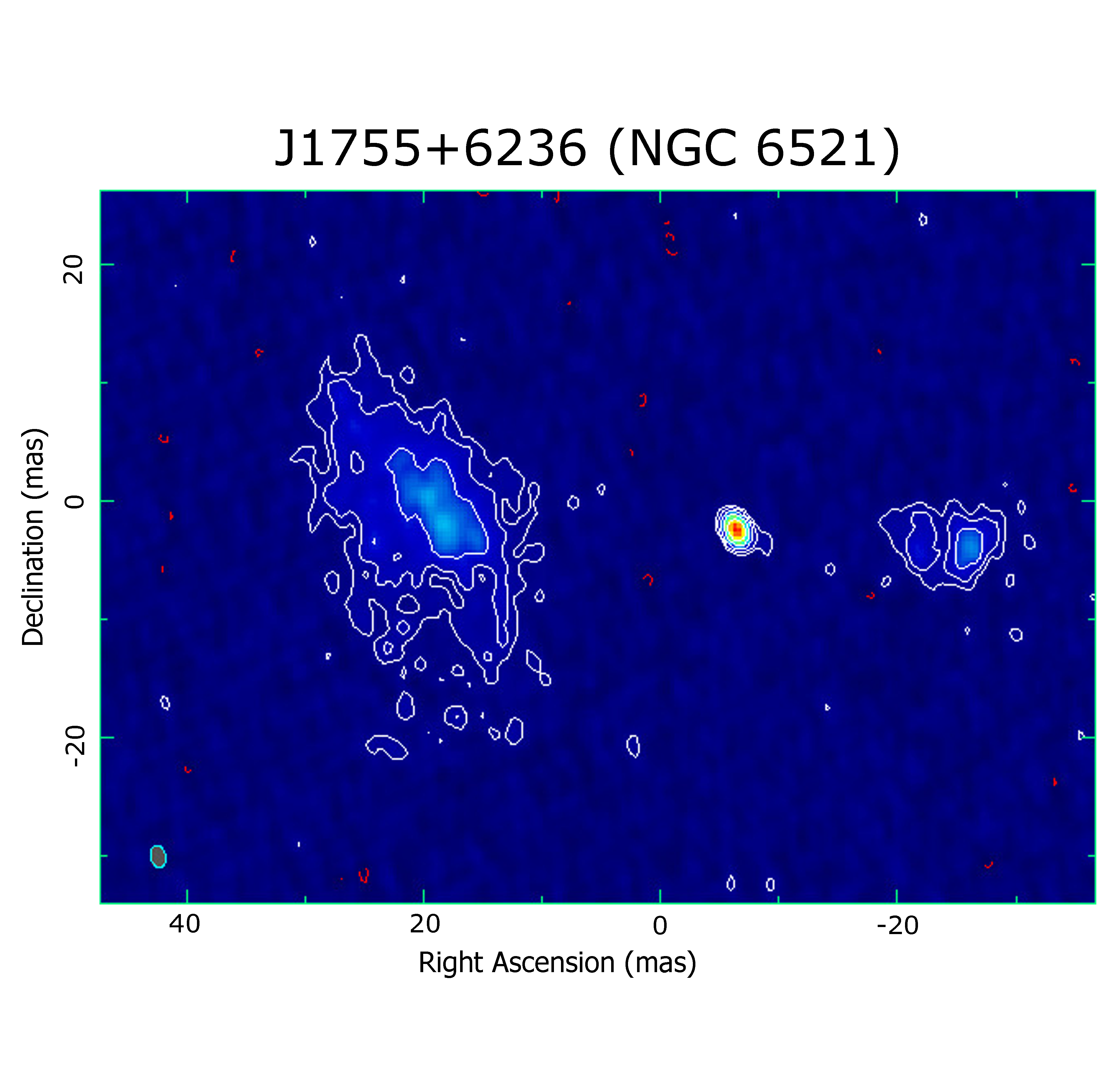}
		\includegraphics[width=5.5cm]{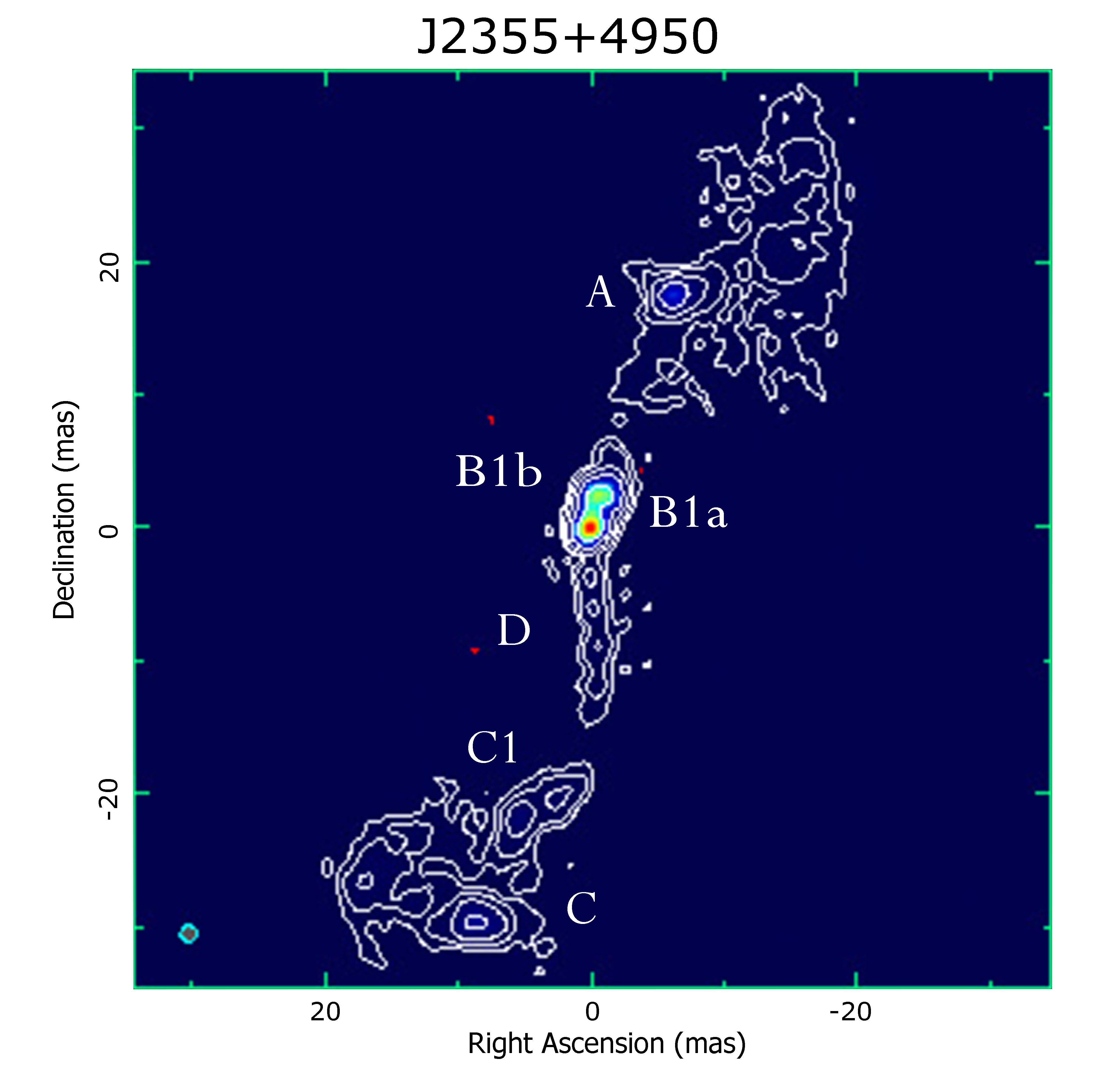}
	}	
	\caption{Examples of structures of CSS/PS sources. Upper left: The CSO 4C31.04 (J0119+3210) which is associated with a galaxy at a redshift of 0.0602, has a peaked spectrum, and shows bright compact hotspots and extended emission on both sides along with a radio core \citep{Giroletti2003}. The projected linear size defined by the outer peaks of emission is 107 pc. Upper right: The CSS quasar 3C48 (J0137+3309) at a redshift of 0.367 which shows a disrupted jet indicating strong interaction with the interstellar medium of the host quasar \citep{An2010}. The total projected size seen in this image is 2.6 kpc from the nucleus, although an extended envelope extends to about 5 kpc from the nucleus, the southernmost component seen in this image \citep{An2010}.  Lower left: The CSS galaxy NGC6521 (J1755+6236) at a redshift of 0.0275 shows diffuse outer lobes with no prominent hot spots indicating that the energy supply may have stopped \citep{Polatidis2009}. {The total projected size of the source defined by the outer lobes is 29 pc.} Lower right: J2355+4950 associated with a galaxy at a redshift of 0.2379 shows hot spots along with diffuse emission on opposite sides and a well-defined radio jet \citep{Polatidis1999,Taylor2000,Polatidis2003,Polatidis2009}. {The total projected size of the source defined by the outer lobes is 240 pc. }   }
 	\label{f:css_structures}
\end{figure*} 

\subsection{Radio structures}\label{s:radiostructure}
A study of the structures of radio sources could provide valuable insights towards understanding the formation, evolution and interaction with the ambient medium of these compact radio sources. For example if the PS and CSS sources are young objects advancing outwards through a dense inhomogeneous and asymmetric external medium on opposite sides of the nucleus, one may find evidence of this in their structural and polarization asymmetries. On the other hand for sources confined to small dimensions by a dense interstellar medium, one may find evidence of disruption of a radio jet as seen in 3C48, or deflection of a jet to form complex structures. To clarify the structures of these sources, one needs high-resolution observations ranging from mas scale for the most compact ones to the sub-arcsec scales. Figure \ref{f:css_structures} illustrates some of the structures observed in CSS/PS sources which include double-lobed sources with or without a detected radio core but well-defined hotspot(s) which could sometimes be significantly misaligned, jets which may be disrupted or appear quite collimated, as well as lobes which appear diffuse with no significant hotspots in them.  

The extended structure of radio sources has traditionally been classified into the Fanaroff--Riley or FR Class I and II sources. The latter are characterized by prominent hotspots at the outer edges and narrow collimated asymmetric jets, while the former have more diffuse lobes of emission with more symmetric jets on kpc scales, and are less luminous than the FRII type sources. \citet{Ledlow1996b} suggested that the dividing luminosity between these two classes is not fixed but increases with optical luminosity of the host galaxy, although \citet{Best2009}, \citet{Wing2011}, \citet{Gendre2013} and \citet{Mingo2019} have questioned this sharp division. From a large sample of 5805 sources from the LoTSS survey \citet{Mingo2019} find FRII sources about 3 orders of magnitude lower in luminosity compared to the traditional FR break. They also find the host galaxies of these 
 low-luminosity FRIIs to be fainter than high-luminosity FRIIs and FRIs matched in luminosity. The FRI/II classification has traditionally been based on radio images of sources initially from the 3CR but generally from bright radio source samples. Deep sensitive surveys at radio frequencies as well as finding radio counterparts from optical surveys such as the Sloan Digital Sky Survey (SDSS) have also revealed sources which have similar core luminosity to those of FRI sources, but the extended emission is weaker by factors of $\sim$100 \citep{Baldi2009}. These have been christened as FR0s, and their relationship to the more extended sources need to be understood \citep{Baldi2015,Sadler2016,Cheng2018}. The FR0 sources, a significant fraction of which are likely to be CSS/PS sources are discussed further in Sect.~\ref{s:fr0_sources}.

Based on their optical spectra radio AGN have been classified into High Excitation Radio Galaxies (HERGs) and Low Excitation Radio Galaxies (LERGs) which appear related to different modes of fuelling the AGN \citep{Hardcastle2007, Buttiglione2010, Best2012, Heckman2014, Tadhunter2016b}. This has been discussed further in Sect.~\ref{s:nuclear}. The LERGs tend to be of lower luminosity than the HERGs, and have an FRI-type structure while the radio structure of HERGS are predominantly of FRII type \citep{Best2012,Heckman2014,Tadhunter2016b}.

Another significant input in our understanding of AGN during the last decade has been the Wide-field Infrared Survey Explorer (WISE) measurements at 3.4 (W1), 4.6 (W2), 12 (W3) and 22 (W4) $\mu$m \citep{Wright2010}. A colour-colour plot W1$-$W2 vs W2$-$W3 \citep[e.g.,][]{Wright2010,Donoso2012} can help distinguish between quiescent galaxies, star-forming ones and different types of AGN. 
The WISE bands provide valuable information on host galaxies, tracing stellar light as well as from the interstellar medium associated with star formation. {While the WISE W1 and W2 bands appear to be good tracers of stellar mass distribution, longer wavelengths trace cooler emission from the interstellar medium \citep[e.g.,][]{Jarrett2013}. The W2$-$W3 colour correlates well with the specific star formation rate with the stronger or more luminous AGN being close to the high star-forming galaxies. Weaker or less luminous AGN extend the relationship to lower specific star formation rate and smaller values of W2$-$W3 \citep[e.g.,][]{Donoso2012}. } 
For those referred to as WISE early-type with W2$-$W3$<$2 most are LERGs in samples of sources of intermediate radio luminosity, while those with W2$-$W3$>$2 are a mixture of both LERGs and HERGs, but a majority appear to be HERGs. The brighter or more luminous source samples although contain both LERGs and HERGs are dominated by the HERGs \citep[e.g.,][]{Hardcastle2007,Tadhunter2016b}. 

Early studies of CSS and PS sources, which includes the CSOs, usually defined to be two-sided with an overall projected linear size $<$1 kpc and the medium symmetric objects or MSOs defined to be $>$1 kpc, were based on bright source samples such as the 3CR \citep{Laing1983}, PW \citep{Peacock1982} and S4 samples \citep{Kapahi1981,Stickel1994} selected at both low and high frequencies. Detailed high-resolution images of these sources with both sub-arcsec and mas resolutions (Fig.~\ref{f:css_structures}) showed that (i) they often, but not always had a double-lobed structure either with or without a detected radio core and could be classified as a CSO/MSO depending on its size. The median values of the relative strength of the radio cores for the CSS/PS radio galaxies and quasars, were similar to those of the larger radio galaxies and quasars respectively. This is consistent with the expectations of the unified scheme with the quasar cores being more prominent. (ii) Large misalignments, where the two oppositely-directed axes are misaligned by more than 20$^\circ$ were often seen which could be due to a combination of interaction with an asymmetric external environment and the effects of projection. While those seen in quasars with strong cores could be due to projection effects, these are also seen in sources with weak cores, including galaxies. In more extreme cases, the source may appear quite complex. (iii) The effects of an ambient medium are also visible in the arm length and flux density ratios of the CSS sources, which are both significantly larger than the corresponding values for the larger sources. The brighter components are often closer to the nucleus suggesting the effects of an asymmetric environment with which the jets interact rather than relativistic beaming effects where the approaching beamed component would appear further \citep[for e.g.][]{Saikia1995a,Saikia2001,Saikia2003a}. 

In addition to these high-luminosity compact sources, it is relevant to investigate whether there is a large population of lower-luminosity compact sources and how do their properties compare with the higher-luminosity ones. 
One of the early lower-luminosity samples, the CORALZ sample, was compiled by \citet{Snellen2004} from the FIRST survey with
S$_{\rm 1400 MHz} >$ 100 mJy, {overall deconvolved angular diameter} $<2$ arcsec and lying in the redshift range $0.005<z<0.16$. These sources which are on average over two orders of magnitude less luminous than the high-luminosity compact sources were also found to predominantly have a CSO or double structure \citep{deVries2009a,deVries2009b}. They also find the turnover frequency in the spectra of these weaker sources to be inversely related to their linear size, and suggest that the turnover in their spectra is consistent with synchrotron self absorption. Kunert-Bajraszewska and collaborators 
\citep{KB2010a,KB2010b,KB2014,KB2016} considered a sample of 44 low-luminosity compact sources (LLCs) with their luminosity at 1400 MHz ranging from $10^{23}$ to $10^{26}\mathrm{\ W\ Hz}^{-1}$ with a median value of $\sim5\times10^{25}\mathrm{\ W\ Hz}^{-1}$. For those with a radio core, they reported significant asymmetries on opposite sides, as seen for the strong source samples. This is also likely to be due to asymmetries in the external environment. They also reported the LLC sources to have more dominant cores compared with both the higher-luminosity FRI and FRII sources, somewhat analogous to the FR0 objects, suggesting their difficulties in forming prominent extended emission. Several CSS/PS sources in the nearby Universe have been studied in some detail. For those within about 100 Mpc detailed studies of the host galaxies are possible \citep[see e.g.,][]{Labiano2007}. An example of a nearby PS source is B1718$-$649 (NGC6328, J1723$-$6500) at a distance of $\sim$60 Mpc with a double-lobed radio structure and an optical galaxy formed possibly by the merger of two systems \citep{Tingay1997,Tingay2015a}. Although NGC1052 (J0241$-$0815) and IC1459 (J2257$-$3627) have been suggested to be nearby GPS sources \citep{Tingay2003b,Tingay2015b}, these sources are variable, with NGC1052 having a nearly flat spectrum and two-sided jets traversing outwards at sub-relativistic speeds \citep{Vermeulen2003}. IC1459 also has similar oppositely-directed jets and a dominant core component \citep{Tingay2015b}. Their possible relationship to the archetypal PS sources need to be understood.

\citet{Sadler2014} have further probed the low-luminosity region. They compiled a sample of 202 sources by cross matching the AT20G sample with the 6dF Galaxy Survey \citep{Jones2009}.
The median redshift of this sample is 0.058, while the median luminosity {at 1.4 GHz} is $3.2\times10^{24}\mathrm{\ W\ Hz}^{-1}$, with 55 of the objects below $10^{24}\mathrm{\ W\ Hz}^{-1}$ \citep{Sadler2016}. Although detailed high-resolution images are not available, based on the ATCA observations as well as comparisons with NVSS and SUMSS images, Sadler et al. find $\sim$68\% to be compact FR0s, most of which are either candidate CSS or PS objects, 8\% are FRIIs and 24\% FRIs. They find their results for FRI and FRII sources consistent with the Ledlow-Owen trend by considering the 20 GHz radio luminosity and the absolute K-band magnitude, with the CSS and PS objects (which are included in their FR0s) spanning about 3 orders of magnitude in luminosity and overlapping with their FRI and FRII sources. For the entire ATG20-6dFGS sample, the majority ($\sim$77\%) are LERGs while $\sim$23\%  are HERGs. This is also true for the CSS and PS (FR0) sources, the percentages being $\sim$75 and 25 respectively. In the WISE colour-colour plot, the FRIs are almost entirely WISE early type with W2$-$W3$<$2.0, the FRIIs of WISE late type with W2$-$W3$>$2.0, while the FR0s 
are largely ($\sim$67\%) of WISE early type, with $\sim$33\%  being of late type. These trends have significant implications in our understanding of the evolution of radio sources \citep{Sadler2014,Sadler2016}.

The evolution scenario in the 1990s revolved around GPS sources evolving into CSS and then on to the larger-sized extended sources \citep{Fanti1995,Readhead1996,ODea1998}. The classification of radio sources into LERGs and HERGs and the recognition of two possible accretion modes, suggest two parallel paths, one for LERGs and one for HERGs: PS$_{\rm LERGs}$ -- CSS$_{\rm LERGs}$ -- Ext$_{\rm LERGs}$, and a similar track at a higher luminosity for HERGs \citep[e.g.,][]{KB2016}. The higher luminosity HERGs could be evolving largely into FRII sources, while the lower-luminosity LERGs into FRIs. Several authors have considered evolutionary models along these lines \citep{An2012a}. However, the large number of low-luminosity compact sources suggests that many of them may not evolve into larger sources, being confined to the small dimensions by the dense interstellar medium of the host galaxies (Sect.~\ref{s:frustrated}), or may represent sources going through intermittent cycles of low-level activity (Sect.~\ref{s:transient}).

\subsubsection{The FR0 sources}\label{s:fr0_sources}
In this section, we discuss possible relationship between the FR0 sources which are weak compact radio sources characterized by the deficit of extended emission and the CSS/PS sources. The FR0s have been seen in large numbers in sensitive radio and optical surveys. Those with a steep or peaked spectrum are analogues of weak CSS/PS sources and pose similar questions related to the formation and evolution of these objects. What fraction of these sources may be young objects and what fraction may evolve into larger objects? Do the FR0s represent transient AGN activity? Is their inability to launch large-scale jets related to their black hole mass and spin?

The FR0 sources are easily discernible in sensitive radio surveys. 
For example, the availability of radio surveys with the VLA and large-area optical surveys such as the SDSS led to the compilation of $\sim$18000 radio sources upto redshifts of $\sim$0.3 and stronger than $\sim$5 mJy at 1400 MHz \citep{Best2012}. These span a range of luminosity from $10^{22}$--$10^{26}\mathrm{\ W\ Hz}^{-1}$ at 1400 MHz, and although these exhibit a wide range of structures including those seen in the stronger 3CR sources, the majority of these are compact. The resolution of the FIRST observations of 5 arcsec would correspond to a size of $\sim$20 kpc at at redshift of 0.3. These objects christened as FR0s have been identified from both low- and high-frequency surveys and dominate the radio source population at low radio luminosities \citep[e.g.,][]{Baldi2009,Baldi2010,Sadler2014,Baldi2015,Whittam2016,Miraghaei2017}. However detailed spectral information would be required to determine what fraction may be similar to CSS/PS sources.

At high frequencies, \citet{Sadler2016} note that the FR0s which make up most of the AT20G-6dfGS sample are a mixed lot with $\sim$75\%  being LERGs and 25\%  HERGs, and $\sim$67\%  being WISE early-type and 33 per cent late-type galaxies. Their luminosity at 1.4 GHz ranges from $10^{22}$ to $10^{26}\mathrm{\ W\ Hz}^{-1}$ and many of these could be PS/CSS type sources. For a complete sample of 96 sources (S $>$ 0.5 Jy) chosen from the Cambridge 10C survey at 15.7 GHz, 65 have been classified as compact or FR0s with 13 as CSS, 10 as PS and the remaining 42 as unclassified \citep{Whittam2016}. In both the samples, the number of FR0s are in the range of 70--80\%, although the mean radio spectral indices and optical properties may exhibit dependence on the flux density of the radio sources, with the radio spectra exhibiting a variety of shapes.

At low radio frequencies which is expected to be dominated by optically thin emission, $\sim$70\% of the radio sources in  LoTSS \citep{Shimwell2017,Shimwell2019} appear to be unresolved with an angular resolution of $\sim$6 arcsec. These sources which dominate the low-frequency radio  Universe are likely to give rise to thousands of FR0 candidates \citep[c.f.][]{Hardcastle2019,Sabater2019,Mingo2019}. Deep radio observations of well-studied fields have also yielded large numbers of compact steep-spectrum sources, as seen for example in the ELAIS-N1 field \citep[e.g.,][]{Sirothia2009,IshwaraChandra2020}. Of the 6400 sources found in a wide-area deep survey of this region at 610 MHz, the vast majority are compact with a median spectral index of $\sim$0.85 between 610 and 1400 MHz \citep{IshwaraChandra2020}.

\citet{Baldi2018} have compiled a catalogue of 108 FR0 sources (FR0CAT) with a redshift $<$0.05 and a size less than $\sim$5 kpc starting with the \citet{Best2012} sample. These FR0s are found to reside in massive luminous early type galaxies ($-21 > M_r > -23$) with black hole masses in the range of $\sim 10^{7.5}$--$10^{9}\,M_\odot$, which are less massive than FRI radio galaxies by a $\sim$1.6. Their mid-IR colours are also consistent with those of elliptical galaxies \citep{Wright2010}. The most striking feature is the deficit of extended radio emission. Considering sources of similar [O{\sc iii}] luminosity which is an indicator for AGN power, the radio luminosity of FR0s is lower than 3CR sources by $\sim$100 \citep{Baldi2018}. \citet{Capetti2019} found no evidence of extended emission in these sources in the TIFR GMRT Sky Survey (TGSS), and noted that although $\sim$75 per cent have a convex spectrum the steepening is more gradual than the high-luminosity PS sources. Besides the deficit of extended radio emission the FR0s appear to reside in environments where the galaxy density is lower than for FRIs by a factor of $\sim$2 \citep{Capetti2020}. The association of a \textit{Fermi} source Tol1326$-$379 with an FR0 galaxy \citep{Grandi2016} has aroused interest in their high-energy properties which has been reviewed recently by \citet{Baldi2019b}.

As a significant fraction of these FR0 objects are likely to be PS and CSS sources, although of much lower luminosity, it is important to understand their nature which may provide insights towards understanding the diverse structures of radio sources and the possible evolutionary scenarios \citep[e.g.,][]{Sadler2016,Miraghaei2017,Baldi2019b}. Possible scenarios for the FR0 sources are as follows. (i) Are these radio sources inclined at small angles to the line of sight, being low-luminosity blazars with highly-relativistic jets? This seems unlikely. High-resolution radio 
observations with both the VLA and VLBI techniques often exhibit reasonably symmetric structure on opposite sides of the nuclear component \citep{Cheng2018,Baldi2019a}. Of the 14 sources observed by \citet{Cheng2018} on parsec scale, 4 show Doppler boosting factors from 1.7 to 6, and 2 with multiple epoch observations indicate velocities between 0.23 and 0.49c. Mild variability is found in only one source, while three others exhibit no significant variability over a few years. These observations are consistent with mildly relativistic jets. This is also consistent with the ratio of [O{\sc iii}] to X-ray luminosity for FR0s and FRIs which have a similar value of $\sim -1.6$, while it is significantly lower for BL Lacs due to a beamed X-ray component \citep{Torresi2018}.
(ii) As has been suggested for the luminous PS and CSS sources, are the FR0s likely to evolve into larger FRI and FRII radio sources? This appears unlikely given the much larger space density of FR0s but it is difficult to rule out a small fraction of FR0s evolving into larger sources. (iii) Are these sources compact because of the intrinsic properties of the supermassive black holes which are unable to launch large-scale jets or these are smothered by dense gaseous environment? The similarity of optical host magnitudes and [O{\sc iii}] luminosity of FR0s and FRIs does not suggest that the jets in FR0s are frustrated by a dense
environment. \citet{Miraghaei2017} suggest that due to a lower black hole mass, FR0s find it difficult to launch large-scale radio jets and may be able to support jet activity for only short periods of time. The launching of jets could be due to a low black hole spin limiting its ability to extract energy to launch the jets, consistent with a relation between black hole spin and bulk jet Lorentz factor \citep[e.g.,][]{Tchekhovskoy2010,Maraschi2012}. This would provide a 
natural explanation for the mild bulk Lorentz factors determined from VLBI observations. \citet{Garofalo2019} explore a theoretical scenario where the evolution of FR0s is governed both by spin of the black hole and availability of fuelling gas.  

\subsection{Polarization observations}\label{s:pol_obs}
As both CSS and PS sources are confined well within the interstellar medium of their host galaxies, multifrequency polarization observations could provide a valuable probe of the magnetoionic plasma in which these are embedded as well as of interactions of the radio jets and lobes with the ambient medium. 
The magnetized plasma causes a rotation of the plane of polarization which is given by $\chi(\lambda) = \chi_0 + {\rm RM} \lambda^2$ where RM = $812 \int_{0}^{l} n_e B_{\parallel} dl$, where $\chi(\lambda)$ is the position angle (PA) of the E-vector at a wavelength $\lambda$, $\chi_0$ is the PA at zero wavelength, RM defined as the Rotation Measure is in units of radian m$^{-2}$, $n_e$ is the electron density in cm$^{-3}$, $B_{\parallel}$ in mG and $l$ in parsec. The degree of rotation is a strong function of the wavelength, leading to depolarization of emission at longer wavelengths. If radio emitting plasma is mixed with the magnetized thermal plasma, the E-vector from regions of greater Faraday depth will be rotated more relative to the regions of smaller Faraday depth, leading to depolarization. Also in the case of a screen of thermal plasma in front of the radio source, adjacent regions within the resolution element of the observations may traverse through regions of differing Faraday depth, again leading to depolarization. As high RMs in the range of 1000s of radian m$^{-2}$ are expected, especially closer to the nucleus \citep[e.g.,][]{ODea1989}, one needs to probe over a wide bandwidth at high frequencies and with high angular resolution to disentangle the different effects. 

One of the early studies of polarization properties of CSS and PS sources based on VLA A-array observations at $\lambda$20 and 6 cm showed that for these sources (i) those associated with galaxies tend to be unpolarized or weakly polarized with the percentage polarization $<$0.5\% at both wavelengths; (ii) the quasars {from this sample of CSS and PS sources} have a wider distribution with a median value of $\sim$2\% at $\lambda$6 cm and show significant depolarization at the longer wavelengths; (iii) several of these compact sources have large rotation measures (RMs) although those with low rotation measures are also known; (iv) strong polarization and total intensity variations were not common, although in some cases flat-spectrum sources could be classified as CSS/PS sources from a few frequency measurements due to variability at the different frequencies \citep{Saikia1987,Saikia1988} (and see Sect.~\ref{s:variability_orientation}). This was broadly the situation at the time of the \citet{ODea1998} review. These studies were largely confined to bright source samples such as the VLA Calibrator Survey and the well-studied 3CR sample.

Since then there have been detailed studies of fainter or lower-luminosity sources, such as the B3-VLA CSS sample with lower resolution using the Westerbork and Effelsberg telescopes to examine integrated properties, and the VLA and MERLIN with higher resolution to examine polarization distribution and relationship to structure and environment. Several well-known 3CR as well as weaker sources have also been studied extensively with high resolution using VLBI techniques to probe polarization properties closer to the nucleus. We summarize these results and possible implications in our understanding of CSS/PS sources.

\subsubsection{Integrated properties}\label{s:pol_int}
The PS sources are more compact than the CSS sources and are expected to show stronger Faraday effects due to larger column densities of the magnetoionic medium. \citet{Cotton2003a} examined the polarization properties of the sample of GPS sources compiled by \citet{Stanghellini1998a} at $\lambda$20 cm using the NVSS survey. They found the galaxies to be more weakly polarized than the quasars {in this sample}, consistent with earlier results. \citet{Cotton2003a} also reported from the dependence of polarization on linear size in a sample of CSS and PS sources \citep{Fanti2001} that sources smaller than 6 kpc tend to be weakly polarized, usually less than $\sim$0.5 per cent, while those larger could be polarized up to several per cent. This suggests a change in the properties of the magnetoionic medium on this scale. A similar trend was also reported by \citet{Fanti2004} for the B3-VLA sample of compact sources which they had observed at 4.9 and 8.5 GHz with the VLA and used the NVSS results at 1.4 GHz. \citet{Rossetti2008} who observed this sample with the Westerbork telescope at 2.64 GHz also reported a similar trend. However, this trend was not seen for the entire sample of 3CR+PW sample by \citet{Mantovani2009} who observed the bright-source sample with the Effelsberg telescope at 2.64, 4.85, 8.35 and 10.45 GHz and also used the NVSS results and previous single-dish measurements from the catalogue by \citet{Tabara1980}. They found that this trend was seen when only galaxies were considered alone. It may be mentioned that the samples considered by \citet{Cotton2003a} and the B3-VLA sample \citep{Fanti2004,Rossetti2008} are dominated by galaxies. Among galaxies too there could be outliers \citep{Rossetti2008}. This difference suggests a polarized component in the quasars, such as a beamed jet, which may be less affected by the magnetoionic plasma. 

A fainter source sample than say the 3CR one is the B3-VLA sample whose polarization properties have been studied extensively by \citet{Fanti2001}, \citet{Fanti2004} and later by \citet{Rossetti2008}, who reported Westerbork observations at 2.26 GHz ($\sim$13 cm) using 8 different intermediate frequencies (IFs), and combined these with earlier VLA observations at 1.4, 4.9 and 8.5 GHz \citep{Fanti2004}. Sources smaller than about 2.5 kpc usually appear depolarized at all frequencies; the critical size where depolarization sets in being 5 kpc at 2.26 GHz, intermediate between those at 1.4 GHz and 4.9/8.5 GHz \citep{Cotton2003a,Fanti2004}. As expected, \citet{Rossetti2008} find the degree of polarization to decrease between 8.5 and 2.6 GHz, which would be expected due to Faraday depolarization, but find a flattening between 2.6 and 1.4 GHz. They explore possible explanations using both the \citet{Burn1966} and \citet{Tribble1991} models, and find that a variation of the Burn model in which the concept of a `covering factor' of the foreground screen is introduced appears to fit the data. \citet{Rossetti2008} also suggest that the `covering factor' they introduced may be due to orientation effects with the receding component being more strongly depolarized compared to the approaching one as in the Laing--Garrington effect. If so one would expect a difference between galaxies and quasars. A more natural origin may be a clumpy, inhomogeneous medium as structural and polarization asymmetries appear to suggest. 

The integrated rotation measure in the source frame, RM$_{\rm sf}$ varies from $\sim$10 to $10^4\mathrm{\ rad\ m}^{-2}$ with the dispersion in RM $\sigma_{\rm RM,sf}$ spanning a somewhat smaller range. The total RM$_{\rm sf}$ suggests a large-scale ordered field as well in addition to a random component, and there are suggestions of both RM$_{\rm sf}$ and $\sigma_{\rm RM,sf}$ increasing with redshift and decreasing with projected linear size. For sources with RM$_{\rm sf} >100\mathrm{\ rad\ m}^{-2}$, for almost all of which RM$_{\rm sf} > \sigma_{\rm RM,sf}$, RM$_{\rm sf} \propto l^{-2.2}$ where $l$ is the projected size in kpc \citep{Rossetti2008}.

Similar results have been obtained for brighter sources. From observations of the 3CR and PW \citep{Peacock1982} sample, \citet{Mantovani2009} report a rapid decrease in polarization at longer wavelengths although the degree of polarization does not usually go to 0 at $\lambda >$49 cm. Their estimated values of RM range from $-20\mathrm{\ rad\ m}^{-2}$ to $\sim3900\mathrm{\ rad\ m}^{-2}$ with sources generally following the $\lambda^2$ law. They also suggest that a variation of the \citet{Tribble1991} model incorporating a covering factor can explain their observations.

\begin{figure*}
 \centering
   \includegraphics[width=5.85cm]{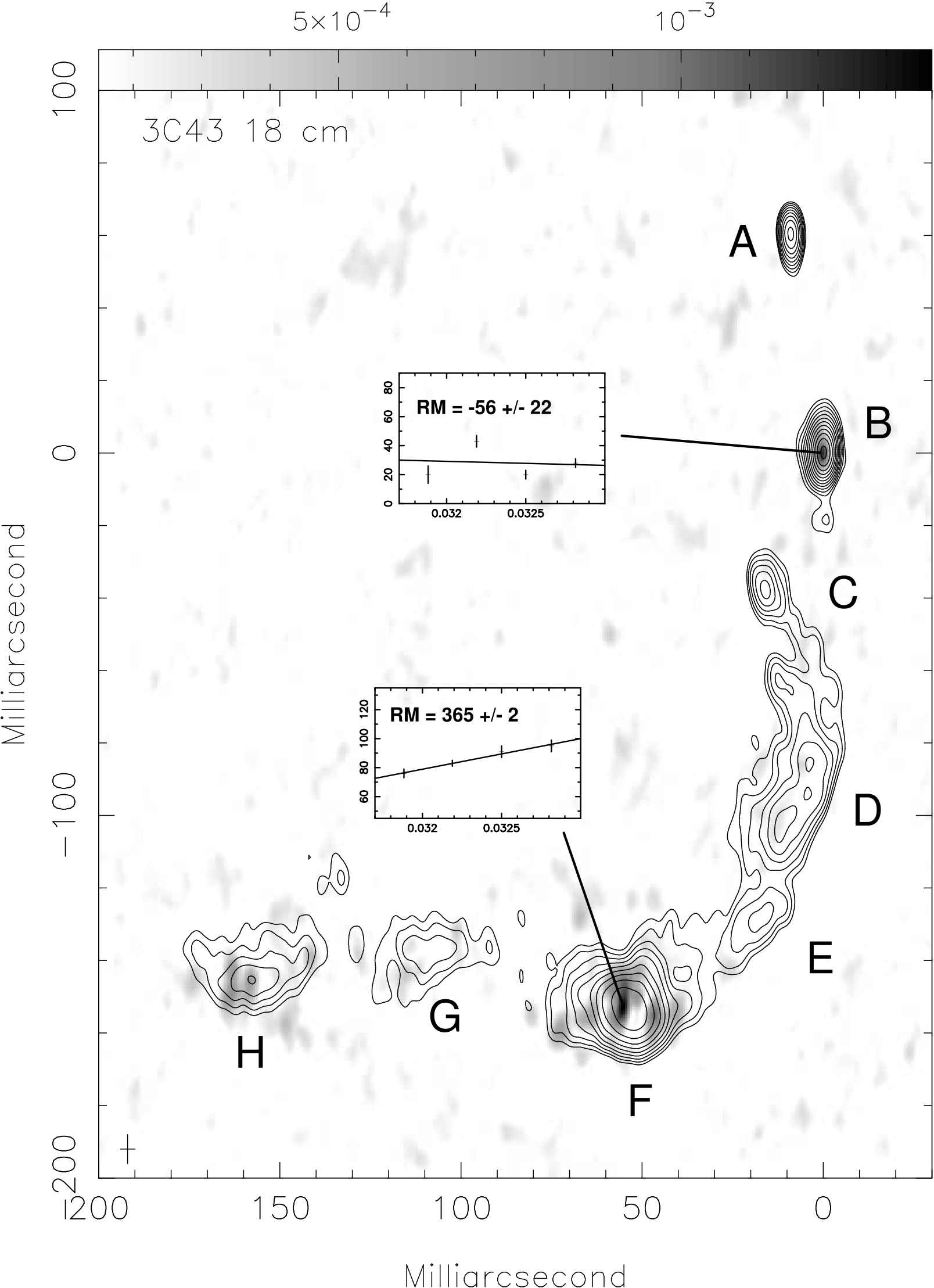}
   \includegraphics[width=5.2cm]{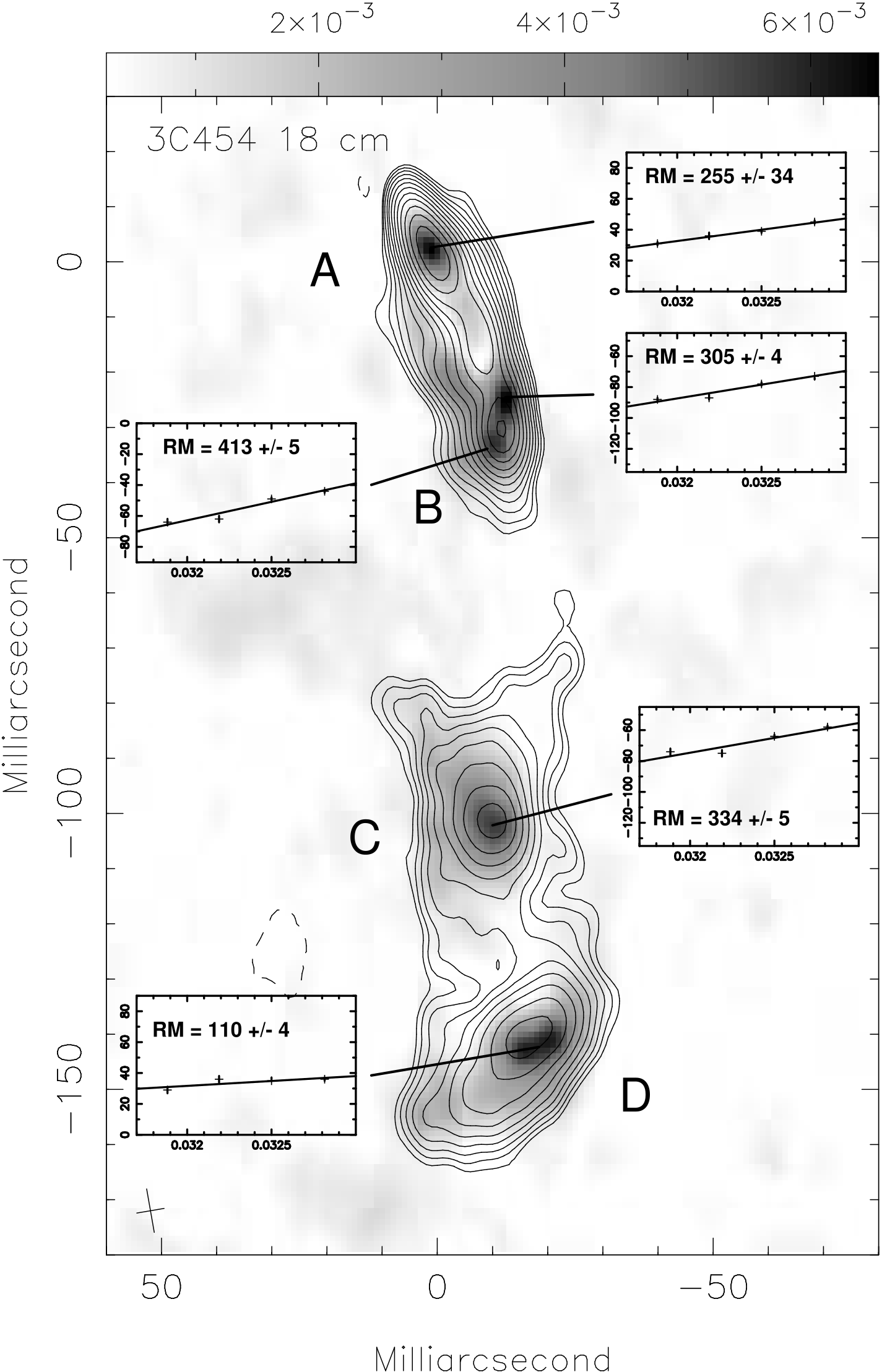}
\caption{VLBA total-intensity contours of the CSS sources 3C43 and 3C454 at 18cm overlaid on the images of the polarized intensity. The values of RM for different regions of the source as observed are shown. These have to be multiplied by 6.1 and 7.6 respectively to be in the source reference frame for the two sources. 3C43 is seen through a thick Faraday screen with the sharp bend being due to jet-ISM interaction, while in 3C454 thermal plasma appears to be mixed along with the radiating relativistic particles \citep{Cotton2003b}. }
\label{f:cssrm}
\end{figure*}

\subsubsection{Higher-resolution observations}\label{s:high-res-poln}
While low-resolution observations give us an overview of the properties of the source as a whole, higher-resolution observations reveal the variation of degree of polarization and rotation measure across the source. It also minimizes the effects of depolarization due to coarser resolution. For example, the median values of polarization at 8.5 GHz for the `low-resolution' images of the B3-VLA sample varies from 0.1 per cent for sources $<$1 kpc, to 1.28, 3.95 and 3.6 per cent for those with sizes of 1$-5$, 5$-$10 and 10$-$20 kpc respectively. The median value for all the sources is $\sim$1.6 per cent compared with $\sim$4 per cent for the components inferred from the high-resolution images at 8.5 GHz \citep{Fanti2001}.

Sub-arcsec resolution images also reveal asymmetries in the polarization properties of the lobes on opposite sides of the nucleus or parent optical object. This has been reported for several of the resolved sources in the B3-VLA sample \citep{Fanti2001}, and also for a sample of largely 3CR and 4C CSS objects whose polarization properties have been well determined \citep{Saikia2003b}. The ratio of the degree of polarization of the oppositely-directed lobes has a median value of $\sim$5 for the 3CR+4C CSS objects, while for the control sample of larger sources observed by \citet{Garrington1991a} is $\sim$1.5, with only two of the sources dominated by one-sided jets having a value $>$5. This is unlikely to be due to effects of orientation as seen in the Laing-Garrington effect \citep{Laing1988,Garrington1988} as about half the sources in the sample are galaxies \citep{Saikia2003b} which should be inclined at large angles to the line of sight. Also the magnetoionic haloes envisaged to explain the Laing-Garrington depolarization asymmetry \citep{Garrington1991b} is likely to have a marginal effect on the scale of the compact objects. These asymmetries are more likely to be caused by interactions with dense clouds of gas which are asymmetrically located relative to the central engine. Structural asymmetries where the brighter component is on the jet side and closer to the nucleus as seen for example in the CSS quasar 3C147 \citep{Junor1999} and in a number of galaxies \citep{Saikia2001}, including the highly-asymmetric radio galaxy 3C459 on a slightly larger scale \citep{Thomasson2003}, are consistent with this scenario. Whether these clouds of gas are related to the fuelling process of the central active galactic nucleus remains unclear.

Detailed imaging at both sub-arcsec and milli-arcsec resolution has been used to study the total intensity and polarization structure to probe possible interactions with the external environment. A number of 3CR sources have been studied in some detail. The CSS quasar 3C147 which was found to have a high observed RM value of $\sim -1600\mathrm{\ rad\ m}^{-2}$ \citep{Kato1987,Inoue1995} exhibited a huge differential Faraday rotation with the RMs in the rest frame being $-3140$ and $+630\mathrm{\ rad\ m}^{-2}$ for the southern and northern components respectively \citep{Junor1999}. With the southern component nearer the core and facing the jet, this was interpreted to be due to jet-cloud interaction. VLBA observations of 3C 147 show large variations in RM with observed values ranging from $\sim -1200$ to $-2400\mathrm{\ rad\ m}^{-2}$ in the inner portion of the jet on parsec scale \citep{Zhang2004,Rossetti2009}.

Besides 3C147, the other CSS sources usually used as primary flux density calibrators are 3C48, 3C138 and 3C286. From VLA observations at the highest available frequencies (20--45 GHz) \citet{Cotton2016} estimate an RM of $\sim10000\mathrm{\ rad\ m}^{-2}$ for 3C48 but with a significant scatter suggesting complex Faraday effects. This is significantly higher than $\mathrm{RM}\sim -64\mathrm{\ rad\ m}^{-2}$ determined by \citet{Mantovani2009} from Effelsberg observations and NVSS data between 1.4 and 10.85 GHz. This suggests that the RM could be frequency dependent affected by opacity and complex Faraday effects. A detailed polarization study of the pc-scale structure of 3C48 between 1.5 and 8.3 GHz has been presented by \citet{An2010} with RM values consistent with those of \citet{Mantovani2009}. The core appears unpolarized at all frequencies, while the most strongly polarized feature is at $\sim$0.25 arcsec north of the core where the jet bends towards the north east. The polarization properties suggest interaction of the jet with the interstellar medium and/or changes in the magnetic field structure. A prominent knot in the jet, labelled B, exhibits a higher RM and no significant proper motion, suggesting a stationary shock, while a feature closer to the core exhibits superluminal motion. 

The CSS quasar 3C138 was earlier found to have a low RM along the bulk of the jet which is strongly polarized and contributes most of the flux density, but high values of RM close to the core \citep{Cotton1997b}. This is consistent with the integrated RM value close to $\sim0\mathrm{\ rad\ m}^{-2}$. More recent VLBA  observations at 5 GHz show the inner jet to have an $\mathrm{RM}\sim -5000\mathrm{\ rad\ m}^{-2}$ (a lower value of $\sim -3000\mathrm{\ rad\ m}^{-2}$ was reported by \cite{Cotton2016} from higher frequency VLA observations between 20 and 48 GHz) in the source frame, with the polarization in the inner jet being seen through holes in a thick Faraday screen \citep{Cotton2003b}. They also monitored time variation in RM and suggest that the observed variations in RM are likely to be due to jet-ISM interaction. Their data were interpreted to be due to a moving component seen behind a Faraday screen with structure on pc or sub-pc scales \citep{Cotton2003b}. The standard polarization calibrator, the CSS quasar 3C286 which has a low integrated RM was shown to have a high degree of polarization of $\sim$8.9 and 11 per cent respectively at 5 GHz for two knots in the inner jet. The E-vector was aligned along the jet indicating a perpendicular magnetic field in the inner jet following a bend in the jet \citep[for e.g. and references in][]{Cotton1997a}. An inverted-spectrum radio core has been identified recently by \citet{An2017}.  

Several well-studied sources show regions of high RM near where the source bends, but sources may also exhibit a high RM without an obvious relationship with structural changes. In the case of 3C43, the observed RM in a component close to the core is small ($-56\pm22\mathrm{\ rad\ m}^{-2}$) while for the polarized region near the bend it increases to $365\pm2\mathrm{\ rad\ m}^{-2}$ or $\sim680\mathrm{\ rad\ m}^{-2}$ in the source frame \citep{Cotton2003b}. However, the jet in 3C454 does not exhibit a brightening near the westward bend, suggesting it may not be due to a collision, and exhibits a high RM all along the jet \citep{Cotton2003b}. The VLBA results for both these sources are shown in Fig.~\ref{f:cssrm}. Similarly \citet{Mantovani2002a} who studied the CSS sources with strongly bent jets B0548$+$165 and B1524$-$136 found high RM ranging from $\sim$1000 to $10^4\mathrm{\ rad\ m}^{-2}$, but enhanced RM only near the bend in B0548$+$165 which also exhibits a brightening in this region. CSSs with complex structure such as 3C119, 3C318 and 3C343 have been found to have RMs in the range of $\sim$3000 to $10^4\mathrm{\ rad\ m}^{-2}$. In the case of 3C119 the highest RM is where the jet bends at a projected distance of $\sim$325 pc from the core \citep{Mantovani2005,Mantovani2010}.

\subsection{Rejuvenated radio sources}\label{s:rejuv}
Over the years there has been evidence of episodic nuclear jet activity from radio observations, based on both structures and spectral index information, and later from X-ray observations as well. For example in the radio galaxy 3C338 a radio jet south of the nucleus has been interpreted to be due to episodic jet activity \citep{Burns1983}, while the extended emission of 3C388 show a sharp discontinuity in spectral index which has been again interpreted to be due to intermittent jet activity \citep{Roettiger1994}. 3C388 has been studied over a much wider frequency range by \citet{Brienza2020} who suggest a duty cycle of $\sim$60 per cent. A sharp gradient in spectral index is also seen in the case of Her A, another example of a galaxy with recurrent activity \citep{Gizani2005}. Evidence of earlier cycle of activity has also been reported from X-ray observations such as in Cygnus A where old electrons are scattered by inverse-Compton scattering to high energies \citep{Steenbrugge2008}. Multiple generations of X-ray cavities in radio-loud AGN suggest a mean outburst interval of $\sim$6$\times 10^7$ yr \citep{Vantyghem2014}. Although evidence of episodic AGN activity may manifest itself in a variety of ways \citep[see also][]{Jurlin2020,Shabala2020}, one of the most striking evidence of episodic activity is when two or more pairs of distinct radio lobes are seen on opposite sides of the nucleus which are due to different cycles of jet activity. Such objects have been christened as double-double radio galaxies or ddrgs \citep{Schoenmakers1999,Saikia2009,Kuzmicz2017}. 

In this review we consider whether PS and CSS sources tend to have diffuse extended emission from an earlier cycle of activity. One of the earliest examples of extended radio emission associated with a CSS/PS source is B0108+388. \citet{Baum1990} reported an extended lobe of emission which could be a relic of an earlier cycle of activity. Although \citet{Kuzmicz2017} mentions a faint optical object in the vicinity of the diffuse lobe, its association with the lobe is unclear, and we consider the diffuse radio lobe to be associated with the PS source as there appears to be bridge of emission connecting the two. In a search for extended emission in a sample of PS sources, \citet{Stanghellini2005} found, in addition to B0108+388 with extended emission $\sim$125 kpc away, the GPS radio galaxy B0941$-$080 to be a good candidate with extended radio emission located $\sim$70 kpc away and B1345+125 (4C12.50), a GPS radio galaxy with double nuclei, to have extended emission on a scale of $\sim$130 kpc in addition to the pc-scale structure. 4C12.50 shows evidence of fast outflows of ionized, atomic (H{\sc i}) and molecular (CO) gas with velocities of $\sim$1000 km s$^{-1}$ largely driven by the radio jet \citep{Holt2003b,Morganti2005b,Holt2011,Dasyra2012,Morganti2013,Fotopoulou2019}.
Interpreting the more extended radio components to be from an earlier cycle of activity, these relics were estimated to be $\sim$10$^7-10^8$ yr ago \citep{Stanghellini2005}. 

\begin{figure}
\centerline{\includegraphics[width=12.5cm]{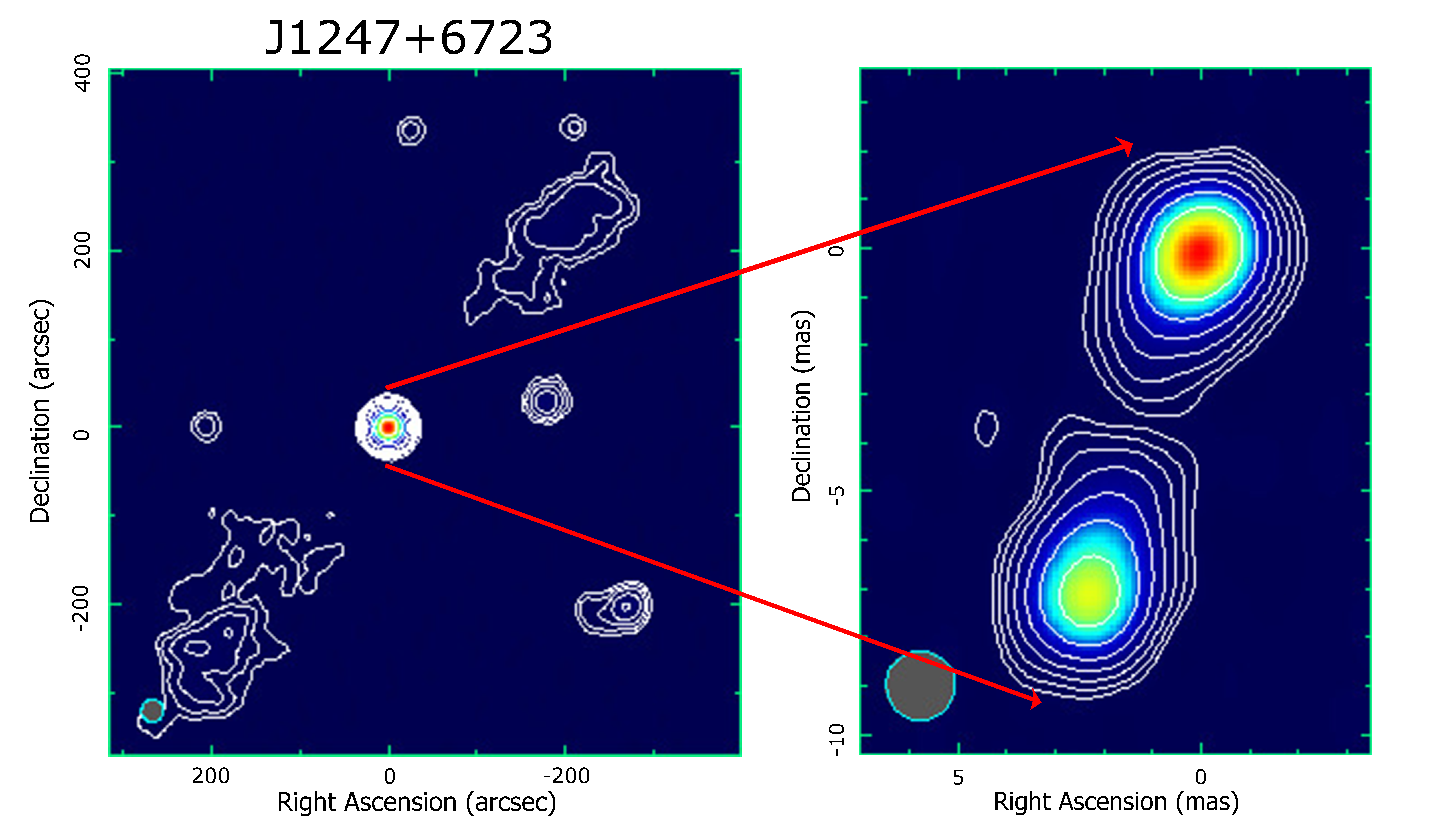}}
\caption{A young PS source is seen in the nucleus of the giant radio galaxy J1247+6723 with the sizes of the inner and outer doubles being $\sim$0.02 and 1200 kpc respectively \citep{Marecki2003a,Bondi2004,Polatidis2009}. }
\label{f:rejuv01}
\end{figure}

\begin{figure}
\centerline{\includegraphics[width=10cm]{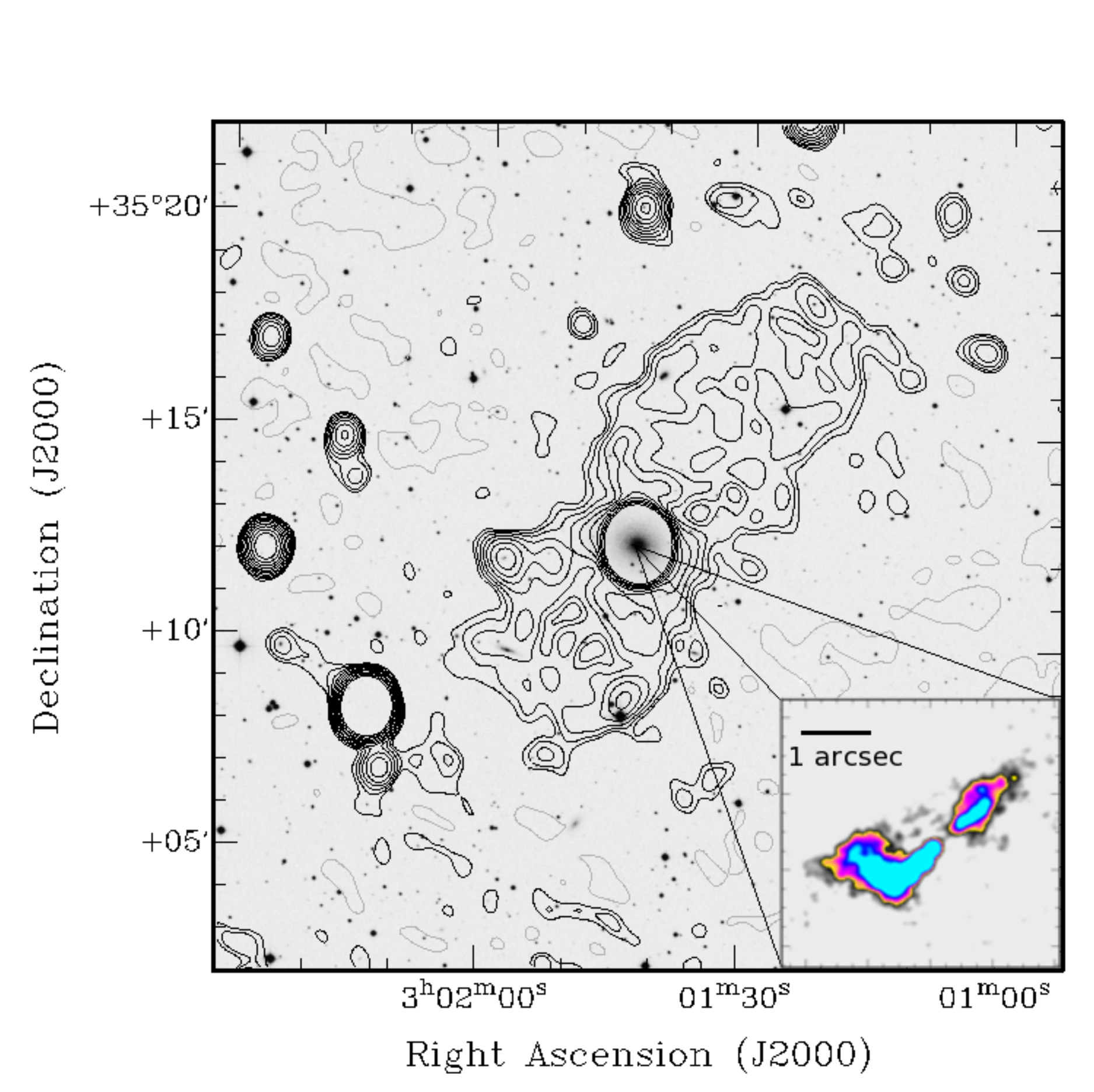}}
\caption{Diffuse radio emission likely to be from an earlier cycle of activity in the CSS source B2~0258+35 which was imaged with the Westerbork Synthesis Radio Telescope is shown overlaid on a DSS2 image \citep{Shulevski2012}. The inset shows a high-resolution VLA image of the CSS source whose structure indicates interaction with the external interstellar medium \citep{Giroletti2005}. }
\label{f:rejuv02}
\end{figure}

On smaller scales \citet{Jeyakumar2000b} found evidence of larger-scale structure compared with the known VLBI-scale structures for 8 CSS and PS sources from interplanetary scintillation observations at 327 MHz using the Ooty Radio Telescope, but the detailed structures remained unclear. Candidate rejuvenated sources have also been suggested from the detection of extended emission around PS sources \citep{Hancock2010}, large radio sources with PS cores \citep{Bruni2019}, and suggestions of both extended emission and PS cores from the integrated spectra of several sources from the GLEAM survey \citep{Callingham2017}. However detailed observations are required to establish the structure of the sources on all scales to firmly establish their episodic nuclear activity \citep{Hancock2009b}. We illustrate here a couple of examples of rejuvenated sources to illustrate the range of structures. Figure \ref{f:rejuv01} shows the inner and outer doubles of the giant radio source J1247+6723 with sizes of $\sim$0.02 and 1200 kpc respectively with the inner double having a peaked spectrum \citep{Marecki2003a,Bondi2004}. Evidence of episodic activity is seen over a large range of size scales. Figure \ref{f:rejuv02} shows the well-studied CSS source B2 0258+35 (J0301+3512) in the nearby galaxy NGC1167 where Westerbork observations have revealed diffuse emission possibly from an earlier cycle of activity but not with very steep spectra \citep{Shulevski2012,Giroletti2005,Brienza2018,Murthy2019}. Here the CSS and outer radio emission have sizes of $\sim$0.7 and 230 kpc respectively.

\begin{figure*}
	\centering
	\hbox{ 
		\includegraphics[width=6.0cm]{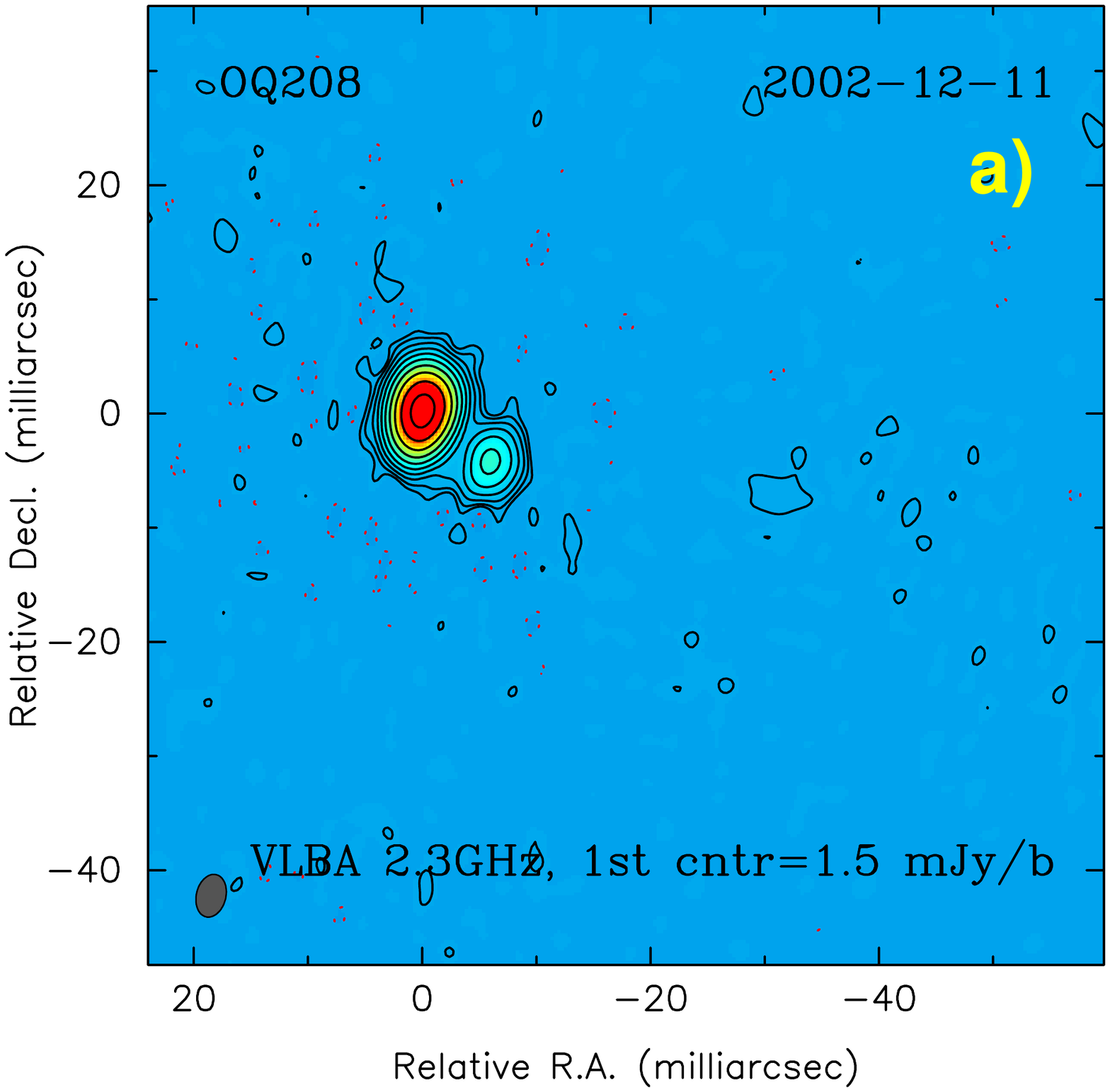}
		\includegraphics[width=6.0cm]{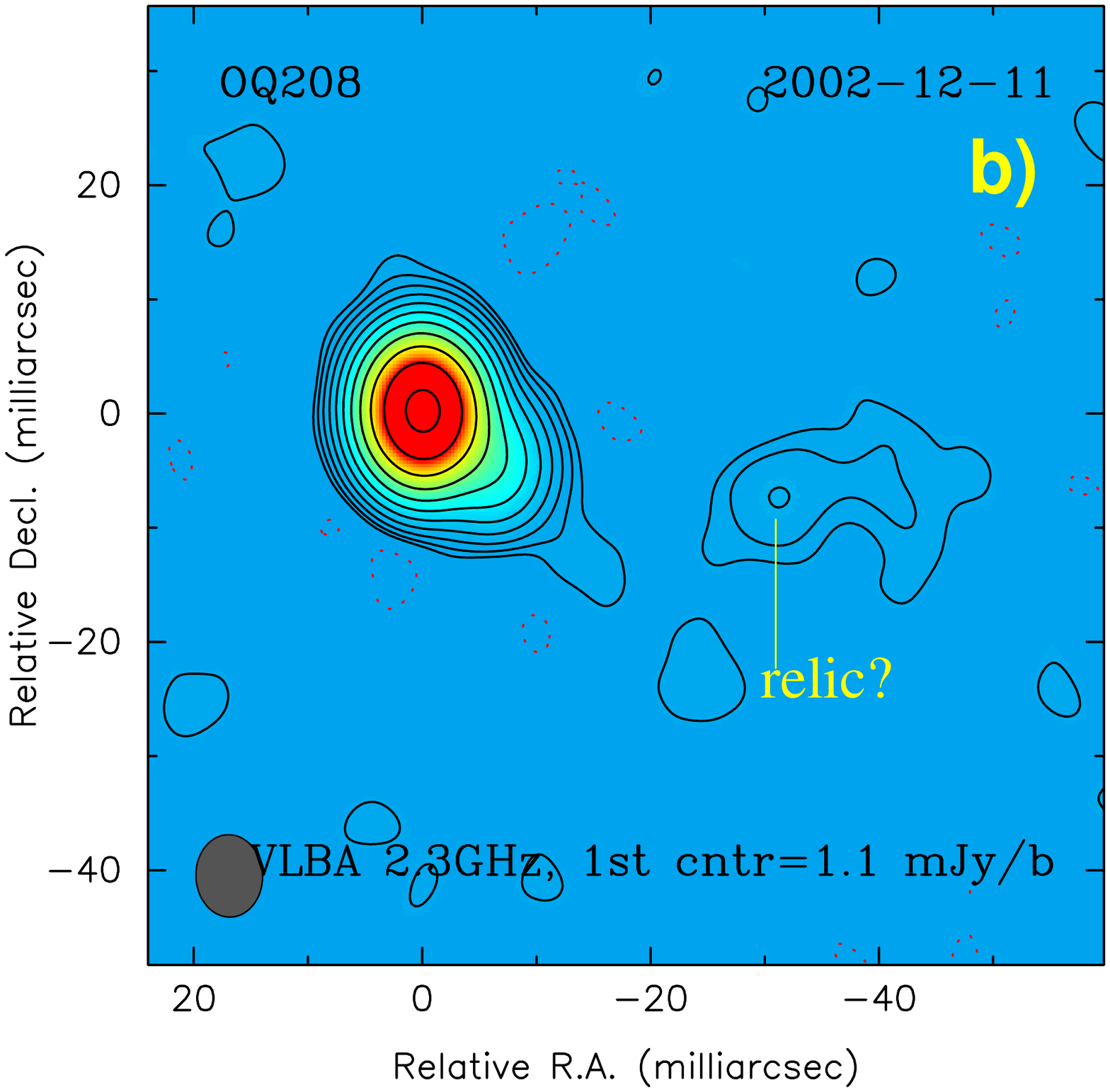}
	}
	\hbox{
		\includegraphics[width=6.0cm]{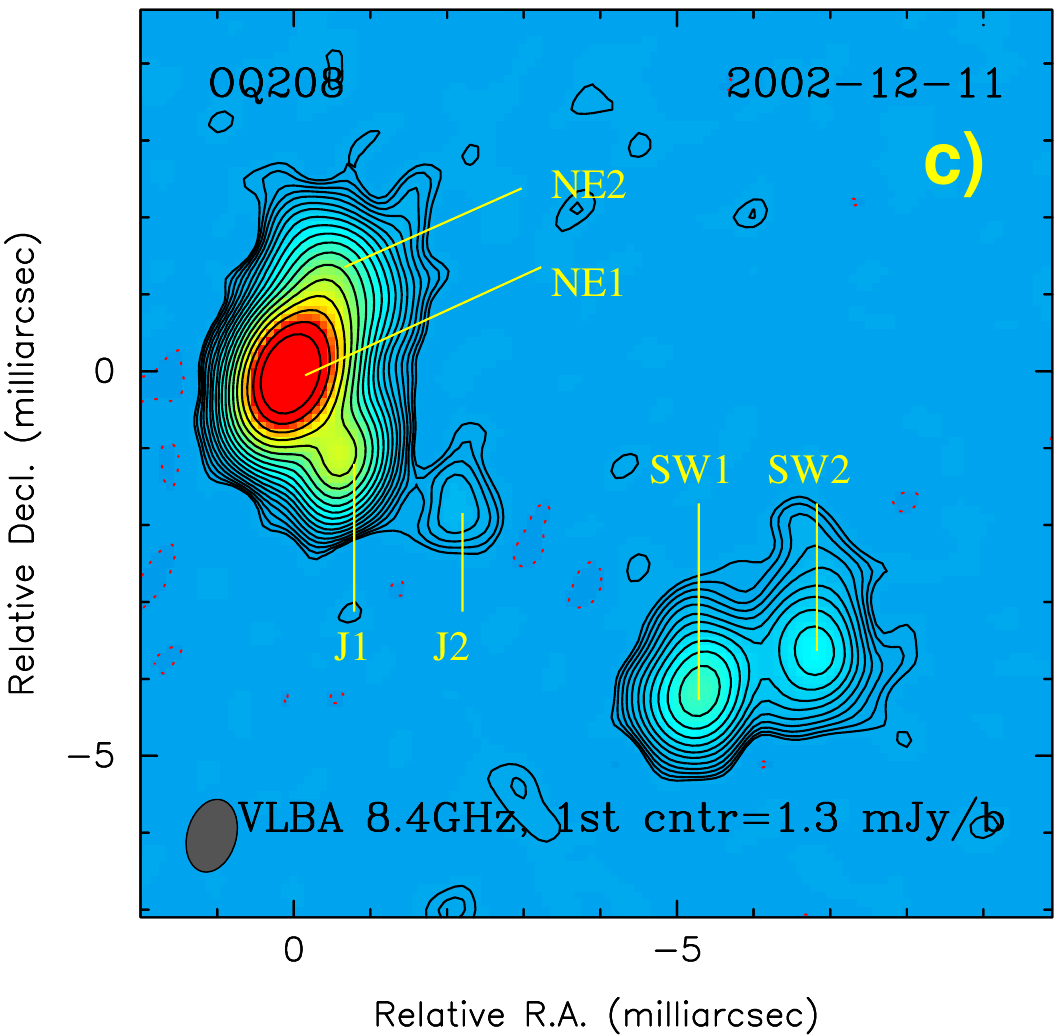}
		\includegraphics[width=6.0cm]{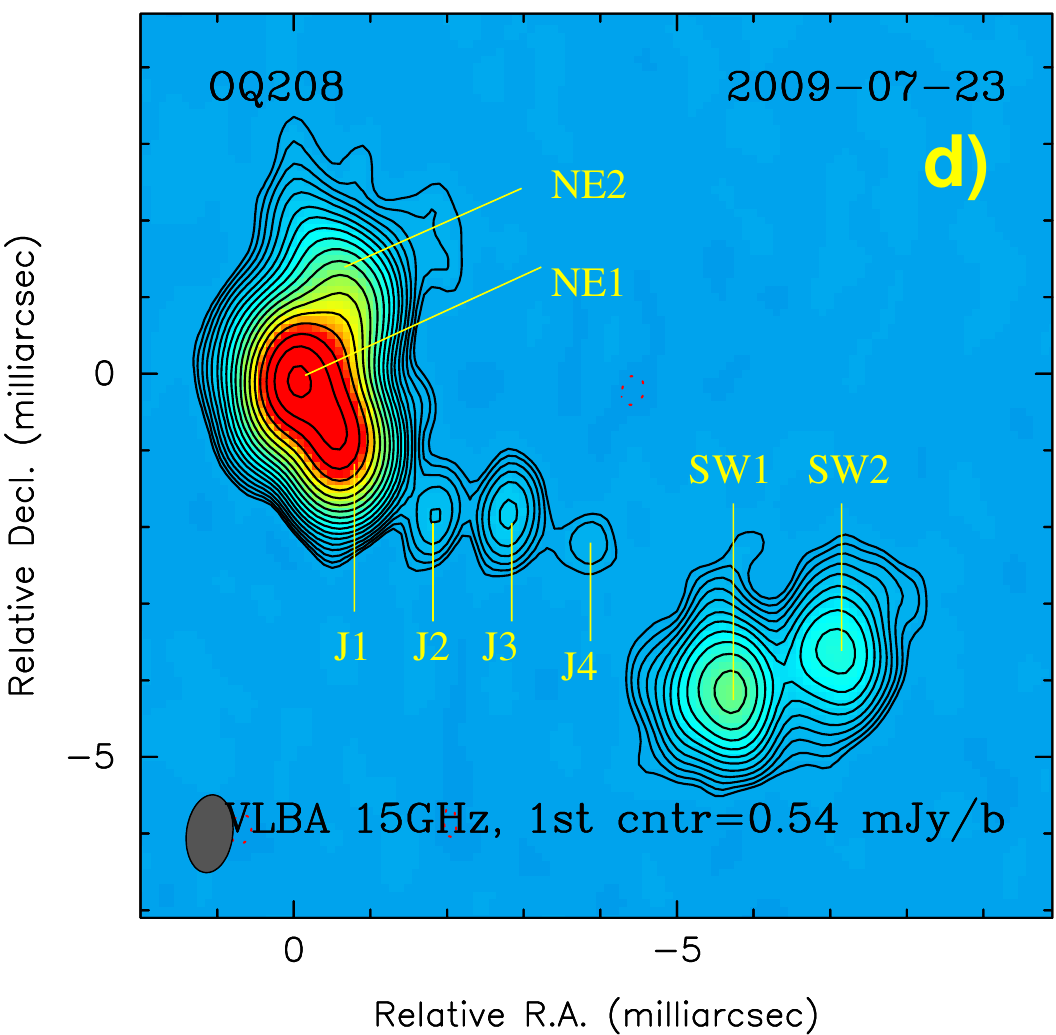}
	}
	
	\caption{VLBI images of the PS source OQ208 from \citet{Wu2013}. The tapered image at 2.3~GHz (panel b) shows evidence of relic emission from an earlier cycle of activity similar to what was seen in B0108+388 \citep{Baum1990}. The lower panels (c and d) show the high-resolution structure revealing each mini lobe splitting into two hotspots. NE1 and SW1 are moving apart at a velocity of $\sim$0.13c implying a hot-spot advance speed of 0.065c, while the knot in the jet is advancing with a significantly higher and mildly relativistic speed of 0.23c relative to the systemic centre. The kinematic age was estimated to be $\sim$255 yr \citep{Wu2013}.}
 	\label{f:rejuv03}
\end{figure*}

There is also direct evidence of diffuse emission on small scales from VLBI observations. For example in OQ208 in addition to the double-lobed structure, there is weak diffuse emission about $\sim$30 mas (40 pc) to the west as shown in Fig.~\ref{f:rejuv03} \citep{Luo2007,Wu2013}. Another good example is the galaxy J1511+0518 which in addition to a well-defined triple source has diffuse extended lobe $\sim$30 mas ($\sim$50 pc) significantly misaligned from the triple \citep{Orienti2008d}. It has also been speculated that the range of VLBI-scale structures in the well-studied GPS source CTA21 (4C16.09) may be due to multiple cycles of nuclear activity \citep{Salter2010}. The VLBI-scale relics are from more recent activity $\sim10^3$--$10^4$ yr ago, suggesting short bursts of activity in the initial phases. An intriguing feature of these diffuse lobes is that they are often one-sided both on kpc and pc scales, somewhat reminiscent of the northern middle lobe of Cen A. In the small scales it may be due to short outbursts or sputtering, while on the larger scales the oppositely directed diffuse lobe may be below the detection threshold due to radiative losses or rapid expansion on one side due to an inhomogeneous, asymmetric environment on opposite sides of the nucleus.


%
%

In a recent compilation of extragalactic radio sources with evidence of recurrent jet activity there are 74 radio sources of which 67 are galaxies, 2 are quasars and 5 are unidentified sources \citep{Kuzmicz2017}. Sensitive low-frequency observations such as with LOFAR are likely to reveal many more ddrgs, as has been demonstrated by \citet{Mahatma2019} in the HETDEX field.
Candidate rejuvenated radio galaxies have also been identified from deep LOFAR images of the Lockman Hole region, constituting $\sim$15 per cent of the sample \citep{Jurlin2020}.
Of the 74 sources compiled by \citet{Kuzmicz2017} only 12 are listed to have a size of the inner double which is less than 20 kpc, falling in the category of either PS or CSS objects. Of these 12 the nuclear structure in the quasar B0738+313 is more likely to be that of a core-jet type rather than a compact double \citep{Stanghellini2001b}. The size of the inner double in Cen A is $\sim$12 kpc \citep{Israel1998,Feain2011} and is included in our list. The CSS/PS sources with evidence of recurrent activity including the ones from \citet{Kuzmicz2017} are summarized in Table \ref{t:rejuv}. 

\begin{table*}
	\caption{Rejuvenated radio sources associated with CSS/PS sources}
	\begin{tabular}{l l  c  l l l l l}
  \hline
 Source & Alt. &  Opt. &   z   & $l_{\rm in}$ &  $l_{\rm out}$ & Refs. & Notes \\
        &      &  Id.  &       &  kpc     &   kpc      &       &       \\
  \hline
  J0111+3906 &  B0108+388 & G & 0.6685 & 0.07  & 126   &  1,4    & GPS; a    \\
  J0301+3512 &  4C34.09 &  G  & 0.0165 & 0.66  & 234   &  2,26,27  & CSS \\
  J0318+1628 &  CTA21   &  Q  & 0.907  & 0.09  & 2.4   &  17,18,19 & GPS; a \\
  J0405+3803 &  B30402+379 & G & 0.0550 & 0.04 & $\sim$10 &  24,25 & CSS; c \\
  J0821+2117 &  B0818+214 & G & 0.4180 & 2.7   & 210   &  3    & CSS    \\
  J0943$-$0819 &  B0941-080 & G & 0.228  & 0.18  &  72   &  4    & GPS; a \\
  J1006+3454 &  3CR 236   & G & 0.101  & 1.8   & 4248  &  5,6  & CSS \\
  J1247+6723 &            & G & 0.107  & 0.02  & 1196  & 7,8   & GPS \\
  J1325$-$4301 &  Cen A     & G & 0.0018 & $\sim$12 & $\sim$600 & 28,29 & CSS \\
  J1347+1217 &  4C12.50   & G & 0.1217 & $\sim$0.3 & 130 & 5,30 & GPS \\
  J1352+3126 &  3CR293    & G & 0.045  & 1.1   & 180   & 9,10,11 & CSS  \\
  J1407+2827 &  OQ+208    & Q & 0.0766 & 0.01  & 0.04  & 20, 21 & GPS, a \\
  J1511+0518 &            & G & 0.084  & 0.01  & 0.05  & 22, 23 & PS, a \\
  J1513+2607 &  3CR315    & G & 0.1083 & 5.9   & 262   & 12  & CSS?  \\
  J1516+0701 &  3CR317    & G & 0.0345 & 0.05  & 51    & 13,14  & GPS; b  \\
  J1528+0544 &            & G & 0.0401 & 15    & 645   &  15     &  CSS  \\
  J1628+3933 &  3CR338    & G & 0.0304 & 7.8   & 48    &  16     & CSS   \\
    \hline
	\end{tabular}
      \label{t:rejuv}
 
 a. extended emission is one-sided; b. \citet{Venturi2004} show the bright central component to have a peaked spectrum with outer steep-spectrum lobes; c. radio source with double nuclei \citep{Bansal2017}
 
 1. \cite{Baum1990}; 2. \cite{Shulevski2012}; 3. \cite{Marecki2009}; 4. \cite{Stanghellini2005}; 5. \cite{Schilizzi2001}; 6. \cite{ODea2001}; 7. \cite{Marecki2003a}; 8. \cite{Bondi2004}; 9. \cite{Bridle1981}; 10. \cite{Akujor1996}; 11. \cite{Joshi2011}; 12. \cite{Saripalli2009}; 13. \cite{Venturi2004}; 14. \cite{Zhao1993}; 15. \cite{Kuzmicz2017}; 16. \cite{Ge1994}; 17. \cite{Jones1984}; 18. \cite{Kellermann1998}; 19. \cite{Dallacasa1995b}; 20. \cite{Luo2007}; 21. \cite{Wu2013}; 22. \cite{Orienti2008d}; 23. \cite{An2012b}; 24. \cite{Maness2004}; 25. \cite{Bansal2017}; 26. \cite{Brienza2018}; 27. \cite{Murthy2019}; 28. \cite{Israel1998}; 29. \cite{Feain2011}; 30. \cite{Morganti2013} 
 
 \end{table*}

Time scales of interruption of jet activity covers a wide range from $\sim10^{7}$ to $10^{8}$ yr for larger sources based on dynamical and spectral ageing arguments as in the case of 3C236 \citep{ODea2001} and 4C34.09 \citep{Shulevski2012} to $\sim10^5$ yr for 3C293 \citep{Joshi2011}. In the case of CTA21, \citet{Salter2010} suggested that there may be repeated cycles of activity based on the radio structure on different scales in VLBI observations of varying resolutions. This appears consistent with the suggestion by \citet{Reynolds1997} that jet activity may be intermittent on time scales on $10^{4-5}$ yr for the CSS/PS sources. 

Statistical studies of samples of radio sources based on their luminosity functions suggest that the time scales of the active phase are dependent on luminosity and mass of the optical host galaxy, with the time scales of jets being off are $\sim10^7$--$10^8$ yr \citep{Shabala2008,Best2005}. In a study of dying radio sources \citet{Parma2007} find the typical age of the active phase to be $\sim10^7$--$10^8$ yr. In the case of double-double radio galaxies the suggestion by \citet{Kaiser2000b} that the formation of the inner double could be due to the dispersion of hot clouds of the inter-galactic medium into the cocoon on time scales of $\sim10^7$ yr may work for the large sources, but not for the small CSS/PS sources. For the smaller time scales one needs to understand the fuelling process in the central regions of these sources. In this context it is interesting to note that the rejuvenated sources appear to have a higher incidence of detection of H{\sc i} in absorption than the general population of CSS/PS sources \citep{Saikia2009,Chandola2010}, although this needs further investigation using larger samples. The H{\sc i} properties of CSS/PS sources are discussed briefly in Sect.~\ref{s:hosts} and have been reviewed recently by \citet{Morganti2018}.

\subsection{Proper motion}\label{s:proper}
Determination of the proper motion of the hotspots from high-resolution observations of these compact sources provides us with estimates of their velocities and kinematic ages.
Among the early measurements, \citet{Tzioumis1989} found no significant evidence of changes in structure of the southern CSS source B1934$-$638 over a 12-year time scale, {although evidence of expansion was reported later by \citet{Ojha2004} over a 32-year time scale. } \citet{Taylor1997} also found no evidence of significant hotspot advance motion in the CSS/PS source with bi-directional jets B1946+708.  However, \citet{Conway1994} reported possible motion in B0108+388 and B2021+614 and transverse motion in B0711+356. The first clear evidences of proper motion of hotspots in CSS/PS sources were reported by \citet{Owsianik1998a} for B0710+439 and \cite{Owsianik1998b} for B0108+388. These were based on VLBI observations stretching over a decade. 
Since then, observations of outward proper motions or upper limits to these for CSS/PS sources have been reported for about 40 sources with the observations being largely at 5, 8--10 and 15 GHz \citep{Polatidis1999,Taylor2000,Tschager2000,Polatidis2002,Polatidis2003,Giroletti2003,Ojha2004,Gugliucci2005,Nagai2006,Giroletti2009,Stanghellini2009b,An2012b,Wu2013,Rastello2016}. In PKS1155+251, the components appear to be moving inwards resulting in the source shrinking in size \citep{Tremblay2008}. \citet{Yang2017} observed the source with the VLBA at 24 and 43 GHz and suggested that the southern component may harbour another black hole. Further observations are required to clarify the nature of this source. Apparent contraction has also been noticed between the core and the southern component in J0650+6001 which could be due to a knot in the jet moving south \citep{Orienti2010}. The CSO J1755+6236 associated with the nearby galaxy NGC6521 has no prominent hotspots in the outer lobes, suggesting that energy supply from the nucleus may have stopped, and exhibits no radial motion \citep{Polatidis2009}. It is relevant to note that the measurements of proper motion and hence velocities could be affected by opacity and resolution, as well as intrinsic variations in the velocities of components or birth of new radio components in the core. 
 
The velocities of separation of the hotspots from the nucleus of the galaxy range from $\sim$0.04c to about 0.4c with a median value of $\sim$0.1c. The corresponding kinematic ages range from as low as 20 yr to several thousand years with a median value of $\sim$600 yr. Estimates of the radiative age of these sources from a spectral break due to synchrotron ageing \citep{Murgia1999,Murgia2003} also yield similar values given the uncertainties, clearly suggesting that most CSS/PS sources are young objects. Considering a few specific examples, the radiative age of B1943+546 \citep{Murgia2003} is very similar to the kinematic age listed here. Among the well-studied sources, \citet{Giroletti2003} find the radiative age of 4C31.04 to be $\sim$4000 yr, but as they note this could reduce to $\sim$1000 yr depending on the assumptions, which is within a factor of 2 of the kinematic age. In the case of CTD93, \citet{Nagai2006} estimate the radiative age only in the northern lobe due to complex spectral break behaviour in the southern one and find the estimated hotspot speed to be similar to that estimated from proper motion. Many of  the CSS/PS sources exhibit prominent radio jets, especially in those associated with quasars. High-resolution observations of knots of emission in the jets yield velocities which are usually higher than those observed for the hotspots as for example in OQ208 \citep{Wu2013}, with some of the features exhibiting superluminal motion as in the inner jet knot of 3C48 \citep{An2010}. Proper motion measurements of CSS/PS sources are summarized in Table \ref{t:proper_motion}. For those with a significantly different value of Hubble's constant, the values have been listed for H$_o$=70 km s$^{-1}$ Mpc$^{-1}$. The notes indicate whether the velocity has been estimated between two hotspots (hs-hs), a core and a hotspot (c-hs) or of components in a jet (c-j, hs-j, j). Evidence of non-radial motion has been denoted by nr and the references are listed at the bottom of Table \ref{t:proper_motion}.

\citet{Polatidis2003} explored possible correlations with other source parameters such as redshift and size. They reported a possible correlation with redshift, but also noted that this could be affected by the difficulty of measuring proper motion in high redshift objects. With the new measurements since, which are often over longer baselines, there does not appear to be a significant correlation with either redshift or size. For a redshift of about 0.5 or above, the lowest radial velocity from proper motion measurements is 0.12$\pm$0.04c for J1324+4048 \citep{An2016b}. The lack of a dependence of velocity of advance on size or separation from the core is not surprising, as large radio galaxies have also been estimated to have similar advancement velocities  \citep{Scheuer1995}. 
\citet{deVries2010} determined the expansion velocities of two low-luminosity radio AGN from the CORALZ sample to be within $\sim$0.1c, and suggested that expansion velocities may be positively correlated with radio source luminosities. 
It would be interesting to determine the expansion velocities for a larger sample of low-luminosity radio AGN for comparison with the higher-luminosity ones. 

{At present over $\sim$75 per cent of the sources monitored show evidence of motion including those which show non-radial motion, with a median expansion velocity of $\sim$0.1c. The time range over which the sources have been monitored ranges from about 3 to 30 years with a median value of about 5 years. Almost all the sources which do not exhibit evidence of motion have been monitored for not more than about 5 years. Considering that B1934$-$638 showed evidence of expansion when observed over about a 32-year time scale \citep{Ojha2004}, the ones with upper limits are also likely to exhibit expansion when observed over a longer time scale. Monitoring over long time scales approaching about a decade or more has often revealed complex and interesting motion of the hot spots \citep[e.g.][]{An2012b}, and it would be desirable to extend such studies to larger samples of sources of varying luminosity. }

\subsubsection{Non-radial motion}
\begin{figure}
\centerline{\includegraphics[width=12cm]{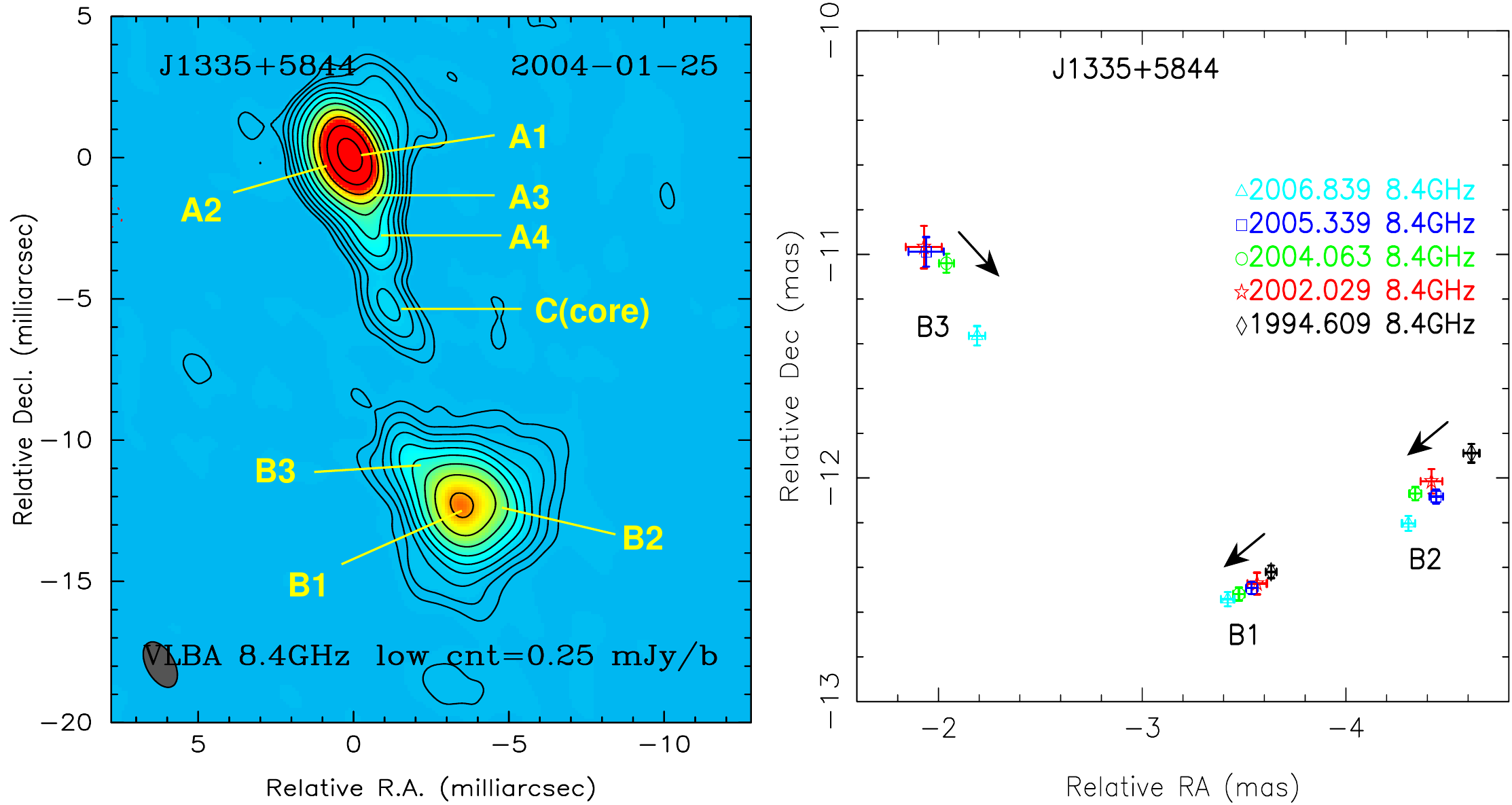}}
\caption{VLBI images of the PS source {J1335+5844 } with a compact symmetric structure showing the motion of the components B1, B2 and B3 relative to A1. The dominant motion of the hot spots B1 and B2 are transverse to the source axis. The slow radial motion of the primary hotspot B1, which suggests a kinematic age of $\sim$1800$\pm$1150 yr, and the dominant transverse motion reflects interaction with the external medium \citep{An2012b}. B3 has the highest radial velocity and is possibly part of a jet. }
\label{f:J1355+5844_tmotion}
\end{figure}

Although velocities of separation are measured along the source axis, a number of sources are seen to exhibit motion transverse to the source axis. \citet{Stanghellini2009b} reported evidence of transverse motion surprisingly in all the three sources they observed, namely J0706+4647, J1335+5844 and J1823+7938. Similarly, \citet{An2012b} find evidence of a wandering hotspot in the CSO quasar J1324+4048 which has a loop-like motion, initially moving towards the north and then abruptly moving towards the south-west, reminiscent of the dentist drill model of \citet{Scheuer1982} for multiple hotspots. \citet{An2012b} also confirm the transverse motion in the CSO galaxy J1335+5844 from measurements spanning about 12.2 years, which was reported earlier by \citet{Stanghellini2009b} from a time range of $\sim$2 year. The dominant transverse motion of the hotspots B1 and B2, and component B3 which is possibly part of a jet are illustrated in Fig.~\ref{f:J1355+5844_tmotion}.

When observed with sufficient angular resolution the lobes on opposite sides of the nucleus of the CSS/PS sources often show multiple hotspots. 
Such features are also seen in larger sources and could arise due to a number of effects. For example the head of the jet could be interacting at different points at different times as in the dentist-drill model \citep{Scheuer1982}, with the primary hotspot being the present location and the secondary hotspot an earlier one. This has been suggested to be the case for J1324+4048 \citep{An2012b}. Alternatively these may be due to precession of the jet axis \citep{Gower1982} or redirected outflows from the primary hotspot to a secondary one \citep{Laing1981,Lonsdale1986}. In the latter case the secondary hotspot is being continuously fed from the primary one, and radiative ageing is not likely to be significant. A deflected outflow has been suggested for the double hotspots in J1511+058 where the secondary one appears to move away from the primary one \citep{An2012b}. Both precession and redirected outflows have been explored for the multiple hotspots of OQ208 (Fig.~\ref{f:rejuv03}) which exhibits complex motion and advances with a significantly smaller velocity than the jet \citep{Wu2013}. While the dentist-drill scenario and redirected outflows suggest inhibition in the advancement of the jet, these do not imply that the source would be confined to small dimensions for its lifetime.

\subsubsection{Structural asymmetries and hotspot velocities}\label{s:asymm}
In sources with a radio core, the velocities could be determined for each of the outer hotspots on opposite sides relative to the core. In such cases it is possible to examine any relationship between hotspot speed and structural asymmetries as the radio jets plough their way outwards through a dense and inhomogeneous external medium. The effects may be more apparent in the most asymmetric sources. \citet{Orienti2007a} considered two very asymmetric sources, B0147+400 and B0840+424, and found from their radiative ages that the hotspot velocities range from 0.005 to 0.05c for the closer one, and from 0.03 to 0.3c for the farther one. This suggests that the closer hotspot is propagating through a denser medium leading to higher dissipation of energy and leading to greater hotspot pressure \citep{Jeyakumar2005,Bicknell2003}, consistent with studies based on structural asymmetries of the sources \citep{Saikia2001,Saikia2003a,Saikia2003b}. A denser medium on the side of the closer hotspot may also be probed by the detection of H{\sc i} absorption towards the brighter and closer lobe \citep[e.g.,][]{Orienti2007a}. \citet{Labiano2006} find the brighter and closer hotspot in 3C49 and 3C268.3 to be detected in H{\sc i} absorption, be depolarized and show evidence of line-emitting gas, although they cannot rule out similar amount of H{\sc i} gas in the opposite lobe. High-resolution H{\sc i} absorption measurements of a large sample of CSS/PS objects would be helpful to examine this further.

Direct measurements of velocities of hotspots relative to the core have also been made in a number of asymmetric sources. The closer and brighter hotspots in J0111+3906 and J1944+5448 are moving with smaller velocities compared with the opposite ones, the one in the more asymmetric source J1944+5448 being lower by more than a factor of 2 \citep{Rastello2016,Polatidis2002}. In sources which are reasonably symmetric, the difference in velocity may not be as striking as seen for example in 4C31.04 \citep{Giroletti2003} and J1511+0518 \citep{An2012b}.

\begin{table*}
	\caption{Proper motions of CSS and PS sources}
	\begin{tabular}{l l l l l l l l}
  \hline
 Source & Alt. &   z   & Sepn. &  v$_{sep}$. & Age & Refs. & Notes \\
        &      &       &  pc  &  c           & yr  &           &       \\
        &      &       &  (mas)  &  (mas/yr)           &   &           &       \\
  \hline
  
  J0000+4054 &  4C40.52    &        & (40) & ($<$0.144) & $>$280 & 6 & hs-hs \\
  J0003+2129 &             & 0.45   &   21  & 0.15 & 500 & 18 & hs-hs \\
  J0003+4807 &             &        & (4.8) & ($<$0.014) & $>$340 & 6 & c-hs \\
  J0005+0524 &             & 1.887  & 15  & 0.7 & 140 & 18 & hs-hs \\
  J0038+2303 & B2 0035+22  & 0.0960 & 31    &  0.17     &  567   & 12,9  & hs-hs   \\
  J0111+3906 & S4 0108+388 & 0.6685 &  32  & 0.26  &  417    &  12,1,3,4 & hs-hs  \\
             &  &  &  41  & 0.04  &  1080    & 13 & hs-hs  \\
  J0119+3210 & 4C 31.04 & 0.0602 & 60  &  0.33 & 550 & 5 & c-hs \\
  J0132+5620 &          &        & (12.2) & ($-$0.005)   & $>$4700 & 15 & hs-hs;nr \\ 
            &          &        & (11.5) & (0.028)   & 410 & 15 & j-hs \\
            &          &        & (10.5) & (0.060)   & 180 & 15 & j-hs \\
            &          &        & (2.4) & ($-$0.073) &  & 15 & j-hs \\
  J0204+0903 &             &        & (18) & (0.070) & 240 & 6 & c-hs \\
  J0405+3803 & B3 0402+379  & 0.0550 & 23  & 0.14  & 502  &  21 & c1-hs \\
            &             &        & 7.3 & 0.0054 &    & 22 & c1-c2 \\
  J0427+4133 &             &        & (1.3) & (0.060) & 20 & 6 & c-hs? \\
  J0518+4730 &             &        & (3.1)   & ($-$0.026)       & $\sim$1200 & 15 & c-hs;nr \\
            &             &        & (1.9)   & (0.030)  & & 15 & c-hs;nr \\
  J0620+2102 &             &        & (27) &  ($<$0.013) & $>$2060 & 6 & hs-hs \\
  J0650+6001 &             & 0.455  & 40   & 0.39 & 360 & 20 & hs-hs \\
            &             &  & 17   & $-$0.37 & & 20 & c-hs \\
  J0713+4349 & B3 0710+439 & 0.5180 & 152 & 0.43 & 932 & 12,2,3,4 &  hs-hs \\
  J0754+5324 &             &        & (20) & ($<$0.009) & $>$2220 & 6 & hs-hs \\
          &               &   & ($\sim$8) & (0.060) & & 6 & j-hs \\
  J0831+4608 &  CORALZ     &  0.1311  & 10 & 0.14 & 245 & 25 & hs-hs \\    
  J1035+5628 & TXS 1031+567& 0.4597 & 198 & 0.27  & 1836 & 12,3  & hs-hs;nr  \\
          &       &  & 190 & 0.86  &  & 12,3  & hs-j?  \\
  J1111+1955 & PKS 1108+201 & 0.2991 & 71 & $<$0.14c & $>$1620 & 6 & hs-hs \\
            &             &        & 78 & $<$0.19c & $>$1360 & 6 & hs-hs \\
  J1143+1834 &             &        & (6.9) & ($<$0.010) & $>$690 & 6 & hs-hs \\
  J1247+6723 & B 1245+676  & 0.1073 & 15  &  0.23     & 190     & 12,9,23  &  hs-hs \\
  J1317+4115 &  CORALZ     &  0.0662  & 4.2 & $<$0.11 & $>$130 & 25 & hs-hs \\    
  J1324+4048 &             & 0.496  & 33 & 0.12   & 870 & 15 & hs-hs \\
            &             &        & 28 &  2.2  &  & 15 & hs-j \\
  J1335+5844 & 4C 58.26    & (0.57) & 85 & 0.16 & 1800 & 15 & hs-hs;nr \\
            &             &        & 84 & 0.45 & 600 & 15 & hs-hs;nr \\
            &             &        & 75 & 2.86 & 90 & 15 & j-hs;nr \\
  J1407+2827 & OQ 208      & 0.0766 &  11     & 0.108  &    & 16,12  & hs-hs;nr  \\
            &             &        &  9.5    & 0.134  & 255 & 16,12 & hs-hs;nr \\
            &             &        &  1.8    & $-$0.168  &   & 16  & hs-j  \\
  J1414+4554 & B3 1412+461 & 0.186  & 89 & $<$0.14 & $>$2030 & 6 & hs-hs \\
  J1415+1320 & PKS1413+135 & 0.2467 & 31 & 0.80 & 130 & 6 & c-j \\
            &             &        & 7.6 & 1.10 & 22 & 6 & c-j \\
  J1511+0518 &             & 0.084  & 3.4  & 0.04  & 300 & 15,18 & c-hs \\
   &             & & 4.0  &   &  & 15,18 & c-hs;nr \\
   &             & & 5.3  & 0.15 &  & 15,18 & c-hs;nr \\
  J1546+0026 &             & 0.55  & $\sim$35  & $-$1.1  & & 6 & c-hs \\
  J1609+2641 & CTD 93  &  0.473 & 300 & 0.34 & 2200 & 7 & hs-hs \\
  J1723$-$6500 & NGC 6328    & 0.0144 &   3    & $\sim$0.1  & 91  & 24,12,9  & hs-hs  \\
  J1734+0926 & PKS 1732+094 & 0.735 & 96 & $<$0.18 & $>$1780 & 6 & hs-hs \\
           & PKS & 0.735 & 101 & 0.25 & 1300 & 15 & hs-hs;nr \\
  J1755+6236 & NGC6521      & 0.0274 & 20 & $<$0.04 & & 19 & hs-hs \\
  J1816+3457 &  B2 1814+34 & 0.245 & $\sim$140 &  & & 6 & hs-hs;nr \\
  J1826+1831 &             &       & (14) & (0.037) & 380 & 6 & c-j \\
            &             &        & (42) & (0.013) & 3000 & 6 & c-hs \\
  J1845+3541 & B 1843+356  & 0.7640 &  32     &  0.57     & 180     &  12,9 &  hs-hs \\
  J1939$-$6342 & PKS 1934-63 & 0.1813 &  130     &   0.26    &  1600    & 8,12  &  hs-hs \\ 
  J1944+5448 & B 1943+546  & 0.2630 &  153  &  0.37 &  1306   & 12,13,4,9  &  hs-hs \\
             &     & &  166 &  0.27 &  1800   & 13  &  hs-hs \\
  J1945+7055 & TXS 1946+708& 0.1008 &  56  &  $<$0.14  &     & 12  & hs-hs  \\
            &      &      &    &  $<$0.57  &  $>$200    & 11  & hs-c  \\
  &      &      &    &  0.29  &      & 12,11  & j  \\
  &      &      &    &  0.7-1.3  &      & 12,11  & j \\
  J2022+6136 & B 2021+614  & 0.2270 &  25     & 0.2 &  440    & 10,9  &  hs-hs \\
  J2203+1007 &        & 1.005 & 82 & 0.53 &  500 & 15,6 & hs-hs;nr \\
            &        &       & 76 & 1.13 & 220  & 15 & hs-j \\
  J2355+4950 & TXS 2352+495& 0.2379 & 190 &  0.17 & 300  & 12,3,4  & hs-hs \\
        &     &  &  &  0.4-1 &   & 3,12  & hs-j  \\
  
  \hline
			\end{tabular}
      \label{t:proper_motion}

{\tiny    
 1. \cite{Owsianik1998b}; 2. \cite{Owsianik1998a}; 3. \cite{Taylor2000}; 4. \cite{Polatidis1999}; 5. \cite{Giroletti2003}; 6. \cite{Gugliucci2005}; 7. \cite{Nagai2006}; 8. \cite{Ojha2004}; 9. \cite{Polatidis2002}; 10. \cite{Tschager2000}; 11. \cite{Taylor1997}; 12. \cite{Polatidis2003}; 13. \cite{Rastello2016}; 15. \cite{An2012b}; 16. \cite{Wu2013}; 17. \cite{Giroletti2009}; 18. \cite{Orienti2008d}; 19. \cite{Polatidis2009}; 20. \cite{Orienti2010}; 21. \cite{Maness2004}; 22. \cite{Bansal2017}; 23. \cite{Marecki2003a}; 24. \cite{Giroletti2009}; 25. \cite{deVries2010}  }
 
 \end{table*}
 

\subsection{Variability, orientation and unification schemes}\label{s:variability_orientation}
The most dramatic examples of variability in radio AGN in both total intensity and polarization are seen in the core-dominated flat-spectrum radio sources, especially the class of blazars. These can be understood via shocks-in-jet scenario in relativistic jets on parsec scales embedded in largely turbulent magnetic fields \citep{Aller2017}. Interstellar scintillation (ISS) which is sensitive to microarcsec scale structure and properties of the interstellar medium (ISM) provides an explanation for the intraday variability (IDV) seen in flat-spectrum AGN. The Micro-Arcsecond Scintillation Induced Variability (MASIV) Survey shows that more than half of their sample of 500 compact flat-spectrum AGN show evidence of IDV \citep[e.g.,][]{Jauncey2020}. Variability may also be seen due to gravitational lensing of for example relativistically moving features in the AGN jets \citep[e.g.,][]{Vedantham2017}.

\begin{figure}
\centerline{\includegraphics[width=8cm]{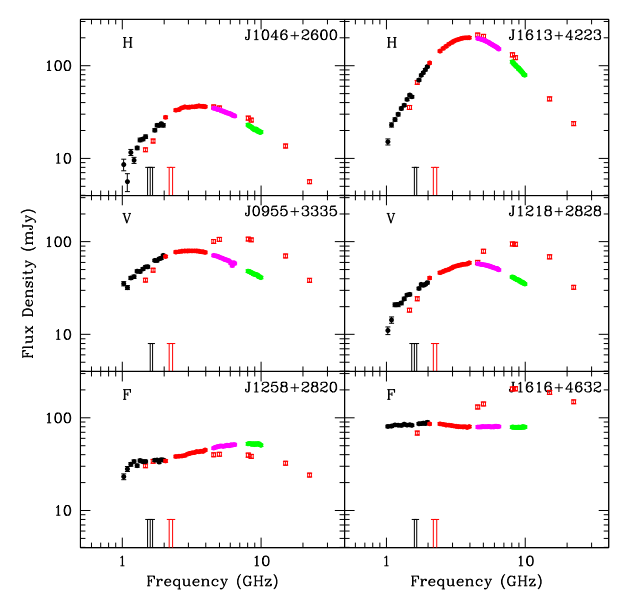}}
\caption{Variability of the integrated flux density of PS sources illustrating those where the spectrum remains convex without showing significant variability (upper panel) and those which show significant variability due to evolution of source components (middle panel). The bottom panel shows flat-spectrum objects which may exhibit a convex or peaked spectrum due to variability \citep{Dallacasa2016}. {The filled circles in different colours denote their observations with the Jansky Very Large Array in 2012 May in different bands: L (1--2 GHz) in black, S (2--4 GHz) in red, C(4.5--6.5 GHz) in violet and X (8--10 GHz) in green. The open squares denote earlier-epoch VLA data. } }
\label{f:int_variability}
\end{figure}

Variability in CSS/PS sources, although low compared with flat-spectrum AGN, may be caused by expansion of the compact outer components as they expand and traverse outwards with the turnover frequency shifting to lower frequencies, variations in optical depth, interstellar scintillation (ISS) and effects of Doppler boosting, although the latter is more likely to be significant in flat-spectrum radio sources (see Fig.~\ref{f:int_variability} for an illustration of variability in PS sources). Although there have been a number of monitoring programmes and several genuine PS sources are known to exhibit variability, the overall trend for genuine CSS/PS sources to exhibit low variability \citep{ODea1998} still holds. For example intensive monitoring of a small sample of 7 CSOs on an average every 2.7 days for about 8 months at 8.5 GHz showed their variability in flux density to be only $\sim$0.7 per cent \citep{Fassnacht2001}. Monitoring of a sample of 33 southern PS sources also showed low incidence of variability with a few exceptions \citep{Jauncey2003}. An interesting PS source in their sample is PKS1519$-$273 which in spite of its variability maintains a PS from 0.4 to 22 GHz and exhibits intraday variability in all Stokes parameters. This has been interpreted to be due to ISS of a 15--35$\mu$as component which is $-3.8\pm0.4$ per cent circularly polarized at 4.8 GHz \citep{Macquart2000}. {ISS has also been invoked to explain longer term variability at low frequencies in a sample of GLEAM peaked spectrum sources \citep{Ross2020}.
 \citet{Aller2002} present flux density and polarization monitoring of the 32 sources in the Stanghellini sample that are observable by the Michigan radio telescope. About 10 sources show flux density variability at some level though generally with longer time scales than found in other AGN studied by the Michigan group \citep{Aller2002}. Even during times of variability, the overall peaked spectrum is generally maintained \citep{Aller2002}.}

 Blazars in the act of flaring can have peaked radio spectra at high frequencies \citep[e.g.,][]{Dent1980,ODea1983,Wills1983,ODea1986,Kovalev2002,Kovalev2005}. This is one of the reasons why blazars can contaminate samples of PS sources. 
Consistent with this, while many of the PS sources with a peak frequency greater than about 3 GHz seem to vary, their spectra usually appear peaked only during an outburst \citep{Jauncey2003}. For a sample of 21 PS sources selected from the ATCA 20 GHz (AT20G) survey, \citet{Hancock2010} suggest only $\sim$60 per cent of the sources showed low-level variability over times scales of $\sim$1-3 yr at 20GHz and are likely to be genuine PS sources. \citet{Tinti2005} also suggested that most PS sources associated with quasars and peaking at high frequencies are likely to be flaring blazars. For a sample of `bright' high-frequency peakers \citep{Dallacasa2000}, multifrequency observations at different epochs showed that only $\sim$56 per cent are genuine PS candidates \citep{Orienti2007b}. Similarly for a sample of `faint' high-frequency peakers \citep{Stanghellini2009a}, about 56 per cent mostly associated with galaxies are candidate young sources, while the remaining largely quasars belong to the flat-spectrum blazar population \citep{Orienti2010c}. Long-term multi-frequency monitoring of high-frequency peaked sources is necessary to determine whether peaked-spectrum sources identified from limited observations are indeed genuine candidates for young sources. From long-term well-sampled data sets spread over about two decades \citet{Torniainen2005} showed that PS quasars are largely flat-spectrum sources having a peaked spectrum during a flare, although a few have a quiescent convex spectrum along with low variability. A similar study for PS galaxies showed that contamination by flat-spectrum objects is far smaller \citep{Torniainen2007,Torniainen2008}. Similar results were obtained from multi-frequency monitoring of a sample of 122 sources with the RATAN600 telescope over four years, with only $\sim$29 per cent identified as PS candidates and the majority of quasars as not genuine PS sources \citep{Mingaliev2012,Mingaliev2013}. These studies underline the importance of long-term multi-frequency monitoring, preferably simultaneously with adequate sampling to identify the genuine PS sources \citep{Tornikoski2001,Tornikoski2009}.

Besides identifying genuine PS sources which are likely to be young objects, long-term monitoring could also provide insights towards understanding the early evolution of these young objects. In the optically thick region of the spectrum adiabatic expansion is likely to play a major role in the case of synchrotron self absorption, and variations in optical depth in the case of free-free absorption. In the case of a self-absorbed component, the opacity decreases as the source expands with the turn-over frequency decreasing and flux density below the turn-over frequency increasing \citep{vanderLaan1966}. In the optically thin region above the turn-over frequency spectral evolution would be determined by both expansion and radiative losses, and injection of fresh particles from the nucleus. The spectral peak in the extreme PS source RXJ1459+3337 has decreased from $\sim$24~GHz in 1996 to $\sim$~12.5 GHz in 2003, accompanied by an increase in flux density below the spectral peak and a decrease above it. \citet{Orienti2008b} explain their observations of RXJ1459+3337 in terms of an adiabatically expanding homogeneous component and estimate the age of the radio emission to be $\sim$50 years. \citet{Dallacasa2016} besides reviewing earlier work present their results on three variable PS sources, J0754+3033, J0955+3335 and J1218+2828, which show similar behaviour for which detailed modelling would be useful. In the well-studied PS source PKS1718$-$649 (NGC 6328) variability was found both below and above the peak, where variations in the optically thick part have been attributed to variations in optical depth while in the optically thin part to adiabatic losses of the synchrotron-emitting lobes \citep{Tingay2003a,Tingay2015a}.

\begin{figure}
\centerline{\includegraphics[width=11.5cm]{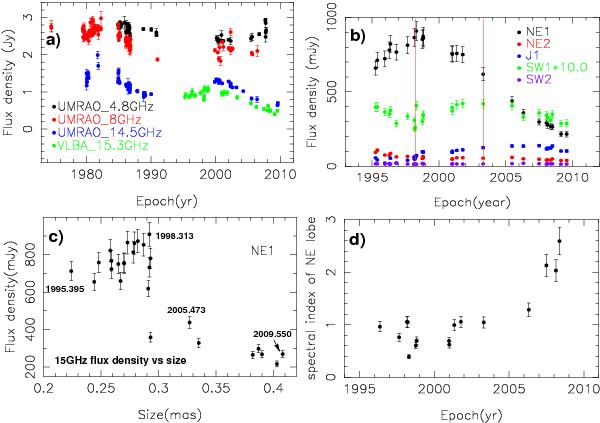}}
\caption{Variability of the integrated flux density and of the components observed by VLBI observations for the PS source OQ208. { Upper left: The total flux density variability of the source as measured with the UMRAO 26-m telescope at 4.8 GHz (black), 8 GHz (red) and 14.5 GHz (blue) is shown along with the sum of the flux densities of the VLBI components at 15.3 GHz (green).  Upper right: The flux density variability of the different VLBI components, with the north-eastern components NE1 and NE2 shown in black and red respectively, while the jet component J1 and the south-western components SW1 and SW2 are shown in blue, green and purple respectively. The flux density of SW1 has been multiplied by a factor of 10. The vertical lines mark the peak epochs of NE1 and SW1. Lower left: This shows the variation of the 15-GHz flux density of hotspot NE1 and its component size. The flux density of NE1 increased till 1998.313 followed by a decline. The component size was found to increase during the entire period. Lower right: This shows the temporal variation of the spectral index of the north-eastern lobe between 8 and 15 GHz }  \citep{Wu2013}.}
\label{f:OQ208_variability}
\end{figure}

In addition, high-resolution VLBI-scale monitoring observations would help reveal variability of individual features such as hotspots and cores. These reveal more complex and interesting variations which we illustrate with a couple of examples. OQ~208 (Mrk 0668 with Seyfert~1 type spectra) consists of two lobes with multiple hotspots and an overall separation of $\sim$10 pc (Fig.~\ref{f:rejuv03}) with both hotspots exhibiting variability, the maximum levels being $\sim$62 and 19 per cent for the north-eastern and south-western hotspots respectively (Fig.~\ref{f:OQ208_variability}). The complex variability pattern requires in addition to expansion and radiative losses, periodic feeding and acceleration of fresh relativistic electrons \citep{Wu2013}. Evidence of feeding is also seen in J1324+4048 a PS source associated with a quasar and with two steep-spectrum, resolved components, A and B, separated by $\sim$32 pc with no flat-spectrum core component. The flux densities of both components appear to decrease from 1993, reaching a minimum at 2005, consistent with expansion losses, and then found to increase by $\sim$40 per cent in 2009. In 1998 a new component was observed at the beginning of hotspot B which appears to be advancing with a superluminal velocity $\sim$2.2c along the axis, possibly feeding energy to the lobe \citep{An2012b}. The radio core and inner jet component in the CSS quasar 3C48 was also found to exhibit significant variability over a 8-year period \citep{An2010}. Long-term VLBI-scale monitoring of a large sample of CSS/PS sources is required to understand their systematics and the rich and diverse phenomena these may reveal.
At radio frequencies, differences in the prominence and variability of the radio core, detection of superluminal motion and source asymmetries could in principle be useful tests of the unified scheme for CSS/PS radio galaxies and quasars. {Superluminal motion is observed when the jet axis is inclined at a small angle to the line of sight with features in the jet appearing to move at velocities greater than than of light. The apparent velocity of such a feature is given by $\beta_{\rm app} = \beta sin~\theta/(1 - \beta cos~\theta) $, where $\theta$ is the angle of inclination of the jet axis to the line of sight. The apparent transverse velocity has a maximal value $\beta_{\rm app,max} = \beta\gamma$ for $sin~\theta = 1/\gamma$ where $\gamma$ is the Lorentz factor corresponding to $\beta$. }

In the unified schemes for radio AGN, the FRII radio galaxies which also tend to have strong emission lines are believed to be inclined at large angles to the line of sight while the quasars are inclined at small angles, the dividing angle being $\sim45^\circ$ \citep{Barthel1989}. Effects of relativistic motion such as Doppler boosting of the flux density of the extended and nuclear jets in the cores, {detection of superluminal motion}, as well as apparent structural asymmetries for intrinsically symmetric sources would be more apparent in those observed at small angles to the line of sight. In this scenario a dusty, clumpy torus obstructs a direct view of the broad-line region in the case of radio galaxies, while the orientation independent parameters such as the extended lobes of radio emission should be similar for both classes of objects. An FRII radio source cannot be changed to an FRI by orientation effects. The relativistically beamed counterparts of the FRI sources which also tend to have weak emission lines have been suggested to be BL Lac type objects. The unified schemes for radio AGN have been reviewed by a number of authors over the years \citep[e.g.,][]{Antonucci1993,Antonucci2012, Urry1995}, and more recently by \citet{Tadhunter2016a}.

\citet{Barthel1989} compared the projected radio sizes of FRII radio galaxies and quasars, and found the radio galaxies to be larger as would be expected if these are inclined at larger angles to the line of sight. However as CSS/PS sources have been defined to be $<$20 kpc, size comparisons may not be very meaningful. Also, asymmetries in both total intensity \citep{Saikia2001,Saikia2002,Saikia2003a,Rossetti2006} and linear polarization \citep{Saikia2003b} are more strongly affected by local inhomogeneities in the immediate environment of the CSS/PS sources rather than the differences in orientation, making it difficult to use these aspects for testing the unified scheme. At radio frequencies one could for example investigate the relative strengths of the cores due to relativistic beaming, core variability, detection of jets and superluminal motion which would depend on orientation.
The cores of radio galaxies have been found to be less prominent than those of quasars in CSS/PS sources as expected in the unified scheme \citep{Saikia2001,Fanti2009}, consistent with earlier studies for both small and large sources. Although there have been reports of core variability in individual sources, a systematic study to examine differences between galaxies and quasars is required even though there could be an unavoidable bias towards observing brighter cores. The candidate PS source J1324+4048 associated with a quasar exhibits superluminal motion of a jet component of 2.2$\pm$0.5c \citep{An2012b}. Clear evidence of superluminal motion of jet components have been reported in the CSS source 3C48 with v$\sim$3.7c \citep{An2010}, 3C138 with velocities of $\sim$2.6c and 7.2c for different components \citep{Cotton1997b,Shen2005} and 3C309.1 with superluminal velocities ranging from $\sim$1.4 to 6.6c for the different components \citep{Lister2013}. 3C216 earlier listed as a CSS has a velocity of $\sim$4c \citep{Paragi2000} but has since been known to have extended emission larger than 20 kpc. These sources are all associated with quasars consistent with the unified scheme \citep{Barthel1989,Antonucci1993,Urry1995}, although there is a natural bias towards monitoring sources with bright cores. In the case of the CSS quasar 3C286, \citet{An2017} have identified a core from multi-frequency VLBA data and determined the jet speed to be subluminal with a value of $\sim$0.5c and an inclination angle of 48$^\circ$. Clearly more data and measurements for a larger sample with varied core strengths are required. {Although only a few CSS and PS sources have been monitored in order to study the motion of the nuclear jets, large samples of sources with prominent cores have been monitored for example by the VLBA \citep[e.g.][]{Lister2013,Lister2019}. These studies show that the apparent speeds range from subluminal ones to over 40c, with the galaxies having small values, consistent with the unified scheme. }

Besides the radio band, unified schemes may be tested across different wavebands using both orientation independent properties which should be similar and also the orientation-dependent ones. Properties which have been studied include their environments, narrow emission line and mid-infrared luminosity, detection of polarized broad emission lines, host galaxy properties, high-energy emission which could be due to Doppler boosting and inverse-Compton scattering of lower energy photons.
The infrared properties suggest that compact sources appear to have similar structure of the central region with black holes, accretion disks and obscuring tori as in the case of  the larger sources (as discussed in Sect.~\ref{s:IR}). However, there may be additional obscuration for the CSS/PS sources if the triggering of radio sources is associated with significant  infall of gas and dust from a merger.
In the context of differences between quasars and radio galaxies \citet{Tadhunter2016a} notes that traditional indicators of bolometric luminosity of AGN such as [O{\sc iii}] emission line luminosity and 24$\mu$ continuum luminosity may also suffer attenuation in narrow-line objects due to circumnuclear dust. In PS and CSS sources, the [O{\sc iii}] emission line luminosity can be affected by a significant contribution from shocks
(Sect.~\ref{s:alignment}).
At X-ray wavelengths observations of a few PS/CSS quasars suggest that at least some of the X-ray emission is due to relativistic beaming (see Sect.~\ref{s:CSSQuasarXray}), consistent with the unified scheme.  \citet{Tadhunter2016a} shows in his review that a wide range of observations can be explained in terms of the unified schemes for radio AGN although there are aspects that need to be understood further. These studies have been done for well-defined samples of luminous radio sources which include CSS/PS sources, but could be extended to larger samples of CSS/PS sources, including the low-luminosity ones. 

\subsection{What causes the turnover in the radio spectrum?}\label{s:turnover}
The peak in the radio spectrum is a defining characteristic of all PS and even many CSS sources. The mechanism responsible for the turnover in the spectrum is of great importance because it will constrain either the internal properties (e.g., magnetic field, thermal electron density) or external environment (e.g., thermal electron density) of the radio source \citep[e.g.,][]{deKool1989,ODea1998,Tingay2003a}. The two mechanisms currently favored are synchrotron self-absorption (SSA) and external free-free absorption (FFA). 

\citet{ODea1998} noted that the magnetic field which would produce the observed turnover in GPS sources due to SSA  was in agreement with the magnetic field calculated assuming minimum pressure. \citet{Snellen2000b} noted that the relation between source size and turnover frequency could be reproduced if the turnover was due to SSA and the sources evolved in a self-similar way while remaining in equipartition. \citet{deVries2009a} confirm that SSA is a better explanation than FFA for the relation between peak frequency and angular size in the CORALZ sample. 
\citet{Jeyakumar2016} found that by adding synchrotron opacity effects into the evolution model of \citet{Kaiser2000a}, it is possible to fit the relation of frequency at the spectral peak vs.\ linear size, and the relation of luminosity of the spectral peak vs.\ linear size. These results are consistent with the turnover in the radio spectrum being due to SSA. Uniform source models are unrealistic and non-uniform source models may give better fits to SSA \citep[e.g.,][]{Artyukh2008}. 

FFA has been suggested to occur in a disk \citep{Marr2001}, a torus \citep{Peck1999,Kameno2000,Kameno2001,Kameno2003}, and in ambient clouds shocked by the expanding radio source, producing a 
 power-law distribution of electron density in the ionized ambient clouds \citep{Bicknell1997,Bicknell2018}.
 There is supporting evidence from individual sources whose spectra are better fit by FFA rather than by SSA \citep[e.g.,][]{Bicknell1997,Peck1999,Kameno2000, Kameno2001,Kameno2003, Marr2001,Marr2014, Tingay2015a,Callingham2015,Mhaskey2019a} and see \citet{Gopal2014}, \citet{Callingham2017} and \citet{Mhaskey2019b} for additional candidates. The FFA is generally consistent with an ionized medium with a temperature $T\sim 10^4$ K, path length $\sim$1--100~pc, and an electron density $n_e \sim 10^3$--$10^4\mathrm{ cm}^{-3}$.

{Thus, typically, the FFA scenario requires high thermal electron density in the environment of compact radio sources. In the models of Bicknell and collaborators, these thermal electrons are generated by shock heating dense ambient clouds.} However,  there is evidence for dense clouds in the environments of some but not all compact radio sources (Sect.~\ref{s:gas_content}, Sect.~\ref{s:frustrated}). The evidence for large gas masses in individual sources suspected of FFA is mixed -- Table \ref{t:gas-mass} shows high molecular gas masses ($> 10^{10}\,M_\odot$) for 0108+388 and OQ208, but not for PKS 1718$-$649 and TXS 1946+708. {Given that the link between dense molecular gas clouds and high thermal electron densities is still uncertain, these limits should not be considered restrictive. }

 \citet{Mutoh2002} suggest that the polarization behavior is more consistent with FFA in some sources. However, the Bicknell model does less well in reproducing the relation between the frequency of the peak and the linear size over the large range (over 3.5 dex) in linear size \citep{Bicknell2018}.

In summary, there is evidence in some individual sources that the turnover is due to FFA. However, the global relationship between frequency of the peak and the linear size is best described by SSA. Thus, we suggest that in most sources the turnover is due to SSA; while in some sources possibly interacting with dense ambient clouds, the turnover is due to FFA.

\section{Infrared properties}\label{s:IR}

The infrared properties of GPS and CSS sources can constrain the AGN bolometric luminosity, the covering factor  of the obscuring material (do they have higher covering factor than large radio galaxies?),  the properties of the circumnuclear region (e.g., torus), and the star formation rate (SFR) averaged over about 200 Myr. See \citet{Tadhunter2016b} for a  discussion of caveats in the determination of SFR in galaxies with AGN.  We list SFR, where available, in Table \ref{t:SFR}.

IRAS and ISO did not detect many GPS and CSS sources and the initial results obtained were consistent with no difference in mid-far Infrared properties between GPS/CSS  and large radio galaxies \citep{Heckman1994, Fanti2000,Hes1995}.

\citet{Willett2010} obtained Spitzer IRS spectra of eight low redshift CSOs. 
{\citet{Willett2010} detect the 11.3 $\mu$m PAH feature in $\sim 87\%$ of the CSOs.}
They find that the sources are heterogeneous in the mid-IR properties, displaying a varying mixture of AGN and dusty star formation properties. Clumpy torus AGN models can fit the silicate features in most of the CSOs.  Based on the PAH strength, silicate features, fine-structure lines, and H$_2$ lines, they find that 4 categories are necessary to classify the CSOs.  A slight majority (5/8) CSOs show moderately dusty environments and moderate ($< 10\,M_\odot$ yr\mone) levels of star formation.

\citet{Dicken2012} compared Spitzer observations of the 2Jy and 3CRR radio galaxies including 8 GPS or CSS objects. Similar to \citet{Willett2010} they find that the GPS and CSS sources are  heterogeneous in their mid-IR properties (PAH strength, silicate features, fine-structure lines, and H$_2$ lines). In addition \citet{Dicken2012}  find a high fraction (6/8) with PAH emission indicating the presence of moderate star formation. Dicken et al. point out that the fraction of compact sources with star formation (75\%) is higher than found for the extended (2 Jy + 3CRR) radio galaxies (21\%). 

Note that \citet{Dicken2012} do not provide any estimates of star formation rate, however, they do provide the flux of the  11.3 $\mu$m PAH feature. \citet{Farrah2007}  give a SFR calibration using the luminosity of the 6.2 $\mu$m  + 11.3 $\mu$m PAH features, 
\begin{equation}\label{eq:PAH}
SFR (M_\odot \ {\rm yr}^{-1}) = 1.18 \times 10^{-41} L_{\rm PAH} ({\rm ergs\ s}^{-1})
 \end{equation} 
where $L_{\rm PAH}$ is the combined luminosity of the 6.2 $\mu$m  and 11.3 $\mu$m PAH features. We note that the 6.2 $\mu$m  and 11.3 $\mu$m fluxes are approximately equal in starburst galaxies \citep{Brandl2006}, and double the 11.3 $\mu$m flux to estimate an approximate SFR which  are included in Table \ref{t:SFR}. Note that the 11.3 $\mu$m PAH is easily excited and  probably gives an upper limit to the SFR \citep{Ogle2010}.

\begin{figure}
\centerline{\includegraphics[width=11cm]{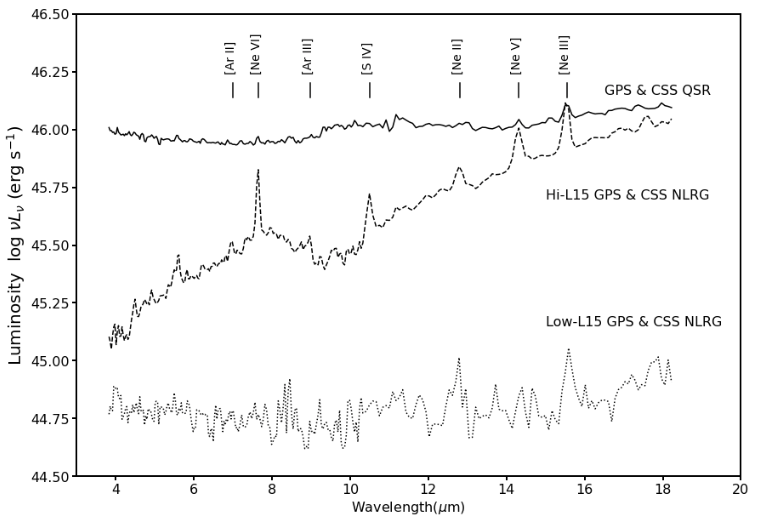}}
\caption{The median MIR spectra of QSRs and NLRGs. These spectra reflect the difference in luminosity seen in Fig.~\ref{f:size_IR}.
 The NLRG population are split into low-luminosity and high luminosity subsamples,  $\nu L_\nu$ (15 $\mu$m) $< 1.3 \times 10^{45}$ erg s\mone,  and $\nu L_\nu$ (15 $\mu$m) $>1.3 \times 10^{45}$ erg s\mone, respectively. The Hi-L15 NLRGs have very red
spectra with strong silicate absorption and high-EW emission lines, consistent with quasars viewed at a high inclination or through a high dust column. The Low-L15 NLRGs have much weaker MIR continuum and emission lines, consistent with their 10x lower luminosity. (Figure from Ogle et al., priv, comm.)
\label{f:mean_IR}}
\end{figure}

Ogle et al. (in preparation) present Spitzer IRS and MIPS observations of 24 GPS and CSS sources at $z=0.4$--$1.0$ and compare them to large FRII radio sources. Median MIR SEDs are shown in Fig.~\ref{f:mean_IR}. CSS and PS quasars have similar mid-IR spectra to the FR II quasars studied by \citet{Ogle2006}. Most of the GPS and CSS radio galaxies show strong high ionization lines such as [Ne V] and [Ne VI] that must be photoionized by obscured AGNs.  The distribution of 15 $\mu$m luminosity is similar between the GPS and CSS and the large FRII sources, with
$\nu L_\nu (15 \mu m) = 10^{45}-10^{46}$ erg s\mone. The Spitzer results from Ogle et al. and the Herschel results from \citet{Podigachoski2016b}  indicate that the compact sources have similar central engines (BHs, accretion structures, obscuring tori) as the large sources. 

There is evidence for additional obscuration in the PS and CSS sources (Ogle et al. 2020). High-luminosity CSS and PS NLRGs (HERGs) have heavily reddened MIR spectra (Fig.~\ref{f:mean_IR}). The greater obscuration compared to FR II NLRGs indicates a larger column density of cold dust in the circumnuclear region or host galaxy.  Lower-luminosity  CSS and PS NLRGs ($\log \nu L_\nu  < 45.0$) may host a mix of obscured and unobscured AGNs, which are difficult to separate based on their low-quality MIR and optical spectra. The MIR spectra of obscured CSS quasars are also heavily reddened and have high-equivalent width emission lines, presumably from the extended narrow-line region.

\begin{figure}
\centerline{\includegraphics[width=11cm]{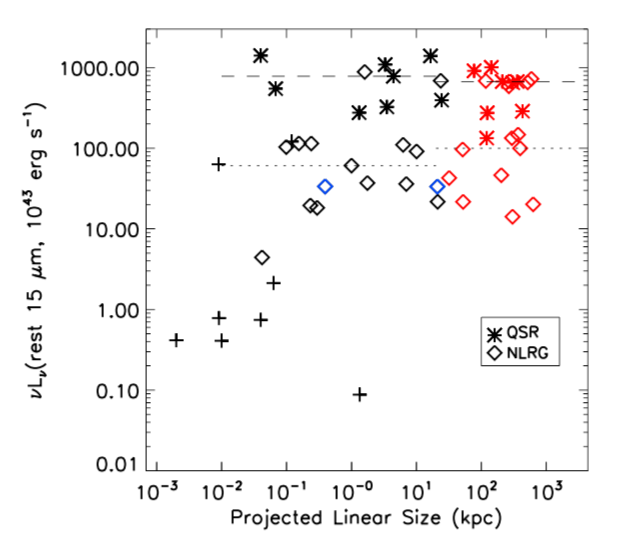}}
\caption{Rest frame MIR luminosities of GPS and CSS quasars (asterisks) and radio galaxies (diamonds) compared to FR II quasars and radio galaxies, vs. projected linear size (Ogle et al. priv, comm). The weakest sources (+ symbols) are nearby CSOs from \citet{Willett2010}. Red = FRII data from \citet{Ogle2006}. Black = Ogle et al. CSS+GPS data.   Blue= CSS/GPS NLRGs with silicate emission.  The luminosities are given as $\nu L_\nu$ (rest 15 $\mu$m), measured from Spitzer IRS spectra.  Quasars of all sizes have high MIR luminosities.  NLRGs, on the other hand show a wider range of MIR luminosity, suggesting that they are a more heterogeneous population.  Regardless of size, there are very few NLRGs (only 4 in the sample) with MIR luminosities as high as quasars. It seems unlikely that the factor of 10 difference in median luminosity (dashed lines for quasars, dotted for NLRGs) is caused by extinction at 15 microns.  Instead, they may be drawn from an intrinsically less luminous population.  
\label{f:size_IR}}
\end{figure}

PS and CSS quasars seem to have systematically higher 15  $\mu$m luminosity than radio galaxies (Fig.~\ref{f:mean_IR}, \ref{f:size_IR}), even after accounting for dust obscuration. This is consistent with known problems with unifying PS and CSS quasars with radio galaxies (Sect.~ \ref{s:variability_orientation}).
It is unclear whether the type 1 fraction increases with luminosity or if it is just more difficult to detect broad lines in the optical spectra of the lower luminosity radio galaxies. A difference in intrinsic luminosity between the quasars and radio galaxies is consistent with the galaxies containing a mixture of intrinsically luminous AGN and a population of intrinsically weaker AGN whose radio power has been increased via interaction with the ambient medium (See Sect.~\ref{S:SFR}).

The detection of 7.7 $\mu$m PAH in 4 compact sources indicates star formation rates of $10$--$60\,M_\odot$ yr\mone. Upper limits on an additional 10 sources do not rule out similar star formation rates. Thus, these data are also consistent with a majority of the compact sources showing moderate star formation as suggested by \citet{Willett2010} and \citet{Dicken2012}.

\citet{Westhues2016} present Herschel photometry  of a sample of 87 3CR sources with $z< 1$. They list 12 CSS, but we remove 3C216 and 3C380 because they have extended emission which makes them larger than 20 kpc \citep{ODea1998} and we add 3C305 \citep[e.g.,][]{Hardcastle2012} for a total of 11 CSS.  \citet{Westhues2016} calculate SFR from the FIR luminosity.  If we adopt SFR $> 10\,M_\odot$ yr\mone\ to indicate substantial star formation, then 11/12 = 91\% of the CSS sources are star forming, and 42/76 = 55\% of the extended 3CR are star forming. This is a higher fraction of extended sources with star formation than found by \citet{Dicken2012}, {but Dicken et al.\ did not use FIR luminosity to calculate SFR because of concerns of an AGN contribution to the FIR.} Nevertheless,
the \citet{Westhues2016} results continue the trend of the GPS/CSS forming stars more often than the large radio sources. 

\citet{Podigachoski2015} present Hershel photometry of $z>1$ 3CR sources, and present SFR and dust mass. This sample includes 10 CSS out of a total  of 62 objects. There are a significant number of 3C sources undetected with Herschel with corresponding upper limits on SFR of up to $100\,M_\odot$ yr\mone,  so we don't merge the statistics with the \citet{Westhues2016} sample. {Star formation rates of at least  $100\,M_\odot$ yr\mone\ are detected in 7/10 (70\%) of the CSS and 16/52 (31\%) of the large radio sources  \citep{Podigachoski2015}.

\citet{Kosmaczewski2020} consider the WISE colors of a sample of 29 GPS and CSOs and compare them to other samples of galaxies and AGN. They find that the WISE colors suggest less star formation in the GPS/CSOs than in starburst and ULIRG galaxies, but more star formation than in  extended FRII radio galaxies. They find that the WISE colors and implied star formation in GPS/CSOs are similar to those in a sample of double-double radio galaxies.\footnote{See Section \ref{s:rejuv} for a discussion of the double-double radio galaxies.} }

Thus, the small samples of GPS/CSS studied so far consistently suggest that compact radio sources exhibit star formation more often than large-scale radio sources. This has implications for the population of compact radio sources  (Sect.~\ref{S:SFR}) and should be investigated with a larger sample of compact radio sources.

\section{Host galaxies}\label{s:hosts}

Here we discuss the host galaxy properties and environments of the PS and CSS sources. See \citet{Holt2009a} for a previous discussion. Star formation is also discussed in Sect.~\ref{s:IR} and ~\ref{S:SFR}.

\subsection{Host types, stellar populations, and magnitudes}

In general, the hosts of PS and CSS sources tend to be large, bright elliptical galaxies dominated by an old stellar population and with magnitudes around M*\footnote{M* is the magnitude at the break of the Schechter luminosity function \citep{Schechter1976}. } \citep[e.g.,][]{deVries1998b,deVries2000b, Snellen1998c,Snellen1999,Snellen2002b, Perlman2001,Drake2004a,deVries2007,Labiano2007,Orienti2010c,Kosmaczewski2020}. There are several examples of sources hosted by  galaxies with significant disk components - the HFP J1530+2705 \citep{Orienti2010c} and the CSS sources MRC B1221$-$423 (\citep{Johnston2010,Anderson2013}, Duggal et al., in preparation), 3C305 \citep{Heckman1982}, PKS1814-637 \citep{Morganti2011}, and the CSS sources B0258+35 and B1128+455 (Duggal et al., in preparation).  The existence of some spiral hosts strengthens the case for a link between CSS and NLSy1s (Sect.~\ref{s:NLS1}). {Note that AGN with lower radio power tend to be found in hosts with less luminous stellar bulges \citep{Best2005,Brown2011,Vaddi2016}  and/or lower bulge/disk ratios \citep{Pierce2019}.}

The hosts tend not to be brightest cluster galaxies \citep[e.g.,][]{deVries2000b,Drake2004a}. These host galaxy properties are consistent with those of the large radio galaxies at lower redshift ($z <0.6$), especially FRII hosts \citep[e.g.,][]{deVries1998b,deVries2000b,deVries2007}. However,  at higher redshift ($z >0.6$) the hosts of the 3C are more luminous than the hosts of the GPS radio galaxies likely due to the additional aligned light in the larger sources \citep{ODea1998,Snellen2002b,deVries2007}.
 
\begin{figure}
\centerline{\includegraphics[width=11cm]{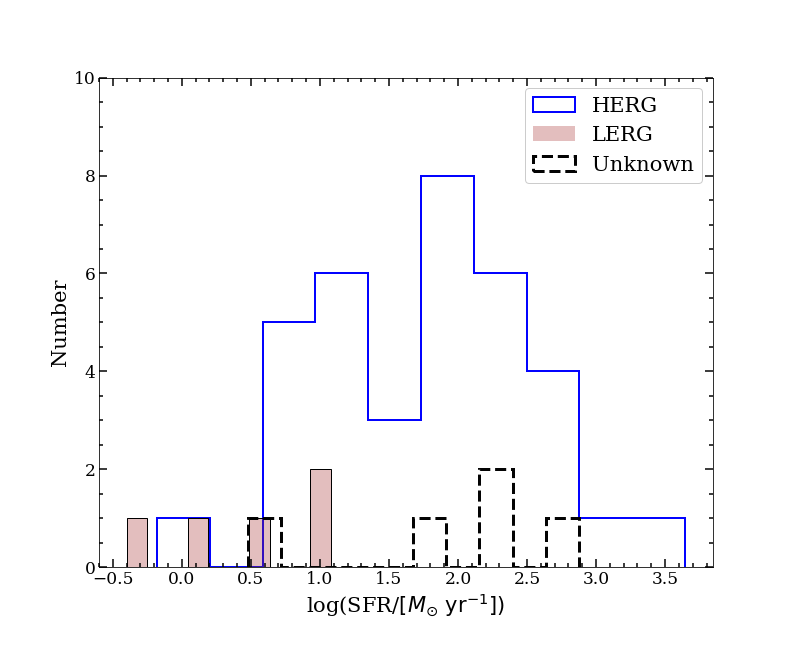}}
\caption{SFR from Table \ref{t:SFR}. If there are more than one estimate for a given source, the estimates were averaged. The values are shaded using the emission line class (HERG/LERG).}
\label{f:SFR_hist}
\end{figure}

 In addition to the dominant old stellar population, there is often a younger stellar population in the host galaxy due to recent or ongoing star formation \citep[e.g.,][]{Holt2007,Holt2009a, Labiano2008a,Hancock2010,Fanti2011,Tadhunter2011}, and Duggal et al. (in preparation).  SFRs are tabulated in Table \ref{t:SFR} and a histogram is plotted in Fig.~\ref{f:SFR_hist}. As discussed in Sect.~\ref{s:IR}, PS and CSS sources can have moderate SFR of a few to tens of $M_{\odot}\ {\rm yr}^{-1}$, but there is also a subset above $100\,M_{\odot}\ {\rm yr}^{-1}$. The number of LERGs with estimates of SFR is small, but there is a possible trend for the LERGs to have lower SFR than the HERGs.

\begin{table*}
	\caption{Star formation rates}
	\begin{tabular}{l l l l l l l l}
  \hline
 Source     & Alt.       &   z       & ID  & H/L    &  Tracer  & SFR           &    Refs.  \\
            &            &           &     &         &       & $M_{\odot}\ {\rm yr}^{-1}$ &    \\
    (1)     &   (2)      &  (3)      & (4) & (5)     & (6)    &   (7)          &    (8)     \\
 \hline
J0025$-$2602   &  PKS  0023$-$26& 0.322   &   G   &   H     &  PAH      &  270      & 1 \\
J0111+3906 & S4 0108+388 &  0.668 &      G   & L   &   PAH     &    12   &        2 \\
J0119+3210 & 4C 31.04    & 0.0602    & G   &    H   &   FIR      &   4.9    &  3  \\
          &            &           &      &       &    PAH     &    6.4    &     4  \\
          &            &           &       &      &     Ne     &    7.8    &      4  \\
 J0137+3309 & 3C48      &  0.3670   &   Q    &  H    &    FIR   &   573   &    7 \\     
 J0141+1353 &  3C49   &   0.6207    &  G    &  H     &    FIR    &  54& 7 \\
J0252$-$7104  & PKS 0252$-$71 & 0.566  &   G  &   H    &      PAH    &   $<$200 & 1 \\
J0405+3803 & 4C 37.11    &   0.055   &    G  &    H   &   Ne     &     11     &  4   \\
          &             &          &       &        &   PAH     &   0.8--1.6 &   4  \\
J0432+4138  & 3C119  &  1.023 & G & \dots   &    FIR & 150  &   8 \\
J0503+0203 & PKS 0500+019  &  0.58457 & G   & H     &    PAH     &   23      &    2  \\
J0521+1638 &  3C138  &  0.7590   &    Q  &  H   &         FIR  &  174 & 7 \\
J0542+4951 &  3C147  &  0.5450   &   Q  &  H    &      FIR  &  125 & 7 \\
J0713+4349 & B3 0710+439  & 0.518   &  G  &   H  &        PAH     &      $ <$ 10  &   2  \\
J0741+3112  &  0738+313   & 0.6314  &   Q  &    H     &      PAH     &    $<$20    &    2 \\
J0745$-$0044  &  0743$-$006   & 0.994    &  Q    &  H     &   PAH      &    $<$100    &   2  \\
J0801+1414  & 3C190   &  1.196 &  Q &  H   &   FIR &  470  & 8 \\
J1035+5628 & TXS 1031+567& 0.4590 &  G  &   H      &   PAH      &    $<$6       &   2 \\
J1148+5924 & 1146+59     & 0.0108  &  G  &       L    &   Ne      &         0.5    &      4 \\
         &  NGC3894           &        &       &           &   PAH      &       0.3     &     4 \\
J1154$-$3505 &  PKS 1151$-$34 & 0.258  &    Q  &    H    &  PAH        &       140&  1 \\
J1206+6413 & 3C268.3  & 0.3717 &   B  &     H  &          FIR  &   17  & 7 \\
J1223$-$4235 & MRC B1221$-$423 & 0.1706 &   G  &    H       &   H$\alpha$  &    $>$54    &     5   \\
J1308$-$0950  & PKS 1306$-$09  &0.464    & G   &   H     &          PAH &        $<$120 & 1 \\
J1331+3030 &  3C286   &   0.8499 &     Q &    H    &          FIR   &   226 & 7 \\
J1347+1217 & 4C 12.50    &  0.1234 &   G &  H      &      Ne    &   159      & 4  \\
          &             &       &      &         &      PAH    &           31.5  &    4 \\
J1352+2126$^a$ & 3C293 & 0.045 & G & L & H$\alpha$ + 24 $\mu$m &  3.2  & 10 \\
J1400+6210   & 1358+624    &  0.431  & G  &    H   &    PAH      &     $ <$4       &     2 \\
J1407+2827 & OQ 208      & 0.07658 &  B   &   H  &         Ne      &       54  &     4  \\
         &             &         &      &       &       PAH      &       47.8   &  4  \\
J1419+0628 & 3C298   &  1.438  & Q &  H   &   FIR & 930  & 8 \\    
J1447+7656 & 3C305.1 & 1.132 & G & \dots  &      FIR & 220  & 8 \\ 
J1449+6316 & 3C305       & 0.042   &   G   &   H    &     PAH      &       14  & 1 \\
          &             &         &       &       &    FIR     & 2.4 & 7 \\
J1459+7140& 3C309.1   & 0.9050  & Q   & H   &             FIR   &   120   & 7 \\    
J1520+2016 & 3C318   &  1.574& Q  &  H   &    FIR & 580  & 8, 9 \\
J1609+2641 & 1607+268    &  0.473  &  G    &   \dots &      PAH       &     $<$3    &   2  \\
J1634+6245 & 3C343     &  0.9880  & G   & H      &            FIR    &  261 & 7 \\
J1638+6234 & 3C343.1  & 0.7500  & G  &  H    &              FIR   &     61 & 7 \\
J1723$-$6500 & PKS 1718$-$649 &  0.014428 & G  &      L &      Ne       &     1.8   &  4  \\
         &              &          &      &        &     PAH      &     0.8   &  4  \\
J1744$-$5144 & PKS 1740$-$517  &  0.44 &  G   &   H    &   [OII] $\lambda$3727  &    $<$0.66  &   6 \\
J1819$-$6345  & PKS 1814$-$63 & 0.063   &  G   &   H      &   PAH &            18 & 1 \\
J1945+7055 & 1946+708  &   0.101   &  G   &   L    &    Ne           &    11.1  &  4  \\
J1939$-$6342 & PKS 1934$-$63 &  0.18129  & G   &   H   &  PAH  &  29 & 1 \\
           &           &          &      &        &     PAH        &    1.7-3.1 &  4  \\
J2130+0502  & 2128+048    &  0.990   &   G  &   \dots     &    PAH        &    $<$60   &   2  \\
J2137$-$2042   & PKS 2135$-$20 & 0.635    &  B    &   H      &   PAH    &    4400  & 1 \\
J2251+1848  & 3C454  &  1.757 &  Q  & H  &    FIR & 620  & 8 \\
J2250+7129  & 3C454.1 & 1.841 & G   & \dots   &   FIR  & 750  & 8 \\
J2255+1313  & 3C455    &  0.5430  &  G &  H   &              FIR   &    17& 7 \\
J2344+8226  &  2342+821   &   0.735   &  Q   &  H  &       PAH        &   $<$40    & 2  \\
 \hline
\end{tabular}
      \label{t:SFR}
      Column 1, J2000 Source name. Column 2, Alternate name. Column 3, redshift. Column 4, host identification (G) Galaxy, (B) Broad Line Radio Galaxy, or (Q) Quasar.  Column 5, Emission line excitation class -- (H) high excitation, (L) -- low excitation. Column 6,  Tracer used to determine the SFR. Equation \ref{eq:PAH} is used to calculate the SFR from the 11.3 $\mu$m PAH fluxes from \citet{Dicken2012}.   Column 7, SFR ($M_{\odot}\ {\rm yr}^{-1}$).  Column 8, Reference for the SFR  or data used to calculate SFR.  
References.  \citet{Dicken2012}  using equation \ref{eq:PAH}. 2. Ogle et al., in preparation. Upper limits are 2 $\sigma$. 3. \citet{Ocana2010}. 4. \citet{Willett2010}.  5. \citet{Anderson2013}. 6.  \citet{Allison2019}.    7. \citet{Westhues2016}.  8. \citet{Podigachoski2015}. 9. \citet{Podigachoski2016a}. 10. \citet{Labiano2014}.
Footnotes. $^a$ 3C293 is the inner CSS in a double-double radio source.
 (Sect.~\ref{s:rejuv}).
\end{table*}
 
\subsection{Merging and interaction} 
There are examples of well-studied mergers, e.g., 3C48 \citep[e.g.,][]{Kirhakos1999,Canalizo2000, Zuther2004,Scharwachter2004,Stockton2007},  MRC B1221$-$423 \citep{Johnston2010,Anderson2013}, and 4C12.50 (PKS 1345+12) \citep[e.g.,][]{Evans1999,Batcheldor2007,Emonts2016}.  In addition, a few systematic studies show that mergers and interactions in PS and CSS soures are common - with a range of measured fractions of sources that have either close companions and/or distorted isophotes from about 40\% \citep{Randall2011} to about 70\%  \citep[e.g.,][]{ODea1996b,deVries1997b,Drake2004a}, and Duggal et al. (in preparation). This is within the range of previous studies of interaction in large powerful radio galaxies \citep[e.g.,][]{Smith1989,Ramos2011,Ramos2012,Chiaberge2015,Storchi2019}. { Based on deep optical observations, all 7 of the CSS/GPS in the $0.05 < z < 0.7$ 2Jy sample of \citet{Ramos2011,Ramos2012}  show evidence for galaxy interactions/mergers, but the rate is also high for the extended radio sources in the same sample (95\% for the HERG sources).} Thus, it appears the compact and extended radio sources have similar merging rates. Additional studies with larger samples of compact sources would be useful.

\subsection{Environments}
\citet{Wold2000} selected 21 radio-loud quasars (including 6 CSS) from the 7C and Molonglo/APM Quasar Surveys. They find that the data are consistent with CSS quasars having the same clustering environments as the other quasars studied. \citet{Ramos2013} studied the clustering environments of the 46 powerful radio galaxies in the  2-Jy sample (including 1 GPS and 6 CSS). They find that the GPS and CSS sources have similar clustering environments as the FRII radio galaxies; and both FRII and GPS/CSS are less clustered than the FRI sources. Optical imaging of the fields of PS sources shows the presence of either candidate or confirmed companion galaxies suggestive of a group environment \citep{Stanghellini1993,Snellen2002b,Orienti2006c,Orienti2010c}. 
Thus, results so far indicate that the powerful GPS and CSS sources seem to be in similar environments as powerful large radio sources.  Both of these studies included relatively powerful sources and included small samples of PS and CSS sources.  Follow-up studies should be done for larger samples and also extended to lower radio power PS and CSS sources. 

 
\subsection{Gas content}\label{s:gas_content}
 
\begin{figure}
\centerline{\includegraphics[width=11cm]{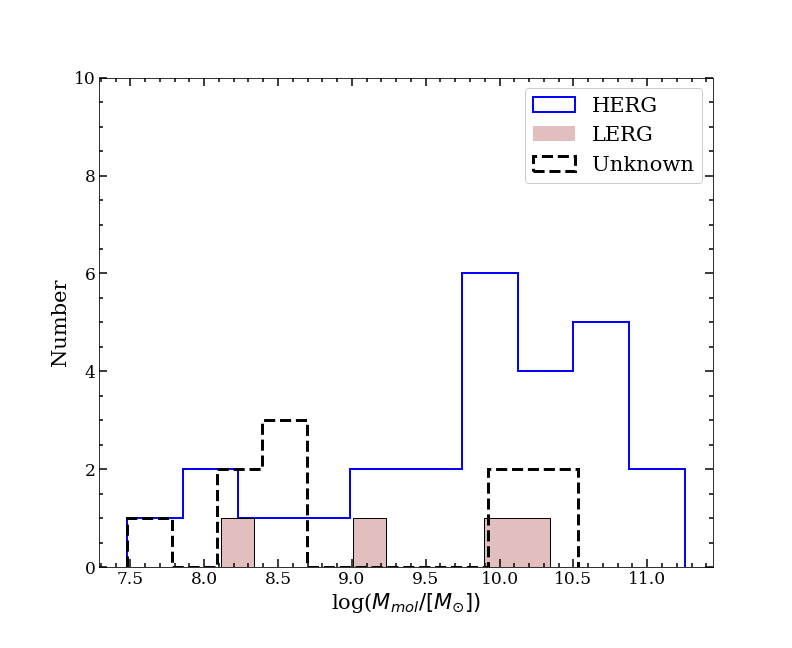}}
\caption{Mass of molecular gas from Table \ref{t:gas-mass}. If there are more than one estimate for a given source, the estimates were averaged. The values are shaded using the emission line class (HERG/LERG).}
\label{f:gas-mass}
\end{figure}

The gas content of the host galaxies of PS and CSS sources is relevant to the issue of the triggering and fueling of activity. It is also relevant to the hypothesis that some subset of the PS and CSS source are confined by dense gas (Sect.~\ref{s:frustrated}). Estimates of the molecular gas masses are presented in Table \ref{t:gas-mass} and plotted in Fig.~\ref{f:gas-mass}.  We caution that the objects in Table \ref{t:gas-mass} are not a complete sample.  There is a broad range of masses  from $10^7$ to $10^{11}\,M_\odot$. The independent estimates of the gas mass in 3C 48, 4C 12.50, and 3C 318 give consistent results. The gas mass estimates based on N$_{\rm H}$ from X-ray absorption (Sect.~\ref{s:xray_column}) tend to be lower than other estimates,  possibly because the estimates using the other tracers (CO, dust) require higher gas masses to ensure a detection. 
 
\subsubsection{Atomic gas}
 
{Overall, the detection rate of H{\sc i} absorption in compact sources is about twice that of the extended sources (32\% vs. 16\%, e.g., \citep{Maccagni2017}). This might be due to the presence of bright radio emission on galaxy scales (where the atomic hydrogen is preferentially located) in the compact sources \citep[e.g.][]{Conway1997,Pihlstrom2003,Gupta2006b,Chandola2011}. }
The properties of H{\sc i} absorption in PS and CSS sources have been reviewed by \citet{Morganti2018} as part of their comprehensive review of H{\sc i} absorption in AGN. Since then a major survey for H{\sc i} absorption in a sample of 145 CSS, PS and flat-spectrum objects over a redshift range of $0.02<z<3.8$ using the Green Bank Telescope was reported by \citet{Grasha2019}. Besides re-detecting H{\sc i} absorption in 6 known systems, they surprisingly did not detect any new system in the  spectra of 108 sources free from RFI, and none beyond a redshift of 0.7. In contrast, for a sample of 11 compact (mostly CSS/PS) sources in the intermediate redshift range $0.7<z<1$, 4 were detected in H{\sc i} absorption using the uGMRT by \citet{Aditya2019}. This suggests that the detection rate at intermediate redshifts of $\sim$30 per cent could be similar to those at lower redshifts, without consideration of WISE colours \citep{Chandola2017}. Beyond a redshift of 1, the detection rate is much smaller. The tendency of compact sources with WISE colours W2$-$W3$>$2.0 having a detection rate of $\sim$60 per cent \citep{Chandola2017} has been confirmed for a much larger sample of $\sim$240 sources combining the results of Westerbork \citep{Maccagni2017} and new GMRT observations \citep{Chandola2020}. \citet{Chandola2020} also find that while LERGs have a lower detection rate compared with HERGs, there is no significant difference between these two classes when they are all chosen to have WISE colours W2$-$W3$>$2.0. HERGs and LERGs are believed to differ in their accretion modes, with the Eddington\footnote{The Eddington ratio (or fraction) is the ratio of the AGN bolometric luminosity to the Eddington luminosity
$\lambda_{\rm Edd} =  L_{\rm bol}/L_{\rm Edd}$. The Eddington luminosity is the maximum possible luminosity due to accretion onto a compact object, assuming spherical accretion and a balance between gravity and radiation pressure.}
ratio being $> 0.01$ for HERGs and $< 0.01$ for LERGS \citep{Heckman2014}. A similar H{\sc i} detection rate for those with W2$-$W3$>$2.0 indicates that detection of H{\sc i} alone does not imply a high accretion mode AGN \citep{Chandola2020}. Besides the properties of the supermassive black hole and the accretion disk, a more detailed understanding of inflow in the circumnuclear region is required to understand the fuelling mechanisms \citep{Martini2004,Storchi2019}.
 
 There have also been new studies of individual sources. H{\sc i} and $^{12}$CO(2-1) absorption towards the young southern PS source PKSB1740$-$517 show that it has $\sim10^7$--$10^8\,M_\odot$ of cold gas \citep{Allison2019}. They argue that this reservoir  of cold gas was accreted $\sim$50 Myr ago, lending support to the hypothesis that luminous radio sources are powered by minor mergers.
 
 The gas content of PKS 1718$-$643 is also discussed in detail \citep{Maccagni2014,Maccagni2016a,Maccagni2016b,Maccagni2018}. The gas has three main  components - an outer (15 kpc) warped disk of atomic and molecular gas, an inner ($<$ 700 pc) circumnuclear molecular disk perpendicular to the outer disk, and individual molecular clouds which are infalling and could contribute to the fueling of the AGN. 
 
 VLA and EVN H{\sc i} observations of the rejuvenated CSS source B0258+350 show that the cold gas in the circumnuclear region is very turbulent, due to interaction of the jet with the interstellar medium of the host galaxy NGC1167 \citep{Murthy2019}, the results being broadly consistent with the expectations of numerical simulations \citep{Mukherjee2018a,Mukherjee2018b}. This provides evidence of the importance of AGN feedback. We discuss H{\sc i} absorption further in Sect.~\ref{S:NH-NHI} and Sect.~\ref{s:feedback}.
\\
 
 \subsubsection{Molecular gas}
 
 
 CO has been detected in a few objects in merging/interacting systems -- 3C48 \citep{Krips2005}, 3C318 \citep{Willott2007},  4C 12.50 
 \citep{Evans1999,Dasyra2012,Fotopoulou2019}, PKS 1549$-$79 \citep{Oosterloo2019}, and B1740$-$517 \citep{Allison2019}, the possible cool-core CSS PKS 0023$-$26 \citep{Morganti2020},
 and two nearby CSOs -- 4C 31.04 \citep{Garcia-Burillo2007,Ocana2010} and PKS B1718$-$649
 \citep{Maccagni2018}. Given the small number of sources with CO detections, {the focus has been on the ``interesting" sources}; thus the current trends are preliminary.  The merging systems tend to have more gas mass ($> 10^{10}\,M_\odot$) and a more irregular distribution of gas, while the CSOs have smaller gas mass ($< 10^{10}\,M_\odot$) with most of the molecular gas in a disk. 
 
 Other tracers of molecular gas have been used to place upper limits on gas content. \citet{Gupta2006b} observed a sample of 17 GPS and CSS sources in OH, and \citet{ODea1994,ODea2005} observed 7 GPS sources in a variety of tracers (CO, OH, CS, and NH$_3$) obtaining upper limits to the molecular gas. 

 Near-IR observations of ro-vibrational transitions of H$_2$ provide a probe of warm molecular hydrogen in 3C305, OQ208, PKS 1549$-$79 \citep{Guillard2012}, 4C 31.04
 \citep{Zovaro2019a}, 4C 12.50 \citep{Dasyra2011, Dasyra2014, Guillard2012}, 3C293 \citep{Ogle2010}, and PKS B1718$-$649 \citep{Maccagni2016b}.  
 In these sources, the warm H$_2$ is heated to temperatures of $\sim 500-1000$ K, likely due to shocks from the radio jet \citep{Dasyra2011, Dasyra2014, Guillard2012}.  In PKS 1549$-$79 \citep{Guillard2012,Morganti2020} and  4C 12.50 \citep{Dasyra2011, Dasyra2014, Guillard2012, Fotopoulou2019} both the warm molecular gas measured with the near-IR lines of H$_2$ and the colder phase traced by CO are  participating in the gaseous outflow.
 Multi-phase outflows are discussed further in Sect.~\ref{s:feedback}.
 
\citet{Baker2002}  suggested that there was evidence for dense gas in the environment of compact radio sources due to the preferential presence of strong CIV absorption in CSS quasars. However, this result was not confirmed by subsequent work \citep[e.g.,][]{Vestergaard2003,Stone2019}.

\subsubsection{Dust}

\citet{Tadhunter2014} summarize the dust masses of the 2Jy sample, but do not give values for individual objects. They adopt a gas-to-dust conversion of $M_{\rm gas} = 140 M_{\rm dust}$ \citep{Draine2007,Parkin2012}.
\citet{Westhues2016} present dust masses for 11 CSS sources based on Herschel photometry  as part of a study of a sample of 87 3CR sources with $z< 1$. \citet{Podigachoski2015} present dust masses for an additional 7 CSS sources from a study of 3C sources with $z> 1$.
We convert dust mass to gas mass using the adopted factor of 140 and list the gas masses in Table \ref{t:gas-mass}.
 
 \subsubsection{X-ray absorption column densities N$_{\rm H}$ } \label{s:xray_column}
 See Sect.~\ref{s:xray} for a discussion of X-ray properties. 
 The column densities found in X-ray absorption in GPS and CSO sources cover a broad range from $10^{21}$ to $10^{23}$ cm\mtwo\ with  values in a few sources approaching the Compton thick limit of $10^{24}$ cm\mtwo\ \citep[e.g.,][]{Vink2006,Tengstrand2009,Siemiginowska2016,Ostorero2017,Sobolewska2019}. 
 There is some evidence that CSOs with higher N$_{\rm H}$ have smaller linear size for a given 5 GHz power \citep{Sobolewska2019}. If this is confirmed with a larger sample, it would be consistent with a denser environment, possibly keeping these CSOs compact (i.e., the frustration scenario, Sect.~\ref{s:frustrated}), or increasing radio power for a given source size  (Sect.~\ref{S:SFR}). The sources with high N$_{\rm H}$ have a maximum linear size of 40 pc, and this might represent the size of the
 region of high density \citep{Sobolewska2019}. 
 Since N$_{\rm H}$ is related to N$_{\rm HI}$ (see Sect.~\ref{S:NH-NHI}), it would be good to look for a similar relation using trends in radio size vs radio power for different values of N$_{\rm HI}$. Also, it would be good to search for independent evidence of a dense ISM in galaxies with high N$_{\rm H}$.  Important questions are: how much gas mass is associated with the high values of N$_{\rm H}$, and how is the gas mass distributed in the host galaxy? 
 
 We obtain a rough estimate of the amount of gas associated with the X-ray absorption by assuming that the absorbing column is distributed uniformly over some surface area. Then, M$_{\rm gas} \sim N_{\rm H} \pi r^2 m_{\rm p}$, where $r$ is the radius of the region containing the X-ray absorbing gas, and $m_{\rm p}$ is the mass of the proton. We choose $ r= 1$ kpc based on the high detection fraction of H{\sc i} absorption in compact radio sources out to sizes of about 1 kpc (Sect.~\ref{S:NH-NHI}). This gives 
 \begin{equation}\label{eq:Mgas}
    M_{\rm gas} \sim 2.5 \times 10^8\,M_\odot \left( { N_{\rm H} \over 10^{22} \  {\rm cm}^{-2} } \right) 
    \left(  {r \over 1\ {\rm kpc} } \right)^2
 \end{equation}
which is given in Table \ref{t:gas-mass}.

\subsubsection{The relation between X-ray N$_{\rm H}$ and HI N$_{\rm HI}$ }\label{S:NH-NHI}

In GPS sources, the column densities measured in the X-rays N$_{\rm H}$ are typically 1--2 orders of magnitude larger than those estimated by H{\sc i} absorption N$_{\rm HI}$ \citep{Vink2006,Tengstrand2009,Ostorero2017}, and  see also the review by \citet{Morganti2018}. N$_{\rm HI}$ is likely measured towards a compact radio lobe \citep[e.g.,][]{Araya2010,Morganti2013}. Recall that N$_{\rm H}$ measures the column through all phases of the gas, and N$_{\rm HI}$ is the column though the atomic hydrogen only. N$_{\rm HI}$ is directly proportional to the Spin Temperature, which if underestimated, could also result in a lower value for N$_{\rm HI}$.

The two column  densities are correlated, but with large scatter \citep{Ostorero2010,Ostorero2017}. Consistent with this, \citet{Glowacki2017} and \citet{Moss2017} note that H{\sc i} absorption is preferentially found in galaxies with high N$_{\rm H}$, but do not explicitly test for a correlation. Also, \citet{Moss2017} find a correlation between H{\sc i} optical depth and X-ray hardness (where harder spectra are missing the softer photons due to absorption).  
In Sect.~\ref{s:xray-radio} we discuss the possibility that in the 
LERG GPS galaxies the X-ray emission comes from the radio lobes (via synchrotron or inverse Compton scattering \citep{Stawarz2008, Ostorero2010}), and in the HERG GPS galaxies, the X-ray emission is dominated by the accretion structure (e.g., disk-corona, \citet{Hardcastle2009}).  This would cause the total path length for X-ray absorption through the galaxy to be different in the LERGs and HERGs. 
 The detection of possible Compton thick absorption in four sources \citep{Guainazzi2004,Tengstrand2009,Ostorero2017} is consistent with the X-ray emission being produced in the nucleus in these four sources.  Is the correlation between
N$_{\rm H}$ and  N$_{\rm HI}$  different for HERGs and LERGs? 

The correlation between N$_{\rm H}$ and N$_{\rm HI}$ suggests that the two absorbing columns are physically connected. 
There is a peak in the H{\sc i} absorption detection rate of over 50\% in compact radio sources in the size range 0.1 to 1 kpc,  suggesting that the size of the absorbing structure is roughly this scale (\cite{Curran2013} and references therein). Upper limits on spin temperature are obtained by setting  N$_{\rm H} = \rm{N}_{\rm HI}$ and assuming a covering factor for the H{\sc i} of unity, obtaining upper limits in the range 100--5000 \citep{Allison2015,Glowacki2017,Moss2017}. The estimated spin temperatures might be more accurate if restricted to LERGs where the X-ray and radio emission likely both originate in the lobes, resulting in the column densities probing the same lines of sight. 

\begin{table*}
	\caption{Molecular Gas Masses}
\begin{tabular}{l l l l l l l l}
  \hline
 Source     & Alt.       &   z       & ID  & H/L    &  Tracer  & Mass       &    Refs.  \\
            &            &           &     &         &           & $M_\odot$  &         \\
  \hline
 J0025$-$2602  & PKS 0023$-$26 & 0.32188 & G & H & CO(2-1) & $5 \times 10^{10}$ & 15 \\
 J0038+2303 & B2 0035+22  & 0.0960 &   G   &   \dots  &     N$_{\rm H}$ & $3.5 \times 10^8$ & 1  \\
 J0111+3906 & S4 0108+388 &  0.66847 &  G   &    L    &     N$_{\rm H}$ & $1.4 \times 10^{10}$ & 1   \\
 J0119+3210 & 4C 31.04    & 0.0602    & G   &    H   &   12CO(2-1)  &  $6 \times 10^9 $ &  2,3    \\
 J0432+4138 & 3C119   & 1.023  &  G & \dots   &    dust &   $9.8 \times 10^9$   & 13 \\
 J0137+3309 &  3C48       & 0.367    &   Q  &    H   &  12CO(1-0) & $1 \times 10^{10}$  &  4   \\
            &             &         &      &        &     dust   &  $3.7 \times 10^{10}$ &  5  \\
 J0141+1353  & 3C49     & 0.6207   &   G    &  H      &     dust  &      $5.7 \times 10^9$   & 5 \\   
 J0503+0203 & PKS 0500+019  &  0.58457 & G   & H    &  N$_{\rm H}$  &  $1.3 \times 10^8$  &  1 \\
 J0521+1638 &  3C138  &  0.7590   &    Q  &  H     &        dust &    $2.0 \times 10^{10}$ &  5  \\ 
 J0542+4951  & 3C147   & 0.5450     &  Q  &  H       &       dust   &       $6.8 \times 10^9$   & 5 \\
 J0713+4349 & B3 0710+439  & 0.518   &  G  &   H  &      N$_{\rm H}$  & $2.5 \times 10^8$ & 6 \\
 J0801+1414 & 3C190   &  1.196 &  Q  & H   & dust  &   $6.9 \times 10^{10}$   & 13 \\
 J1035+5628 & TXS 1031+567& 0.4590 &  G  &   \dots      &  N$_{\rm H}$  & $1.3 \times 10^8$ &  1  \\
 J1206+6413 & 3C268.3  & 0.3717 & B  & H       & dust   &    $2.0 \times 10^9$   & 5 \\   
 J1326+3154 & B2 1323+32  &  0.370  &  G &       &       N$_{\rm H}$  & $3 \times 10^7$  &   1 \\
 J1331+3030 & 3C286    &  0.8499  & Q &  H      &       dust &      $6.1 \times 10^{10}$ &  5  \\ 
 J1347+1217 & 4C 12.50    &  0.1234 &   G &  H      &    N$_{\rm H}$  & $1.2 \times 10^9$ &   1 \\
           &             &          &     &        &    12CO(1-0)    &  $1.0 \times 10^{10}$  &  7  \\
 J1352+2126$^a$ & 3C293 & 0.045 & G & L & 12CO(1-0) &  $2.2 \times 10^{10}$  & 16, 17 \\
 J1400+6210 & 4C 62.22    &  0.431   &  G   &  H    &      N$_{\rm H}$  &  $7.3 \times 10^8$ &     1 \\
 J1407+2827 & OQ 208      & 0.07658 &  B   &   H  &  N$_{\rm H}$  &  $2.2 \times 10^{10}$ & 1 \\
 J1419+0628 & 3C298  &   1.438  & Q  & H   & dust  &  $5.3 \times 10^{10}$   & 13 \\
 J1447+7656 & 3C305.1 & 1.132 & G & \dots  & dust &     $2.0 \times 10^{10}$   & 13 \\
 J1449+6316 & 3C305    &  0.0416   & G & H   &          dust &      $1.2 \times 10^9$   & 5 \\
 J1459+7140 & 3C309.1  & 0.9050  & Q &  H     & dust &       $9.0 \times 10^{10}$ &  5  \\ 
 J1511+0518 &             &  0.084  &  G    &       &      N$_{\rm H}$  &  $9.5 \times 10^9$  &  6 \\
 J1520+2016 & 3C318      & 1.574    &   Q  & H    &      12CO(2-1)    &  $3 \times 10^{10}$ &  8,12 \\
           &            &          &      &       &  dust & $2.4 \times 10^{10}$   & 13, 14 \\
 J1556$-$7914 & PKS 1549$-$79 & 0.1521   &  G    &   H   &  12CO(1-0)     & $1 \times 10^{10}$   & 9   \\
 J1634+6245 & 3C343      & 0.9880  & G  & H      &         dust &     $1.8 \times 10^{11}$ &  5  \\ 
 J1638+6234 & 3C343.1  & 0.7500 &  G  &  H    &    dust &  $2.6 \times 10^9$   & 5 \\
 J1723$-$6500 & PKS 1718$-$649 &  0.014428 & G  &      L &        H$_2$   &  $2 \times 10^9$  &  10   \\
           &              &          &      &       &    N$_{\rm H}$ &  $1.3 \times 10^8$  &  6  \\
           &             &           &      &       &     12CO(2-1) & $1.9 \times 10^9$  & 18 \\
 J1744$-$5144 & PKS 1740$-$517  &  0.44 &  G   &   H    &     12CO(2-1)    & $1-10 \times 10^7$  &  11 \\
 J1845+3541 & 1843+356     & 0.764   & G    &        &    N$_{\rm H}$  & $2 \times 10^8$  &   1 \\
 J1939$-$6342 & PKS 1934$-$63 &  0.18129  & G   &   H   &     N$_{\rm H}$ & $3 \times 10^7$    &  1 \\
 J1944+5448 & 1943+546  &  0.263    & G    &        &     N$_{\rm H}$ &     $2.8 \times 10^8$   &  1  \\
 J1945+7055 & 1946+708  &   0.101   &  G   &       &      N$_{\rm H}$ &   $4.0 \times  10^8$  &   1 \\
 J2022+6136 &  2021+614 & 0.227 &   G  &        H   &    N$_{\rm H}$   & $9.0 \times 10^9$   & 1  \\
 J2251+1848  & 3C454 &   1.757  & Q &  H   &     dust &   $3.2 \times 10^{10}$   & 13 \\
 J2250+7129  & 3C454.1 & 1.841 &  G & \dots  &     dust &    $3.4 \times 10^{10}$   & 13 \\
 J2255+1313 & 3C455    &  0.5430 &   G & H      &          dust &        $4.0 \times 10^9$   & 5 \\
 J2355+4950 & TXS 2352+495& 0.2379 & G    &     H  &   N$_{\rm H}$  &   $1.7 \times 10^8$  &  6 \\
  \hline
	\end{tabular}
      \label{t:gas-mass}
      Column 1, J2000 Source name. Column 2, Alternate name. Column 3, redshift. Column 4, host identification (G) Galaxy, (B) Broad Line Radio Galaxy, or (Q) Quasar.  Column 5, Emission line excitation class -- (H) high excitation, (L) -- low excitation. Column 6,  Tracer used to determine the gas mass. Equation \ref{eq:Mgas} is used to calculate the  mass from the X-ray column density N$_{\rm H}$.  Column 7, Mass of gas ($M_\odot$).  Column 8, Reference for the gas mass or data used to calculate gas mass.  Cold Hydrogen is the default mass. However, N$_{\rm H}$ is sensitive to all phases of the gas along the line of sight to the X-ray emission. 
References. 1. \citet{Ostorero2017}. 2. \citet{Garcia-Burillo2007}. 3. \citet{Ocana2010}. 4. \citet{Krips2005}.
5. \citet{Westhues2016}. 6. \citet{Sobolewska2019}. 7. \citet{Dasyra2012}. 8. \citet{Willott2007}. 9. \citet{Oosterloo2019}. 10. \citet{Maccagni2016b}. 11. \citet{Allison2019}. 12. \citet{Heywood2013}.
13.  \citet{Podigachoski2015}. 14. \citet{Podigachoski2016a}. 15. \citet{Morganti2020}. 16. \citet{Evans1999}. 17. \citet{Labiano2014}. 18. \citet{Maccagni2018}. 
Footnotes. $^a$ 3C293 is the inner CSS in a double-double radio source (Sect.~\ref{s:rejuv}).
 \end{table*}

\subsection{Emission lines} \label{s:em-lines}

\subsubsection{Nuclear emission lines, HERG and LERG} \label{s:nuclear}
A significant development in the last decade or so has been the realization of a dichotomy in their optical spectroscopic properties and relationship to the nature of fuelling the AGN \citep{Heckman2014,Tadhunter2016a}. Starting with the early work by \citet{Hine1979}, \citet{Laing1994}, and \citet{Baum1995}, and more recent work by, for example, \citet{Ogle2006}, \citet{Hardcastle2007}, \citet{Leipski2009}, \citet{Buttiglione2010} and \citet{Best2012} and references therein, radio AGN could be divided into High Excitation Radio Galaxies (HERGs) and Low Excitation Radio Galaxies (LERGs). As summarized by \citet{Heckman2014}, in the high-excitation or radiative mode AGN, the Eddington ratio is greater than 1 per cent, while in the low-excitation or jet-mode AGN the Eddington ratio is less that 1 per cent. Accretion in the radiative mode AGN is characterized by the classical optically thin, geometrically thick accretion disk, the ultraviolet radiation from which gives rise to prominent emission lines in both the broad- and narrow-line regions. In this class of objects, namely the HERGs, an obscuring torus of dust and molecular gas could obstruct a view of the broad-line region giving rise to Type 2 AGN, while closer to the jet axis one would get a direct view of the broad-line region giving rise to Type 1 AGN. 
On the other hand for LERGs, the inner disk is not there resulting in weak or no emission lines, with the inner region being dominated by a ``geometrically thick  advection-dominated accretion flow'' \citep{Heckman2014}. LERGs tend to have higher black hole masses, occur in more massive early-type galaxies and have lower specific star-formation rates approaching the red sequence, compared with the HERGs \citep{Heckman2014}. Recent studies of the LOFAR-Bo\"otes sample by \citet{Williams2018} and the Stripe 82 1--2 GHz JVLA survey by \citet{Whittam2018} are consistent with these trends. Most of the HERGs have an FRII morphology while LERGs predominantly appear to have an FRI structure. LERGs are on average of lower luminosity than the HERGs \citep{Best2012,Heckman2014}. \citet{Whittam2018} find mechanical feedback to be significant for both LERGs and HERGs, although there is a scatter of about 2 dex in the fraction of accreted power deposited in the ISM.

Optical spectroscopy is available for PS \citep[e.g.,][]{Snellen1999,deVries2000a,deVries2007,Labiano2007} and CSS  \citep[e.g.,][]{Gelderman1994,Morganti1997,Hirst2003,Labiano2005} sources. 
Current results suggest that the nuclear emission line spectra are very similar for compact radio sources and larger radio sources \citep[e.g.,][]{Morganti1997,Hirst2003,KB2010b}.  The CSS and PS sources can be classified as HERGs and LERGs similar to the large radio sources \citep{Buttiglione2010,KB2010b}.  The compact and large HERGs lie on the same correlation of [OIII] $\lambda 5007$ luminosity with radio power \citep{KB2010b}. A parallel correlation is found for the compact and large LERGs. In addition, as discussed here in Sect.~\ref{s:xray-radio}, \citet{KB2014} show that the compact and large HERGs lie on the same correlation of X-ray luminosity with radio power. And a similar correlation is found for the compact and large LERGs. These optical - radio - X-ray correlations for HERGs and LERGs argue that the compact and large radio sources have similar central engines. They also make it plausible that at least some of the compact sources evolve to become large radio sources. They also imply (but do not require) continuity, i.e., compact sources start as HERG or LERG and then evolve to become  large radio sources while maintaining their HERG or LERG nature. If some of the compact sources are short-lived (Sect.~\ref{s:transient}), then they must fade without disrupting the observed correlations. 

\citet{ODea1998} showed that GPS radio galaxies tend to have lower [OIII] $\lambda 5007$ luminosities than CSS radio galaxies.  \citet{Labiano2008b} confirmed this for a larger sample of sources. \citet{Vink2006} find that the ratio of [OIII]/X-ray luminosity is lower for the GPS sources than for larger sources.  \citet{Vink2006} suggest that if the emission line gas is photo-ionized, the young age of the GPS sources means that the Stromgren\footnote{The Stromgren sphere is an idealized representation of the size of a nebula which has been ionized by a compact source of high energy photons.} sphere which determines the size and luminosity of the ionized NLR is still small and growing. Alternately, if there is a strong contribution to the emission line luminosity from jet-driven shocks, the small size of the GPS radio source means that only a small volume of gas has been shocked \citep[e.g.,][]{Moy2002}.

The smaller radio sources have larger [OIII] $\lambda 5007$ line FWHM than the larger radio  sources \citep{Gelderman1994,Best2000, Labiano2008b, Holt2008}. As discussed in Sect.~\ref{s:alignment} and \ref{s:feedback}, this is consistent with strong jet interaction with the clouds producing the emission line gas. 

\subsubsection{Extended emission lines and the alignment effect}\label{s:alignment}
GPS radio sources are generally too small to be resolved in the optical, thus, the detection of the alignment effect has been mostly restricted to CSS radio sources (but see \citet{Batcheldor2007} for an example of two nearby GPS sources slightly resolved by HST). 
Extended emission lines in CSS  are almost always aligned with and co-spatial with the radio source \citep{deVries1997b,deVries1999,Axon2000,Privon2008}. The alignment effect is seen in CSS sources at all redshifts \citep{deVries1997b,Privon2008}, while extended radio sources display the alignment effect only at $z>0.6$ \citep{McCarthy1993}.
Spatially resolved spectroscopy reveals that the kinematics of the emission line gas are consistent with the emission line clouds having been shocked by the radio source and accelerated outwards \citep{Chatzichristou1999, ODea2002, Holt2008, Shih2013, Reynaldi2013}. 
The excitation of the aligned emission line gas is generally consistent with shocks combined with AGN photoionization \citep{debreuck2000,Best2000,Moy2002,Inskip2002, Inskip2006, Labiano2005, Holt2009b, Shih2013,Reynaldi2013, Reynaldi2016}. Thus, the alignment effect in CSS sources is a signature of radio-jet feedback to the host galaxy ISM and is discussed further in Sect.~\ref{s:feedback}.

{ 
\subsection{Jet-induced Star Formation}\label{s:jet-SF}

\begin{figure}
\centerline{\includegraphics[width=11cm]{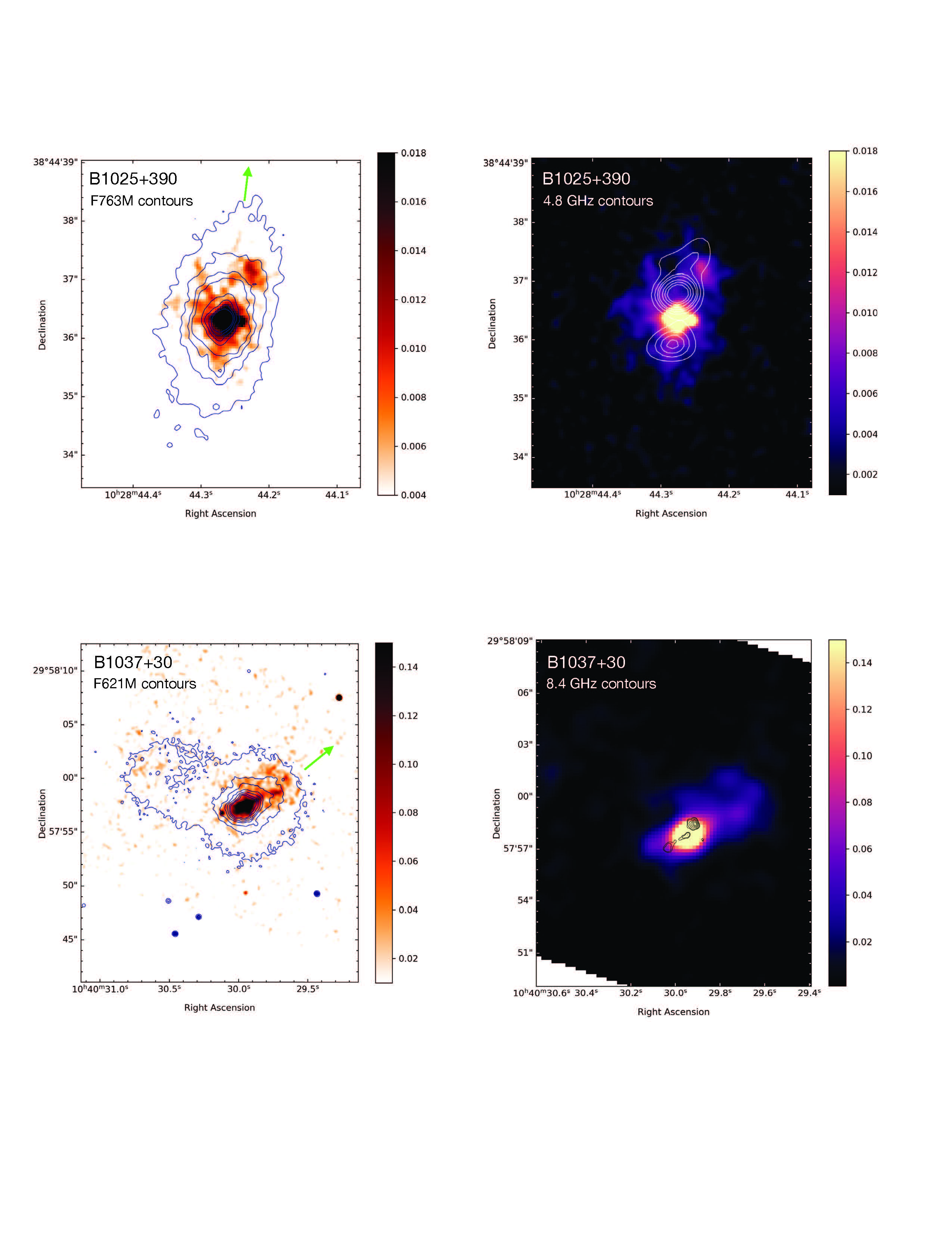}}
\caption{Possible jet-induced star formation in 2 CSS radio galaxies. (Left Panels) HST WFC3/UVIS data. Contours of R-band superposed on image of UV continuum.  The green arrow shows the orientation of the radio source axis. (Right Panels). 
HST WFC3/UVIS UV continuum image on contours of radio image. (Top) J1028+3844 (4C 39.32). (Bottom) J1040+2957 (4C 30.19).  (Figure from Duggal, private communication).
 }
\label{f:radio-SF}
\end{figure}

Jet-induced star formation\footnote{This  process is commonly referred to as jet-induced star formation, though it is expected that in powerful radio sources, the bow shock driven by the over-pressured cocoon is responsible for compressing the ambient clouds and triggering star formation \citep[e.g.,][]{Rees1989,Begelman1989, DeYoung1989, Daly1990}. However, there is also evidence for star formation triggered by direct jet impact, e.g., Minkowski's Object \citep{Brodie1985, vanBreugel1985, Croft2006, Salome2015, Lacy2017, Zovaro2020}. }
is expected to be a natural consequence of radio jets propagating though the ISM 
\citep[e.g.,][]{Rees1989,Begelman1989, DeYoung1989, Daly1990, Tortora2009, Gaibler2012, Dugan2014,Dugan2017, Fragile2017}.  Because of this, jet-induced star formation  should be a key signature of radio jet feedback \citep[e.g.,][]{Gaibler2012,Fragile2017}.

There is observational evidence for jet-induced star formation associated with several {\it extended\/} radio sources, e.g., Minkowski's Object \citep{Brodie1985, vanBreugel1985,Croft2006,Salome2015, Lacy2017,Zovaro2020}, Centaurus A \citep{Graham1998,Mould2000,Rejkuba2002,Oosterloo2005, Crockett2012,Santoro2015,Santoro2016,Morganti2016b, Salome2017}, 3C285 \citep{vanBreugel1993,Salome2015}, 4C 41.17 \citep{Dey1997,Nesvadba2020}, and 3C441 \citep{Lacy1998}.
Jet-induced star formation is also supported by statistical studies of large samples of radio sources \citep[e.g.,][]{Zinn2013, Kalfountzou2017}. 

GPS/CSO radio sources are too small for current facilities to show jet-induced star formation on the scale of the radio source. However, there is growing evidence for jet-induced star formation in CSS radio sources. \citet{Labiano2008a} obtained HST ACS/HRC observations through the F330W filter of 3 CSS sources and detected extended UV light in one object, 3C303.1. The UV light in 3C303.1 is along the radio axis and seems associated with the brighter southern radio lobe.  Duggal et al. (in preparation) obtained HST WFC3/UVIS imaging of the UV continuum in the region between rest-frame [CIII]$\lambda$1909 and MgII$\lambda$2798. This region is free of bright emission lines \citep[e.g.,][]{McCarthy1993}. Duggal et al. find extended UV continuum aligned with the radio source in 6 of 9 CSS sources observed. Two examples are shown in Figure \ref{f:radio-SF}. Combining the two studies, extended UV light aligned with the radio source is found in 7/12 (58\%) of CSS radio sources. 
If the UV light is confirmed to be due to star formation, this would indicate that jet-induced star formation is common in CSS radio galaxies.
There may be a contribution to the UV light from  scattered AGN light, which is identified by its polarization  \citep{diSerego1988,diSerego1989,diSerego1993,Antonucci1984, Fabian1989,Dey1996,Cimatti1997, Tran1998,Tadhunter1992,Tadhunter1996,Tadhunter2002}. There may also be a contribution from  nebular continuum  \citep{Dickson1995,Cimatti1997,Wills2002,Tadhunter2002,Inskip2006}.  
}

\subsection{Black-hole Masses, Eddington ratios, the black-hole fundamental plane, and jet production efficiency}\label{s:BH}

Estimates of the masses of the central supermassive black holes in PS and CSS sources have been made  \citep{Wu2009b,Son2012,Liao2020a,Wojtowicz2020}.
The mass estimates range from $\log_{10} (M/M_\odot) \sim 7.32$ to 9.84, with a median value of 8.72 \citep{Liao2020a}. \citet{Liao2020a} compare the black hole mass estimates with those of low redshift large-scale radio galaxies studied by \citet{Hu2016}. There is good overlap of the black masses at the high end. However, none of the large-scale radio galaxies (and 11\% of the PS and CSS sources) have estimated black hole masses less than $\log_{10} (M/M_\odot) = 8$. The subset of PS and CSS sources with low black hole masses ($\log_{10} (M/M_\odot) < 8$) may represent a population which will not evolve to become large-scale radio galaxies (Sect.~\ref{s:what}). 

Estimates of the Eddington ratio in PS and CSS surces have also been made \citep{Holt2006b,Holt2011,Wu2009b,Son2012, Liao2020a,Wojtowicz2020}. \citet{Son2012} show that the high Eddington ratio sources tend to be HERGs and the low ratio sources tend to be LERGs as found for large-scale radio sources \citep[e.g.,][]{Best2012}. The Eddington ratios for the PS and CSS sources are in a broad range $R_{\rm Edd} \sim 10^{-4.93}$--$10^{0.37}$ \citep{Liao2020a}.  The Eddington ratios tend to be a bit higher for the PS and CSS sources 
(mean value $R_{\rm Edd} \sim 10^{-2.26}$) than the \citet{Hu2016} large-scale radio galaxies (mean value $R_{\rm Edd} \sim 10^{-3.05}$) \citep{Liao2020a}. If the PS and CSS evolve into these large-scale radio sources, the accretion rates must decline as the sources age. 

Note the following two cavaets.  (1) The black hole masses,  bolometric luminosities, and Eddington ratios have considerable uncertainty. (2) The PS and CSS sample is heterogeneous and may not be well matched with the \citet{Hu2016} sample of low redshift large-scale radio sources.  Thus, these interesting results deserve further study. 

AGN with estimates of black hole mass can be placed on the black hole fundamental plane which is defined by nuclear radio luminosity, nuclear X-ray luminosity, and black hole mass \citep{Merloni2003,Falcke2004a}. The existence of the fundamental plane is consistent with the radio luminosity being produced in a jet whose power is related to black hole mass and accretion rate, and the X-ray emission being produced in a disk-corona system (whose power is related to black hole mass and accretion rate) and/or the jet \citep{Merloni2003, Falcke2004a}.   PS and CSS sources have been placed on the black hole fundamental plane  \citep{Fan2016,Liao2020b,Wojtowicz2020}. Their results are consistent with the X-ray emission containing a contribution from the radio source, e.g., synchrotron or inverse Compton \citep{Stawarz2008, Ostorero2010,Migliori2012} (and see discussion in Sect.~\ref{s:xray}).

{ 
The jet production efficiency, $\eta_{\rm jet}$, is defined to be the ratio of the jet kinetic power to the accretion power
\citep[e.g.,][]{Wojtowicz2020}. \citet{Wojtowicz2020}
find a broad range ($ 10^{-3} < \eta_{\rm jet} < 0.2$) for jet production efficiency in a sample of 17 CSOs. \citet{Wojtowicz2020} point out that these values for $\eta_{\rm jet}$ are below those expected for magnetically arrested disks around maximally spinning black holes and consistent with those of a sample of extended FRII quasars. \citet{Wojtowicz2020} suggest that the jets in these CSOs are produced most efficiently in high/hard states in analogy with Galactic XRBs \citep{Homan2005,McClintock2006}.
}

\section{High-energy properties}\label{s:highenergy}

\subsection{X-ray properties}\label{s:xray}

 At the time of the previous review \citep{ODea1998}, some PS and CSS quasars had been detected in X-rays at luminosities consistent with those of other quasars. However, few PS and CSS galaxies had been detected.   X-ray studies of PS and CSS sources have been previously  reviewed  by \citet{Siemiginowska2009b} and \citet{Migliori2016a}. { Currently, nearly all powerful extragalactic radio sources (both compact and extended) are detected in X-rays in the 2Jy \citep{Mingo2014} and 3CR samples \citep{Stuardi2018,Jimenez2020}. Nearly all (91\%) of CSOs are detected \citep{Siemiginowska2016}, while detection rates for CSS galaxies are lower with 4/7 (57\%) of low power CSS galaxies detected \citep{KB2014}.}

\subsubsection{X-rays from PS and CSO galaxies}\label{s:xraycso}

Recent studies with ASCA, BeppoSAX, XMM-Newton and Chandra have greatly increased the detections of PS and CSO galaxies \citep{ODea2000,Risaliti2003,Guainazzi2006, Vink2006, Tengstrand2009, Siemiginowska2016,Beuchert2018,Sobolewska2019}.  
 \citet{Tengstrand2009} discussed a sample of 16 PS galaxies detected in X-rays.  \citet{Siemiginowska2016} presented new Chandra observations of 10 CSOs and archival Chandra observations of another 6.  \citet{Sobolewska2019}
 presented XMM-Newton observations of 3 CSOs thought to have high column densities and compiled a list of 24 CSOs with X-ray data. 
 In general, the X-ray spectra of PS and CSO galaxies can be described by an absorbed power-law with photon index in the range $\Gamma \sim 1.4 - 1.7$. The absorbing column in the PS and CSO galaxies lies in the range  N$_{\rm H} \sim 10^{21} - 10^{24}$ cm\mtwo\ and \citet{Sobolewska2019} confirm that 5 sources have  N$_{\rm H} > 10^{23}$ cm\mtwo. 
 \citet{Sobolewska2019} show that the 5 heavily absorbed sources have a smaller radio size for a given 5 GHz luminosity than the less absorbed sources and suggest that the small, heavily absorbed sources are confined by a dense environment.
 
 There are two main hypotheses for the origin of the X-ray emission in PS and CSOs.  It is possible that the X-ray emission in HERGs is produced primarily by the accretion disk and corona \citep[e.g.,][]{Tengstrand2009}. Note that the attribution of the X-ray emission to a disk/corona is not meant to imply that these systems are well understood or well modelled \citep[e.g.,][]{Blaes2007}. 
 However, it is  also possible that the X-ray emission is produced in the radio source via inverse Compton scattering \citep{Stawarz2008, Ostorero2010}. Both mechanisms may contribute in individual sources. Depending on the origin of the X-ray emission, the measured  N$_{\rm H}$ samples a different line of sight through the host galaxy. 
 Demanding that the X-ray derived N$_{\rm H}$ matches that derived from the 21-cm line in absorption against the radio source  \citep[e.g.,][]{Vermeulen2003} requires a ratio of spin temperature to covering factor of $T_s/c_f \sim 100$--5000~K in the atomic gas \citep{Ostorero2010,Allison2015,Glowacki2017,Moss2017}. (See also Sect.~\ref{S:NH-NHI}.)

\subsubsection{X-rays from CSS galaxies}\label{s:CSSx-ray}

 Because the radio source size is well matched to the size of the  host galaxy, CSS galaxies probe the process of energy transfer from the radio source to the host galaxy ISM (Sect.~\ref{s:feedback}). X-ray emission is expected from hot gas which has been shocked by the expanding radio source \citep{Heinz1998,Bicknell2006,Sutherland2007}.

 \citet{ODea2006} present XMM observations of 3C303.1. There is no evidence in the X-ray spectrum for absorption at the redshift of the galaxy, suggesting that the emission is  not dominated by the nucleus. They detect a thermal component with a temperature kT $\sim 0.8$ keV which they attribute to the ISM of the host galaxy. There is also evidence for an additional component contributing to the spectrum. The component could be synchrotron self-Compton from the southern radio lobe if the magnetic field is below the equipartition value by  a factor of $\sim 3.5$. Alternately, the second component could be hot gas which has been shocked by the expansion of the radio source. If due to hot gas, the fit to the spectrum gives a temperature of kT $\sim 45$ keV which corresponds to a shock Mach number M$\sim 13$. The kinematics of the optical emission line gas are consistent with shock acceleration \citep{ODea2002,Reynaldi2016} and the excitation is consistent with shocks contributing to the heating of the emission line gas \citep{Labiano2005,Holt2009b,Shih2013,Reynaldi2016}.

Chandra observations show the  soft X-ray emission in 3C305 is spatially extended over $5^{\prime\prime}$ (4 kpc) and  has a closer correspondence with the optical emission line gas than with the radio emission \citep{Massaro2009,Balmaverde2012,Hardcastle2012}. This implies the X-ray emission is thermal rather than non-thermal  \citep{Massaro2009}.  Deeper Chandra observations confirm that the X-ray gas is shock-heated 
\citep{Hardcastle2012}. The data are consistent with the hypothesis that the X-ray gas is part of a gaseous outflow driven by the radio source \citep{Hardcastle2012}. The kinematics and excitation of the optical emission line gas are also consistent with acceleration and heating by shocks \citep{Reynaldi2013}.

Chandra observations of the double-double source 3C293 (Sect.~\ref{s:rejuv}) are presented by \citet{Lanz2015}. The Chandra data show evidence for hot ($10^7$ K) shock-heated gas associated with the nucleus and the inner jets in the central CSS source. Combined with the observations of warm H$_2$ \citep{Ogle2010}, this suggests that the radio jet is driving shocks into both the hot and cold gas in the host galaxy \citep{Lanz2015}.

Observations of 7 low radio power CSS galaxies detected 4 and showed that one of them (1321+045) is in a cool core cluster \citep{KB2013,KB2014}.  1321+045 (z=0.263) has a projected size of 17 kpc and a radio power L$_{5 GHz} \sim 10^{25}$ W Hz\mone\ \citep{KB2013}.  This source does not show evidence for strong interaction with the environment; e.g., the optical emission lines are consistent with AGN photoionization, and the extended X-ray emission does not show any distortions which could be attributed to interaction with the radio source  \citep{KB2013}.

\citet{ODea2017}  compiled selected radio and X-ray properties of the nine CSS radio galaxies with X-ray detections so far. They found that 2/9 CSS radio galaxies show X-ray spectroscopic evidence for hot
shocked gas (3C303.1 \citep{ODea2006} and 3C305 \citep{Massaro2009,Hardcastle2012}) and 3 CSS sources (with 2 overlap)  show X-ray emission aligned with the radio source (3C237 \citep{Massaro2018}, 3C303.1 \citep{Massaro2010}, and 3C305 \citep{Massaro2009,Balmaverde2012,Hardcastle2012}). These detections of hot shocked gas and aligned emission in the X-rays are consistent with the results of numerical simulations \citep{Heinz1998,Bicknell2006,Sutherland2007} and suggest significant jet-mode feedback to the ISM (see Sect.~\ref{s:feedback}).  \citet{ODea2017} suggest that hot shocked gas may be typical of CSS radio galaxies due to their propagation through their host galaxies.
More data are needed (deep Chandra and XMM-Newton observations on a sample of CSS galaxies) in order to improve our understanding of the interaction of CSS radio galaxies with their host galaxies. 

\subsubsection{X-rays from PS and CSS quasars}\label{s:CSSQuasarXray}

About 13 PS and CSS quasars have been studied so far with Chandra and XMM \citep[e.g.,][]{Worrall2004,Siemiginowska2008,Salvesen2009,Migliori2012}.
The Chandra study  by \citet{Siemiginowska2008} includes new observations and summarizes previous work on 6 objects classified as PS quasars and  7 classified as CSS quasars. Two of the ``GPS quasars''  have large scale jets detected by Chandra (B2 0738+313 \citet{Siemiginowska2003}, although note that \citet{Stanghellini2001a} suggest it to be a core-jet source, and PKS 1127$-$145 \citet{Siemiginowska2002,Siemiginowska2007}) indicating that these are extended quasars with GPS-shaped nuclei. The X-ray spectra of the PS and CSS quasars are generally  well fit by an absorbed power-law \citep[e.g.,][]{Worrall2004,Siemiginowska2008,Salvesen2009}.
\citet{Siemiginowska2008} note that the power-law indexes of the PS and CSS quasars are steeper (median $\Gamma = 1.84$) than those of typical radio loud quasars ($\Gamma = 1.5$) \citep{Elvis1994,Richards2006}. If this is confirmed with a larger sample, it would suggest that there is an additional component to the X-ray emission in the PS/CSS sources that is not seen in the typical radio loud quasars (or vice versa). \citet{Siemiginowska2008} also find that the absorbing columns towards the GPS/CSS quasars are no more than $\sim$ few $\times 10^{21}$ cm\mtwo, suggesting that the higher absorbing columns seen in two high z GPS quasars \citep {Elvis1994} are not common. Observations of some PS/CSS quasars are consistent with relativistic beaming of at least some of the X-ray emission (e.g., 3C48 \citet{Worrall2004},  3C287 \citet{Salvesen2009}, and 3C186 \citet{Migliori2012}).

\subsubsection{X-ray--radio relations and the origin of the X-ray emission}\label{s:xray-radio}

See Sect.~\ref{s:nuclear} for an introduction to HERGs and LERGs.
In extended radio galaxies, at a given radio luminosity, the HERG galaxies have more luminous X-ray emission than the LERG galaxies \citep[e.g.,][]{Hardcastle2009}. This is interpreted to mean that the X-rays in the HERGs are produced in the accreting gas (e.g., a disk-corona system), while the X-rays in the LERGS (which are missing a bright accretion structure) are produced in the radio source, possibly in the base of the jet \citep[e.g.,][]{Hardcastle2009}. Though in the CSOs, the X-ray emission might be produced in the radio lobes \citep{Stawarz2008, Ostorero2010}.

\begin{figure}
\centerline{\includegraphics[width=11cm]{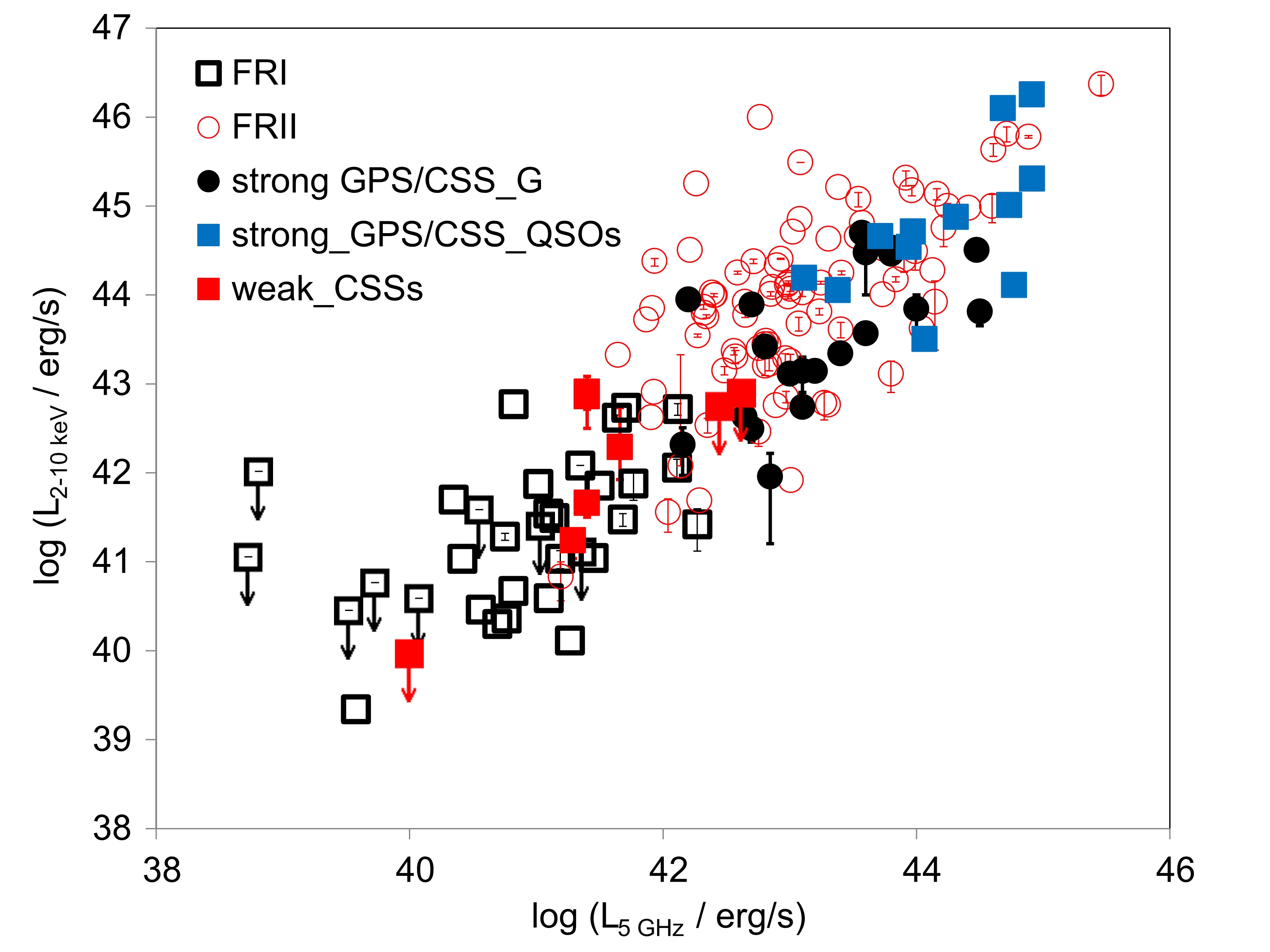}}
\caption{2--10 keV luminosity vs.\ 5 GHz luminosity for GPS, CSS, FRI and FRII radio sources from \citet{KB2014}. Strong and weak refer to high and low radio power, respectively.
\label{f:xray_radio}}
\end{figure}

\citet{KB2014} compiled X-ray and radio fluxes for GPS, CSS, FRI, and FRII radio sources in order to study relations on the radio-X-ray plane. The GPS and CSS sources lie on the same radio X-ray correlation as the FRI and FRII radio sources (extending 6 decades in X-ray power and 5 decades in radio power, Fig.~\ref{f:xray_radio}) consistent with \citet{Tengstrand2009}. Thus, there is continuity in the properties of the small and large sources.  

\begin{figure}
\centerline{\includegraphics[width=11cm]{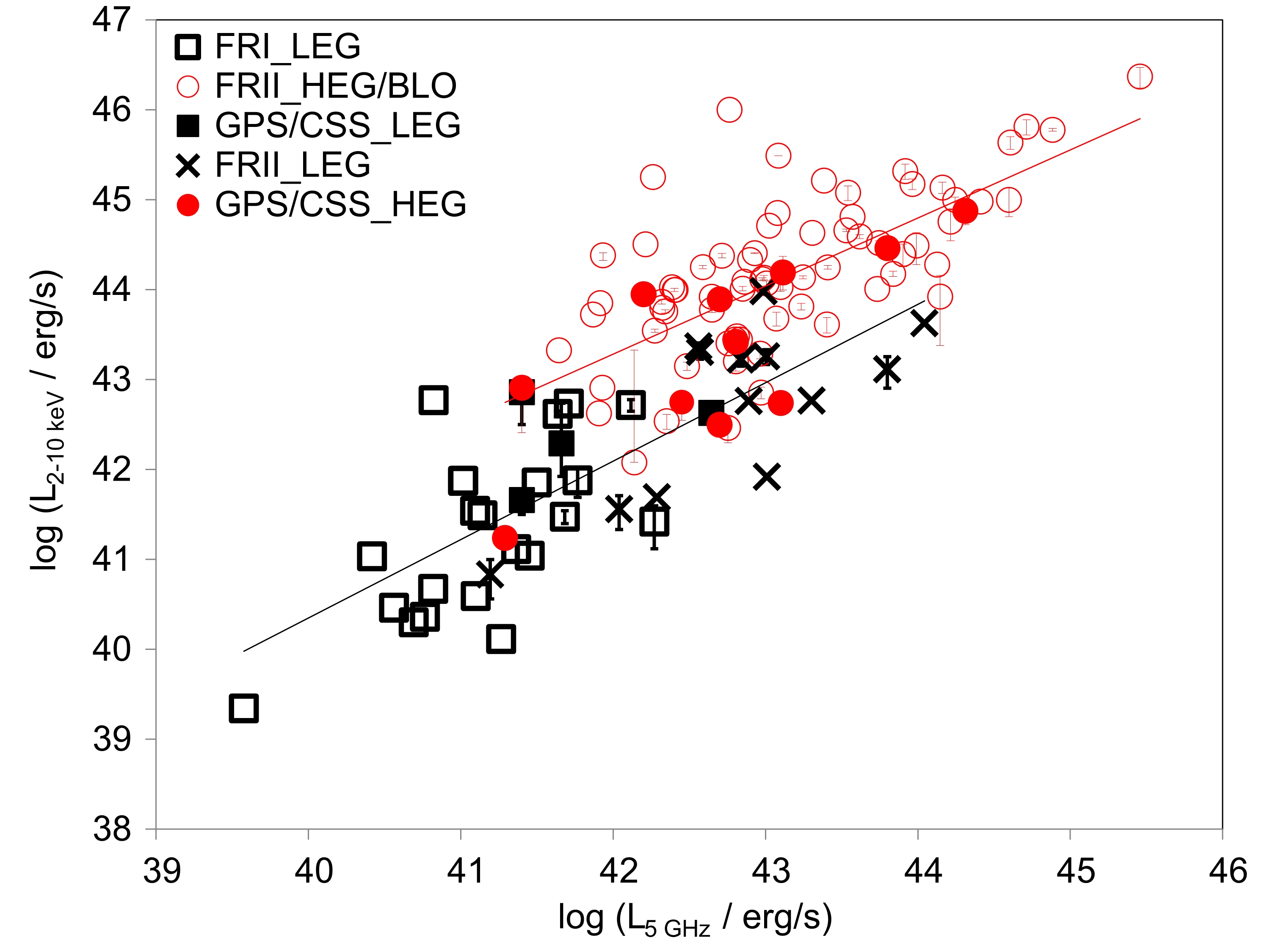}}
\caption{2--10 keV luminosity vs.\ 5 GHz luminosity for GPS, CSS, FRI and FRII radio sources from \citet{KB2014}. High Excitation (HERG, in this figure called HEG) and Low Excitation (LERG, in this figure called LEG) emission line objects are noted. 
\label{f:heg_leg}}
\end{figure}

In addition,  at a given radio power, the objects with high excitation emission lines have X-ray powers about an order of magnitude higher than the objects with low excitation emission lines (Fig \ref{f:heg_leg}). 
This indicates the presence of an additional source of X-rays in the HERG sources and is consistent with the X-ray emission being dominated by the accretion disk-corona in the high excitation galaxies  \citep[e.g.,][]{Hardcastle2009}. { Note that there are only 3 GPS/CSS LERGs in Figure \ref{f:heg_leg}, so it remains to be confirmed that the GPS/CSS LERGs show the same trend as the extended radio galaxies which are LERGs.}

Based on the above results, we suggest the following  scenario for X-ray emission in GPS and CSS sources. In LERG galaxies the X-ray emission comes from the radio source (via synchrotron or inverse Compton scattering), in HERG galaxies, the X-ray emission is dominated by the accretion structure (e.g., disk-corona). In the quasars, because the radio emission is beamed, but the accretion flow is not, the radio source will make a relatively larger contribution to the X-ray emission. If this scenario is correct, it will assist the interpretation of the relation between X-ray N$_{\rm H}$ and H{\sc i} N$_{\rm HI}$ (Sect.~\ref{S:NH-NHI}).

\subsection{ $\gamma$-ray emission}\label{s:gamma}

The status of $\gamma$-ray observations of PS and CSS sources was previously reviewed by \citet{Migliori2016a}.
Inverse Compton (IC) scattering of ambient photons by the relativistic electrons in lobes and cocoons \citep[e.g.,][]{Stawarz2008,Ito2011,Kino2013}, and jets \citep{Migliori2014} of compact radio sources are predicted to produce luminous $\gamma$-ray emission. Other mechanisms for producing $\gamma$-ray emission include bremsstrahlung \citep{Kino2007,Kino2009}   and synchrotron emission from protons and secondary electrons/positrons produced via the photo-pion cascade \citep[e.g.,][]{Kino2011}.
Thus, $\gamma$-ray emission is potentially an important diagnostic of conditions in the compact radio sources. 

\citet{Migliori2014} report upper limits on $\gamma$-ray emission from a sample of 6 GPS quasars and 7 CSS quasars which have X-ray observations from \citet{Siemiginowska2008}. \citet{DAmmando2016} report upper limits on a sample of 51 HFPs from the sample of \citet{Dallacasa2000}  from which obvious blazars have been removed as discussed by \citet{Orienti2008b}. 
Likely $\gamma$-ray detections include the CSOs PKS 1413+135 \citep{Gugliucci2005}  and 2234+282 \citep{An2016b}, and the CSS quasars 0202+149 \citep{An2016a}, 3C138 \citep{Zhang2020}, 3C286 \citep{An2017, Yao2020, Zhang2020}, and 3C309.1 \citep{Zhang2020}.  However, some may require confirmation that they are either true (unconfused) $\gamma$-ray sources or true (non-blazar) PS and CSS sources \citep[e.g.,][]{DAmmando2016,Migliori2016b}. 

 The detection of $\gamma$-ray emission from the nearby CSO PKS 1718$-$649 is consistent with IC scattering of ambient emission by the relativistic electrons in the compact lobes \citep{Migliori2016b}. The  $\gamma$-ray detection of the CSO PMN J1603$-$4904 also seems reliable \citep{Muller2014,Muller2016,Goldoni2016, Krauss2018}. {Two additional $\gamma$-ray detections in nearby, young CSOs have been reported - NGC 3894 \citep{Principe2020} and TXS 0128+554 \citep{Lister2020}. \citet{Principe2020} and \citet{Lister2020} note the young  kinematic ages ($< 100$ yr) of the CSOs with $\gamma$-ray detections and suggest that in these two CSOs, the $\gamma$-rays are produced in the inner jet rather than the lobes.  \citet{Kosmaczewski2020} reach similar conclusions.}

\subsection{Ultra-high energy cosmic rays from PS and CSS sources?}

Ultra-High Energy Cosmic Rays (UHECR) have energies above $10^{18}$ eV. UHECR in large-scale radio sources are discussed by \citet{Hardcastle2020}. Here we focus on compact radio sources. 
\citet{Elbert1995} noted that the CSS quasar 3C147 was in the direction of a UHECR event with energy $> 10^{20}$ eV. This led \citet{Farrar1998} to suggest an association between the highest energy UHECR and compact (e.g., GPS and CSS) quasars based on 5 events. This association was challenged by \citet{Hoffman1999} (but see the reply by \citet{Farrar1999}) and by \citet{Sigl2001}, but supported by \citet{Virmani2002}. We regard this result as requiring confirmation. \citet{Takami2011} show that UHECR protons might be accelerated to energies up to $10^{20}$ eV in the hot spots of CSOs.  If compact quasars (or any distant source) was found to be the origin of the  highest energy UHECR, this would have profound implications. The interaction of UHECR protons or heavy nuclei with the cosmic microwave background results in loss of the particle energy due to production of pions (the GZK effect, \citep{Greisen1966,Zatsepin1966}). So, UHECR with energies $> 10^{20}$ eV cannot have an origin more than about 100 Mpc from Earth  \citep[e.g.,][]{Dermer2009}. Thus, the association of  UHECR of energy $> 10^{20}$ eV with distant AGN would require a new particle or new physics. 
 
\section{Discussion}\label{s:disc}

\subsection{Implications for PS and CSS sources in selected AGN populations}

Compact and/or peaked spectrum radio sources are found in many populations of AGN. Here we discuss the presence and implications of these sources in several important populations. 

\subsubsection{PS and CSS sources in the infrared faint radio sources}
The Infrared Faint Radio Sources (IFRS) are selected to have a ratio of 20 cm flux density/3.6$\mu$m flux density $> 500$ and a 3.6 $\mu$m flux density $< 30 \mu$Jy \citep[e.g.,][]{Norris2006,Zinn2011}. This selection tends to find powerful radio sources at high redshift ($z \sim 2-4$) \citep[e.g.,][]{Collier2014,Singh2017,Orenstein2019}. Note that at these redshifts, the rest frame is closer to 1 $\mu$m and so 3.6 $\mu$m is not a rest-frame IR selection. Some of the IFRS are found to be GPS and CSS sources \citep[e.g.,][]{Middelberg2011,Collier2014,Herzog2015}. Since the IFRS are a subset of the high-z radio galaxies, it is not clear how representative the relative fractions of GPS and CSS sources are. \citet{Herzog2016} find that at least 18\% of the IFRS are GPS and CSS sources and suggest that this is consistent with the fractions in the general radio source population noted by \citet{ODea1998}. This could also be consistent with a lack of strong evolution in the relative fractions of PS and CSS sources in the radio source population out to redshift $z \sim 4$.

\subsubsection{The population of PS sources at mm wavelengths}\label{s:mm}
There are now several wide field surveys at high frequencies, e.g., AT20G \citep{Murphy2010} and Planck \citep{Planck2011I}. This allows us to address the issue of the contribution of PS sources to the population of sources at mm wavelengths (See Sect.~\ref{s:samples}). PS sources at mm wavelengths are of interest because of their potential contribution to observations of the cosmic microwave background \citep[e.g.,][]{DeZotti2005}. In addition, sources which peak at mm wavelengths are likely to be extremely compact and/or have extreme physical properties. 
As discussed in Sect.~\ref{s:samples}, there is a population of PS sources with spectral peaks above 5 GHz - the high-frequency peakers  \citep{Dallacasa2000,Dallacasa2002c,Hancock2009a,Orienti2012,Orienti2020}.  Many candidate HFPs  identified with quasars turn out to be blazars based on  variability of the shape of the radio spectrum (Sect.~\ref{s:variability_orientation}), 
 very compact radio size suggesting Doppler boosting \citep[e.g.,][]{Bolton2006,Orienti2012}, or high polarization from jet-dominated emission \citep[e.g.,][]{Orienti2008b}.

\citet{Hancock2009a} presents a sample of 656 candidate PS sources with observed peaks above 5 GHz, drawn from a sample of 4404 AT20G sources  with data at three frequencies. 
Follow-up studies on a sample of 21 candidates suggest that $75\%$ of the sources identified with galaxies are non-blazar PS sources, and $25\%$ of the sources identified with stellar objects are non-blazar PS objects \citep{Hancock2010}. The likely fraction of non-blazar PS sources in the AT20G depends on the identifications, but may be something like half of the 656 sources or $\sim 7\%$ of the AT20G sample of 4404 sources.

About $10\%$ of the compact extragalactic sources detected by Planck and listed in the Early Release Compact Source Catalog (ERCSC) have a peaked spectrum \citep{Planck2011I}.  However nearly all of these are identified with known blazars \citep{Planck2011I,Volvach2016}; though \citet{Massardi2016} suggest nine Planck sources are good HFP candidates. One of the nine, J114553$-$695404, is contained in a sample of Planck point sources studied by Rocha et al. (in preparation). The source shows a factor of $\sim 2$ variability in the Planck data and is probably a blazar.

The general lack of non-blazar PS sources in the Planck ERCSC is puzzling given their presence in the AT20G survey. One explanation could be the much poorer sensitivity of Planck. The flux density limit at 20 GHz for the AT20G sample is 40 mJy \citep{Murphy2010}, while the flux density limit for the ERCSC at 30 GHz is at least an order of magnitude higher \citep{Planck2011VII}. This would be consistent with the beamed emission from the blazars dominating over the unbeamed emission from the non-blazar PS sources. At this time it appears that study of the high frequency population of PS sources should focus on the AT20G sample. 


\subsubsection{Are CSS sources related to radio loud Narrow-Line Seyfert 1s?}\label{s:NLS1}

The properties of Narrow Line Seyfert 1s (NLS1s) are reviewed by \citet{Komossa2008} and \citet{Dammando2019}, and the latter has an emphasis on $\gamma$-ray emitting NLS1s. 
NLS1s have narrower Balmer lines (FWHM (H$\beta$) $< 2000$ km s\mone ) than regular Sy 1s \citep[e.g.,][]{Osterbrock1985, Osterbrock1987,Goodrich1989}. Other emission line properties tend to include weak  [OIII] $\lambda 5007$/H$\beta$ 
 and strong Fe II/H$\beta$ \citep[e.g.,][]{Osterbrock1985,Goodrich1989}.
NLS1s have less massive host galaxies \citep{Krongold2001} and less massive black holes \citep[e.g.,][]{Peterson2011,Mathur2012,Xu2012} than regular Sy 1s, but appear to be accreting near the Eddington limit \citep[e.g.,][]{Boroson2002,Xu2012}.
About 4.5 - 7\% of NLS1s are radio loud \citep{Komossa2006,Singh2018}.  The radio properties of RL NLS1s have been compared to those of CSS radio sources \citep{Oshlack2001,Komossa2006, Gallo2006,Yuan2008, Gu2010,Gu2015, Gu2016, Caccianiga2014, Caccianiga2017, Schulz2016, Berton2017, Singh2018,Yao2020}. It has also been suggested that the NLS1s are young AGN \citep[e.g.,][]{Grupe2000,Mathur2000} similar to a subset of PS and CSS sources.  

\citet{Caccianiga2014,Berton2016,Liao2020a} consider the possibility that the CSS/HERGs are the parent population for the beamed (flat spectrum) RL NLS1s. As \citet{Berton2016} note, the two populations are nominally hosted by different galaxies. CSSs are predominantly in elliptical galaxy hosts though there are a few examples of hosts with a disk component (Sect.~\ref{s:hosts}) and NLS1s are found in disk galaxies, sometimes  with a pseudobulge \citep[e.g.,][]{Crenshaw2003, Mathur2012, Kotilainen2016,Olguin2017}. However,  \citet{Berton2016} note that the host galaxies of the flat spectrum RL NLS1s are not yet well studied. { \citet{Dammando2019} note that the hosts have been determined for four $\gamma$-ray emitting NLS1s and two are hosted by elliptical galaxies. } If the flat spectrum RL NLS1s turn out to be in elliptical galaxies, similar to the CSS, this would support the hypothesis that the CSS/HERGs are the parent population of the flat spectrum RL NLS1s. It would also suggest that the flat spectrum RL and RQ NLS1s are different populations of objects. { Based on radio and X-ray properties,  \citet{Dammando2019} argues against a relationship between CSS and the $\gamma$-ray emitting NLS1s. }
Thus, the issue of whether the CSS sources are the parent population of the flat spectrum  RL NLS1s is still an interesting open question. Nevertheless, the similarity in the radio properties between CSS sources and RL NLS1s makes it clear that compact radio sources can form in a variety of AGN.

 
\subsection{What are the PS and CSS sources?}\label{s:what}
 
Here we provide updates on the main hypotheses for PS and CSS sources since the review of \citet{ODea1998}.
 
\subsubsection{Transient or episodic sources}\label{s:transient}

A strong motivation for considering short lifetimes for PS sources is the excess number of sources with sizes below 1 kpc over an extrapolation from large sizes \citep{ODea1997,Reynolds1997, Alexander2000,Marecki2003b,KB2010a}.  In addition, there is recent evidence that some compact sources are turning off and fading  \citep[e.g.,][]{KB2005,KB2006,Giroletti2005,Giroletti2008,Orienti2010e,Callingham2015}. \citet{KB2010a} suggest that many of the low luminosity CSS sources could be short-lived objects. A possible link between short-lived CSS sources and changing look AGN has  been suggested by \citet{Wolowska2017}. The double-double sources also provide strong support to the hypothesis of episodic activity in radio sources (Sect.~\ref{s:rejuv}). In addition, the existence of AGN with X-ray nuclei but no extended emission-line nebulae argues for AGN time scales of $\sim 10^5$ yr \citep{Schawinski2015}. 

Radiation pressure instabilities in accretion disks may provide the timescales needed for intermittent activity  \citep{Czerny2009}.  In this scenario,  activity lasts $\sim 10^3$--$10^4$ yr and episodes are separated by  $\sim 10^4$--$10^6$ yr \citep{Czerny2009, Wu2009a, Siemiginowska2010}. Jet disruption is another way of producing short-lived radio sources  \citep[e.g.,][]{Deyoung1991,Higgins1999, Wang2000, Wiita2004,Sutherland2007, Wagner2011,Mukherjee2016, Mukherjee2017, Mukherjee2018a,Mukherjee2018b,
Bicknell2018,Aditya2019}.  
If the jets in some sources disrupt (and the source fades quickly)  before the radio source propagates out of the core, the distribution of number vs source size \citep{ODea1997} can be reproduced \citep{Alexander2000,Kaiser2007,KB2010a,An2012a}.


\subsubsection{Frustrated sources}\label{s:frustrated}

One of the possible scenarios for the PS and CSS radio sources is that they are confined to their host galaxies through interaction with dense clouds in the host galaxy ISM \citep{vanBreugel1984a,Wilkinson1984,ODea1991}. This has been called the frustration scenario and is motivated by the observation that PS and CSS radio sources are much more asymmetric than the large-scale radio sources, suggesting interaction with a dense medium (Sect.~\ref{s:radiostructure}, Sect.~\ref{s:asymm}). As discussed in Sect.~\ref{s:xraycso}, \citet{Sobolewska2019} show that the 5 CSOs with very large X-ray derived N$_{\rm H}$ have a smaller radio size for a given 5 GHz luminosity than the less absorbed sources and suggest that the small, heavily absorbed sources are confined by a dense environment. Confinement of radio sources on small scales could also help to explain the excess of sources at small sizes. 

Table \ref{t:gas-mass} presents the available molecular gas masses. The objects in the Table are not a complete sample.  However, there is a broad range of masses  from $ 10^7$ to $10^{11}\,M_\odot$. The total gas mass required to confine a radio source is still uncertain and depends on the  mass and distribution of the individual dense clouds, though  rough estimates are that masses of at least $10^9$ to $10^{10}\,M_\odot$ would be required  \citep{DeYoung1993,Carvalho1994,Carvalho1998,ODea1998}. These gas masses are present in the PS and CSS sources, so it remains possible that some fraction of the sources are confined to the ISM of their host galaxies by interaction of the radio source with dense gas.

Simulations show that interaction of jets with large dense clouds can disrupt the jets and/or impede their progress \citep[e.g.,][]{Deyoung1991,Higgins1999,Wang2000,Wiita2004}. In  simulations of jets propagating through a clumpy ISM, the weaker jets can be confined for a long time, while the more powerful jets can break through (or around) the clouds \citep[e.g.,][]{Bicknell2006,Wagner2011, Mukherjee2016,Mukherjee2017,Mukherjee2018a,Mukherjee2018b,
Bicknell2018}. These simulations suggest that in sufficiently dense environments with mean densities of $\sim 300\ {\rm cm}^{-3}$, jets can remain frustrated for 1 to 2 Myr. It is also  possible that the sources which show intense star formation are in dense environments and might be frustrated Sect.~\ref{S:SFR}.
Note that some GPS sources show gaseous outflows suggesting that the jet-cloud interactions are removing gas from the galactic nucleus (Sect.~\ref{s:feedback}). 

\begin{figure}
\centerline{\includegraphics[width=11cm]{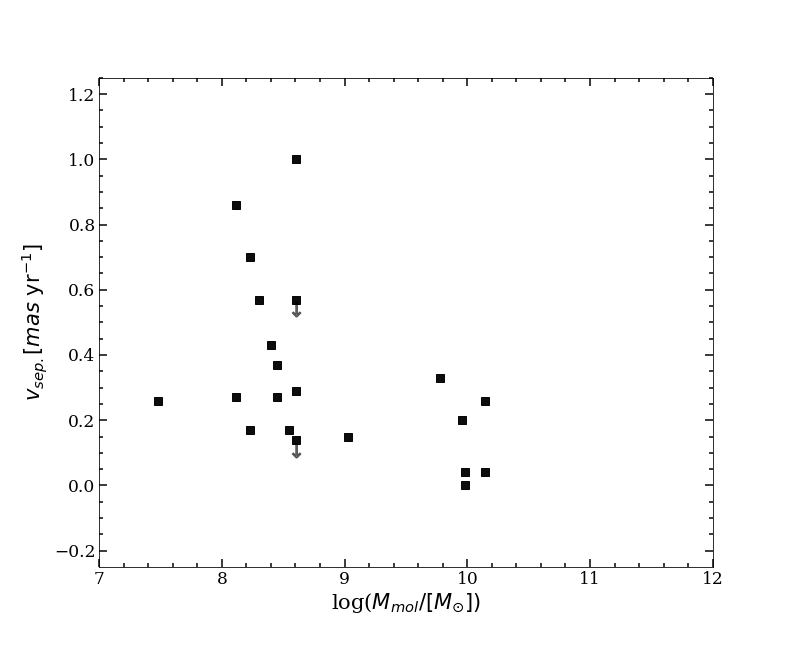}}
\caption{Proper motions of CSOs (Table \ref{t:proper_motion}) are plotted against the molecular gas masses (Table \ref{t:gas-mass})   for sources in common. For sources with multiple measurements of proper motion, all measurements are plotted. The CSOs tend to have higher proper motions in hosts with less molecular gas.
}
\label{f:proper-gas-mass}
\end{figure}

The proper motions of CSOs (Table \ref{t:proper_motion}) are plotted against the molecular gas masses (Table \ref{t:gas-mass}) in Fig.~\ref{f:proper-gas-mass} for sources in common. There is a trend for the CSOs to have lower proper motions in galaxies with larger molecular gas mass. The numbers are still small, so this should be confirmed with a larger sample. 
If confirmed, this would be consistent with the frustration scenario, i.e., the interaction with ambient gas clouds slows the propagation of compact radio sources.

\subsubsection{Imposters: Star-forming, radio-enhanced, compact sources?}\label{S:SFR} 

\citet{GW1991} suggest that intrinsically weak radio sources propagating through a dense ISM (possibly from a merger) will have their radio luminosity enhanced. \citet{ODea1998} noted that PS and CSS sources which are interacting with dense gas from a merger should also have significant star formation. This could result in  a population of intrinsically weak, but radio-enhanced, compact radio sources in hosts which are forming stars. 

\citet{Morganti2011,Tadhunter2011,Dicken2012} have noted the evidence for enhanced star formation in compact radio sources (see discussion in Sect.~\ref{s:IR}, and \citet{ODea2016}) and suggested that this scenario is indeed occurring.  They suggest that the enhanced radio emission lifts the compact sources above the sample flux density limits, so that  compact radio sources in star forming galaxies are  over-represented in radio selected samples. Here we call these sources imposters.  The radio enhancement  is qualitatively consistent with the excess of compact radio sources compared to an extrapolation of the number vs size relation from the extended radio galaxies  \citep{ODea1997}. Thus, the imposter scenario competes with the transient and frustrated source scenarios to explain the excess of compact sources. If the radio-enhanced sources are really intrinsically weaker AGN, then ratios of AGN bolometric luminosity to radio luminosity (for HERGs) should reveal this.  \citet{KB2014} show that the compact and large HERGs lie on the same correlation of X-ray luminosity with radio power. And a similar correlation is found for the compact and large LERGs. Thus, so far there is no evidence for enhancement of radio luminosity relative to AGN bolometric luminosity (as probed by X-ray emission). { One caveat to this argument is that in sources interacting strongly with their environments, it is possible that the X-ray emission will also be enhanced (Sect.~\ref{s:CSSx-ray}). If the contribution to the X-ray emission is significant, the X-ray emission is no longer a reliable proxy for the AGN luminosity.} 

{ 
As discussed in Sect.~\ref{s:BH}, the black hole fundamental plane plots nuclear radio power against a combination of nuclear X-ray luminosity and black hole mass. 
\citet{Wojtowicz2020} find that some CSOs have higher radio power than expected on the black hole fundamental plane, which they attribute to a higher radiative efficiency in the radio. }
 
Note that optical emission line (e.g., [OIII]) luminosity is problematic as a proxy for AGN luminosity if there is a significant contribution from jet-induced shocks \citep[e.g.,][]{Labiano2005,Labiano2009, Holt2009b, Shih2013,Reynaldi2013, Reynaldi2016} or if the emission line luminosity increases with time \citep[e.g.,][]{Vink2006}.

%

\begin{figure}
\centerline{\includegraphics[width=11cm]{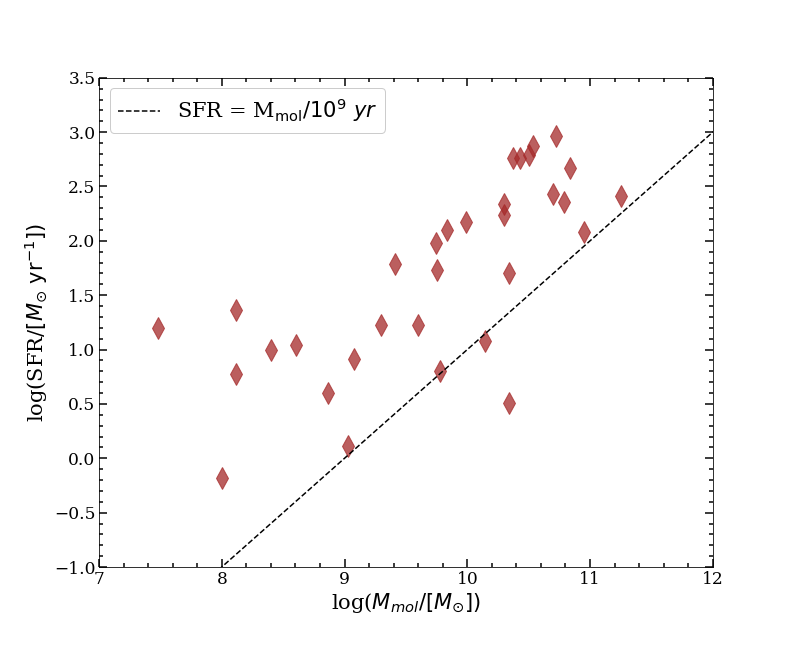}}
\caption{SFR from Table \ref{t:SFR} is shown against the mass of molecular gas from Table~\ref{t:gas-mass}. If there are more than one estimate for a given source, the estimates were averaged. The SFR lie above a line which assumes a gas depletion time scale of $10^9$ yr. This indicates that the PS and CSS sources have enhanced star formation. }
\label{f:SFR-gas-mass}
\end{figure}

We show SFR vs mass of molecular gas in Fig.~\ref{f:SFR-gas-mass}. Above a mass of $10^9\,M_\odot$, the SFR is correlated with the total gas mass as is generally found in star forming galaxies \citep[e.g.,][]{Young1986, Gao2004,ODea2008}. The gas depletion time scale is defined to be the total molecular gas mass divided by the SFR. The dotted line shows the SFR expected with a gas depletion time scale of $10^9$ yr which is typical in a broad range of star forming galaxies \citep[e.g.,][]{Young1986,Rownd1999,ODea2008}. The SFR of the PS and CSS sources lie above the line, showing that they have a relatively high star formation efficiency. Thus, star formation is enhanced in PS and CSS sources. We suggest that this is due to star formation triggered via interaction with the radio source. 

\subsubsection{Young, dynamic sources}
 
It is clear that while some PS and CSS sources may be transient (Sect.~\ref{s:transient}), radio-enhanced (Sect.~\ref{S:SFR}), or frustrated
 (Sect.~\ref{s:frustrated}), there is very likely a subset which are young, dynamic sources which will grow to become large radio galaxies and quasars. We list the arguments here. (1) The PS and CSS sources live in the same host galaxies as the large sources (Sect. \ref{s:hosts}). (2) There is continuity in the properties from small to large sources (e.g., Sect. \ref{s:xray-radio}). (3) There is  evidence so far for a dense clumpy ISM which could confine the radio source in  just a subset of PS and CSS (Sect.~\ref{s:hosts}). (4)  Models for compact radio source evolution provide a plausible scenario for growth of CSOs to become CSS and large scale FRIs and FRIIs and are consistent with currently available data (Sect.~\ref{s:models}). (5)  Observed proper motions of CSOs show the radio sources are indeed propagating through the host ISM (even in galaxies with substantial masses of molecular gas, Figure \ref{f:proper-gas-mass})  and give plausible dynamical ages which are consistent with spectral aging (Sect.~\ref{s:proper}).
  Below we discuss analytical models for propagation of young radio sources.
 
\subsection{Analytical models for compact source propagation}\label{s:models}
 
Numerical simulations of compact radio sources are discussed in Sect.~\ref{s:feedback}. Here we discuss analytical models because they can make predictions for radio source populations which can be compared with observations. 
Assuming that dense clouds in the ISM do not confine the radio source (Sect.~\ref{s:hosts},\ref{s:frustrated}), and the jet remains stable and long-lived, the compact radio sources will propagate through their host galaxies \citep[e.g.,][]{Begelman1999}.
 
Self-similar analytical dynamical models with a constant density 1 kpc core and a power-law decline in density on larger scales results in  radio power that increases with distance from the nucleus out to 1 kpc and then declines on larger scales  \citep{Snellen2000b,Alexander2000}.  Note, that while the assumption of self-similar behavior is very useful, numerical simulations suggest significant departures from self-similarity \citep{Carvalho2002,Cielo2014}. 
Improvements to the models include  adding a non-self-similar phase to the evolution on the smallest scales \citep{Alexander2006}, and  adding radiative losses \citep{Carvalho2003,Maciel2014}. Synchrotron losses are significant in compact sources and modify the evolution of the source \citep{Maciel2014}.  
Since the observational properties of hot spots are relatively easy to measure, some analytical models focus on the evolution of the hot spots \citep{Perucho2002,Kawakatu2006,Kawakatu2008}. 
An alternate set of models has been used to estimate the GPS luminosity function and the contribution of HFPs to high frequency surveys \citep{DeZotti2000,Tinti2006}.  In these models, the radio power always decreases with increasing size.

\citet{An2012a} compare the results of the dynamical models with data on a sample of 24 CSOs (radio power, separation between two hotspots, hotspot separation velocity, and kinematic age of the source). Overall, the properties of CSOs are consistent with the models. However, the excess of small ($<100$ pc) and young ($< 1000$ yr) CSOs is consistent with some sources not surviving to become large sources. 

\citet{Stawarz2008} include radiative processes in their dynamical model in order to calculate the broad-band SED of the sources. \citet{Ostorero2010} fit these models to eleven GPS sources with CSO radio sources. The fits are consistent with the X-rays being produced by IC scattering of ambient photons by the relativistic electrons in the radio lobes. \citet{Ostorero2010} note that correlation between X-ray absorbing column $N_{\rm H}$ and 21 cm absorption column density $N_{\rm HI}$ (Sect.~\ref{s:hosts}) is consistent with the X-ray emission originating in the radio lobes.

\subsection{The role of compact radio sources in AGN feedback}\label{s:feedback}
We note that AGN feedback is a broad topic, including different mechanisms (accretion disk-driven and radio jet-driven) and a wide range of size scales from galaxies to clusters of galaxies \citep[e.g.,][]{Croton2006, McNamara2007,McNamara2012, Fabian2012,Somerville2015, Harrison2018}. Here we focus on feedback to the host galaxy ISM from compact radio sources (CSO and CSS) (and see also \citet{Tadhunter2016a,Tadhunter2016b,Wagner2016,Hardcastle2020}).

We have previously discussed several lines of evidence for AGN feedback in CSO and CSS sources: (1) The asymmetric radio morphology of CSO and CSS sources suggests interaction with dense clouds in the host galaxy ISM (Sect.~\ref{s:radiostructure}, Sect.~\ref{s:asymm}). (2) The kinematics (and to some extent the excitation) of aligned optical emission line gas is consistent with shocks being driven by the radio source (Sect.~\ref{s:alignment}). (3) X-ray observations of shocked and aligned hot gas are consistent with shocks driven by the radio source (Sect.~\ref{s:CSSx-ray}). { (4) There is evidence for jet-induced star formation (Sect.~\ref{s:jet-SF}).}

There are  clear examples of multi-phase outflows in CSO and CSS radio sources. The outflows are seen  in neutral atomic hydrogen \citep[e.g.,][]{Morganti2003a,Morganti2005a, Morganti2005b, Morganti2013, Holt2008, Gereb2015, Maccagni2017,Morganti2018} (and possibly also \citep{Labiano2006, Vermeulen2006}). About 23\% of compact sources with an H{\sc i} absorption detection show a blue wing in the H{\sc i} with an outflow velocity $>$ 100 km s\mone\ \citep{Maccagni2017,Morganti2018}.  

In addition, outflows are seen in warm ionized gas seen in optical (mainly [O{\sc iii}] $\lambda 5007$)  emission lines \citep[e.g.,][]{Tadhunter2001, Holt2003a, Holt2006b, Holt2008,Buchanan2006, Nesvadba2007, Roche2016, Santoro2018, Liao2020a}, 
and IR lines \citep{Dasyra2011, Guillard2012}, and in absorption in the UV \citep{debreuck2000,Gupta2005}. About 22\% of a set of 68 PS and CSS sources with SDSS data show blue shifted wings in the [O{\sc iii}] line \citep{Liao2020a}. { \citet{Holt2008} detect warm outflows in 10 (71\%) of a sample of 14 compact sources,  and 8 of the 10 detections are in sources with outflows in H{\sc i}.  Thus, the detection rate for warm outflows seems to be higher in objects with outflows in H{\sc i}. }

The H{\sc i} outflows are seen in galaxies with a rich supply of gas which  can be transported outwards \citep{Morganti2005b,Guillard2012,Maccagni2017}.
Outflows of molecular gas are rare so far in compact radio sources, but are detected in PKS 1549$-$79 \citep{Guillard2012,Oosterloo2019} and in the ULIRG 4C 12.50 (PKS 1345+12) \citep{Dasyra2011, Dasyra2012, Dasyra2014, Guillard2012,Spoon2013,Fotopoulou2019}. A significant fraction of the H$_2$  in the outflow in 4C 12.50 has been heated to a temperature of about 400 K \citep{Dasyra2014}. In other compact radio sources with H{\sc i} outflows, the near-IR H$_2$ lines do not display the same kinematics as the out-flowing H{\sc i} and warm gas  \citep{Guillard2012}. However, the symmetric line width and excitation diagrams of the H$_2$ lines suggest that the radio source is shocking the molecular gas instead of entraining it in an outflow \citep{Ogle2010,Guillard2012}. 

There is evidence that these outflows are driven by the radio source rather than by radiation pressure. 
In CSS sources, the close spatial relationship between the [O{\sc iii}] $\lambda 5007$ emission and the radio source (alignment effect, Section 
\ref{s:alignment}) suggests that the radio source is driving the outflow. In some sources the out-flowing emission line gas is seen on scales which are too small to be driven by a kpc scale starburst wind \citep{Batcheldor2007,Santoro2018}. In 4C 12.50,  a radio component (hotspot?) 100 pc from the nucleus appears to be pushing an  H{\sc i} cloud with a blue-shifted velocity of 1000 km s\mone\  providing the direct link between the radio jet and the outflow \citep{Morganti2013}. In addition, the high velocity [O{\sc iii}] $\lambda 5007$ gas in 3C48 is associated with the base of the radio jet \citep{Stockton2007,Shih2014}. 

\citet{Best2000} showed that smaller radio sources had emission line nebulae with lower ionization, higher luminosity, and broader line widths than in larger radio sources, consistent with shocks driven by the radio source in the smaller sources.  These results are confirmed by subsequent studies \citep{debreuck2000,Moy2002, Inskip2002}, though \citet{Moy2002} suggest that the emission line region in the very smallest sources ($< 2$ kpc) are dominated by AGN photoionization because the size of the region shocked by the radio source is still very small.  Note that \citet{Best2000} defined small to be $< 125$ ($< 90$, correcting for cosmology) kpc, suggesting that the importance of shocks (and thus radio mode feedback) is not limited to the scale of CSS sources ($< 20$ kpc) and/or that the effects of feedback continue even after the radio source has propagated beyond the host galaxy. If feedback remains active while the radio source propagates for an additional 35 kpc beyond the CSS phase at a lobe advance speed of $10^4$ km s\mone\ \citep[e.g.,][]{ODea2002}, the additional time corresponds to $\sim 3 \times 10^6$ yr. 

{ The observed correlations between Eddington ratio or AGN bolometric luminosity and outflow properties such as outflow velocity or kinetic power in large samples of AGN 
\citep[e.g.,][]{Greene2005,Woo2016,Fiore2017,Rakshit2018}
support the hypothesis that the outflows are driven by radiation pressure. There is also a correlation between AGN bolometric luminosity and radio power  \citep[e.g.,][]{deBruyn1978,Baum1989,Xu1999}.  Thus, we might expect to see a correlation of outflow properties and radio power. However, the relationship with the radio source appears to be complicated  (\citet{Mullaney2013} and references therein).  }

\citet{Mullaney2013} studied [O{\sc iii}] $\lambda 5007$ line profiles using SDSS spectra of a sample of 24,264 optically selected AGN. They find the broadest [O{\sc iii}] $\lambda 5007$ line widths (i.e., fastest outflows) in the subset of AGN with compact ($< 2$ arcsec in size) radio sources with moderate radio power $P_{1.4 {\rm GHz}} \sim 10^{23}$--$10^{25}$ W Hz\mone\ \citep{Mullaney2013}. Using FWHM $> 500$ km/s as the criterion, 25\% of all AGN and 50\% of sources with radio power P$_{1.4} > 10^{23}$  W Hz\mone\ have  broadened [O{\sc iii}] lines indicative of outflow. However, see \citet{Woo2016} and \citet{Rakshit2018} for differing conclusions regarding the influence of radio power.  The detection of ionized outflows in a large sample of compact radio sources by \citet{Mullaney2013} supports earlier work that showed that compact radio sources display more extreme [O{\sc iii}] $\lambda 5007$  kinematics than larger radio sources \citep{Gelderman1994,Best2000,Holt2008} and indicates that compact radio sources do indeed drive outflows. Singha et al. (in preparation) point out that  the preference for outflows in {\it compact\/} radio sources found by \citet{Mullaney2013} might be because at least 80\%  of the radio sources in the low redshift ($z < 0.4$) sample considered by \citet{Mullaney2013}  are compact (see also Sect.~\ref{s:fr0_sources}). { At this point the statistics of outflows in compact vs.\ extended radio sources needs to be clarified.}

 The mass outflow rates and kinetic powers of the {\it ionized\/} outflows are generally too low to have a significant effect on star formation in their host galaxies \citep[e.g.,][]{Holt2006b,Holt2011,Tadhunter2007, Santoro2018,Santoro2020} { or even to shock-ionize the optical emission lines \citep{Santoro2020}.  However, the other components of the multiphase outflows can carry more mass and dominate the kinetic power of the outflows \citep[e.g.,][]{Holt2011,Hardcastle2012,Dasyra2014}.}
 The outflows of molecular gas tend to be more massive and carry more kinetic power, but do not seem to extend beyond the inner kpc of the host galaxy \citep[e.g.,][]{Dasyra2014,Morganti2020}.
 We note that the sources discussed here are at relatively low redshift ($z < 0.5$).  The outflow energetics scale with AGN luminosity \citep[e.g.,][]{Mullaney2013,Woo2016}. We speculate that during Cosmic Noon, when AGN luminosity is much higher, the outflows would be capable of significantly suppressing star formation. However, confirmation requires observations of compact radio sources at high redshift \citep[e.g.,][]{Nesvadba2007,Kim2013,Lonsdale2015,Patil2020}. 

Arguments based on numerical simulations and estimates of jet power also suggest that compact radio sources can drive outflows out to size scales of several times the radio source size  \citep{Zovaro2019a,Zovaro2019b}.
Numerical simulations which include radiative cooling of the gas show that ISM clouds which are shocked by the jet can collapse and cool and may form stars, consistent with the positive feedback required for jet-induced star formation \citep[e.g.,][]{Mellema2002, Fragile2004, Fragile2017, AD2008, Gaibler2012}. Nevertheless, AGN feedback must be globally negative in order to suppress star formation in galaxies. 
Numerical simulations have been used to explore the interaction of jets in compact radio sources with the ISM and the impact on both the jet and the ISM \citep[e.g.,][]{Steffen1997,Bicknell2006, Sutherland2007, Wagner2011, Cielo2014, Bicknell2018}. When the jet interacts with a distribution of small, dense clouds, the jet tends to split and follow the path of least resistance between the clouds producing a diffuse, spherical radio source \citep{Sutherland2007,Wagner2011,Bicknell2018}. If the jets are sufficiently powerful, they are able to accelerate and disperse clouds \citep{Wagner2011, Cielo2014, Mukherjee2016,Mukherjee2017, Mukherjee2018a,Mukherjee2018b, Bicknell2018}. Independent simulations find that a significant fraction of the jet kinetic energy and momentum are deposited in the galaxy ISM \citep{Wagner2011, Cielo2014}. { In practice, the observations indicate much lower fractions for the outflows of warm gas \citep[e.g.,][]{Holt2006b,Holt2011, Tadhunter2007, Santoro2018}. A higher fraction 
 of the jet energy (more comparable to the simulations) is deposited in the ISM in the full
 multiphase outflows  in three well-studied sources \citep{Hardcastle2012,Dasyra2014, Oosterloo2019}.}

\section{Future work}\label{s:future}

The motivation for studies of PS and CSS sources has been presented for ASKAP \citep{Norris2011,Allison2016}, LOFAR \citep{Snellen2009,Shimwell2019}, ngVLA \citep{Patil2018b}, and SKA \citep{Falcke2004b,Kapinska2015,Afonso2015}. \citet{Norris2011} note that the EMU survey with ASKAP will detect millions of GPS and CSS sources down to very faint levels, allowing the creation of large, complete samples. Deep surveys with MeerKAT will also add to the samples of CSS/PS sources at faint flux density levels \citep{Jonas2016,Mauch2020}. \citet{Allison2016} discuss systematic H{\sc i} absorption studies with ASKAP of PS/CSS sources which will probe the interaction of compact radio sources with their environments and the properties of H{\sc i} outflows driven by the radio source. The H{\sc i} observations will be extended  with the more sensitive SKA \citep{Morganti2015a}.
\citet{Snellen2009} highlight 3 areas to be addressed with LOFAR: determining the origin of the spectral turnover (FFA/SSA), the search for extended emission from previous cycles of activity, and discovery of $z>6$ GPS sources which are compact because of a dense environment (as suggested by \citet{Falcke2004b}). \citet{Shimwell2019} discuss the LOFAR detection of low luminosity peaked-spectrum sources which might be short-lived radio sources. \citet{Patil2018b} discuss the potential contributions of the ngVLA to three areas: sub-arcsec resolution radio imaging, determining spectral ages, and constraining the spectral turnover. \citet{Kapinska2015} highlight the ability of SKA to detect many radio sources (over a large range of redshift and luminosity) at all stages of the life cycle, allowing comprehensive statistical studies. \citet{Afonso2015} discuss the detection of powerful, young AGN at high redshift (the first generation of AGN) with SKA and the effects of inverse Compton scattering of the hot, bright CMB. 

Here we list some interesting areas for future research.
\begin{itemize} 
\item {\bf Origin of PS and CSS Sources} In this review, we discuss four scenarios for the origin of the PS and CSS sources. It is possible that they each describe a subset of the population. We would like to know the relative contributions to the population, and if possible determine the origin of individual objects.  It will be important to measure SFR and molecular gas mass in a large sample of objects. The ratio of AGN bolometric luminosity to radio luminosity is also an important diagnostic. The distribution of sizes and proper motions for a large sample of CSOs will help constrain radio source propagation models.

\item {\bf Radio Sources Driving Shocks} Deep Chandra and XMM-Newton observations of CSS sources to search for hot shocked gas will constrain the impact of the radio source on the host galaxy ISM. 

\item {\bf Jet Induced Star Formation} High resolution UV imaging to search for jet-induced star formation should also be done. 

\item {\bf $\gamma$-rays} There are  confirmed $\gamma$-ray detections of several CSOs and CSS quasars at the present time. It is important to continue the search for  $\gamma$-ray emission from PS and CSS sources. 

\item {\bf AGN Feedback} Compact radio sources may contribute to AGN feedback by driving  gaseous outflows. Given the high detection rate for H{\sc i} absorption, this is a promising approach. Detections of H{\sc i} in outflow by e.g.,  ASKAP, SKA, WSRT, uGMRT and MeerKAT should be followed up by VLBI H{\sc i} absorption measurements to locate the components which are driving the outflow, and combined with VLBI measurements of proper motion of the radio components. This will provide a unique data set for advancing our understanding of the interaction of the radio sources with their environments. 

\item {\bf Numerical Simulations} High resolution, 3-D numerical simulations may help to clarify the propagation of compact radio sources and their interaction with their environments.

\item {\bf Mechanism for the Spectral Turnover} Determinations of the radio spectrum at multiple wavelengths near the spectral peak will allow the nature of the spectral turnover to be determined (FFA vs.\ SSA) and its relationship to the environment of the radio source. 

\item {\bf Larger Samples} There are currently many samples of PS and CSS sources selected in different ways. Current and future deep radio surveys  over a broad range of wavelengths by, e.g., ASKAP, AT, uGMRT, LOFAR, SKA, (ng)VLA, WSRT and MeerKAT will allow construction of large, uniform samples of PS and CSS sources. 

\item {\bf Low-luminosity sources and formation of jets} {The lower end of the luminosity distribution of CSS and PS sources, including those classified as FR0s, overlaps with those of Seyfert galaxies. The radio structures of Seyfert galaxies also tend to be of sub-galactic dimensions. It would be important to clarify the precise properties of the central engine and its environment that give rise to a wide range of observed properties of these compact sources of varying luminosity and different host galaxies. }

\end{itemize}


\begin{acknowledgements}
We are very grateful to colleagues who provided thoughtful comments on the manuscript and/or figures for the review, especially T.~An, G.~Bicknell, J.~Conway, W.~Cotton, D.~Dallacasa, C.~Duggal, M.~Giroletti, Y.~Gordon, M.~Kunert-Bajraszewska, A.~Marecki, G.~Migliori, R.~Morganti, D.~Mukherjee, P.~Ogle, M.~Orienti, T.~Oosterloo, A.~Polatidis and L.~Stawarz. We thank our referee, C.~Tadhunter, for helpful comments which improved the paper. 
CO is grateful to M.~Singha   for help with some of the figures and to R.~Antonucci for interesting discussions. 
CO is grateful to the Natural Sciences and Engineering Research Council (NSERC) of Canada for support. This research has made use of NASA's Astrophysics Data System Bibliographic Services. This research has made use of the NASA/IPAC Extragalactic Database (NED), which is funded by the National Aeronautics and Space Administration and operated by the California Institute of Technology.
\end{acknowledgements}

\bibliographystyle{spbasic-FS}      
\bibliography{css-review}   

%
%

\end{document}